\documentclass[11 pt]{article}%
\usepackage{bbm}
\usepackage[longnamesfirst]{natbib}
\usepackage{amssymb}
\usepackage{amsmath}
\usepackage{amsthm}
\usepackage{setspace}
\usepackage[colorlinks]{hyperref}
\usepackage{graphicx}
\usepackage{caption}
\usepackage{subcaption}
\usepackage{enumitem}
\usepackage[margin=1in]{geometry}
\usepackage{verbatim}
\usepackage{sectsty}
\usepackage[svgnames]{xcolor}
\usepackage{subfiles}
\usepackage{xargs}
\usepackage{titlesec}
\usepackage[colorinlistoftodos,textsize=tiny]{todonotes}
\usepackage[capitalize,nameinlink]{cleveref}
\usepackage[normalem]{ulem}
\usepackage[font={small,it},justification=justified]{caption}
\usepackage{xcolor}
\usepackage{pgf}
\usepackage{tikz}
\usepackage{amsfonts}%
\providecommand{\U}[1]{\protect\rule{.1in}{.1in}}
\setlist{noitemsep,parsep=4pt,partopsep=0pt,topsep=0pt}
\hypersetup{
pdftitle={Q-learning with biased policy rules},
pdfauthor={Olivier Compte},
citecolor=DarkBlue,
bookmarksnumbered=true,
urlcolor=Indigo,linkcolor=DarkBlue
}
\theoremstyle{remark}

\theoremstyle{plain}

\renewcommand{\epsilon}{\varepsilon}
\usetikzlibrary{patterns, decorations.pathreplacing, arrows.meta,calc}
\let \savenumberline \numberline
\def \numberline#1{\savenumberline{#1.}}
\makeatletter
\renewcommand\@seccntformat[1]{\csname the#1\endcsname.{\hskip.7em\relax}}
\makeatother


\let\oldfootnote\footnote
\renewcommand\footnote[1]{\oldfootnote{\hspace{.5mm}#1}}
\titlespacing\section{0pt}{10pt plus 2pt minus 2pt}{4pt plus 2pt minus 2pt}
\titlespacing\subsection{0pt}{6pt plus 2pt minus 2pt}{2pt plus 2pt minus 2pt}
\titlespacing\subsubsection{0pt}{6pt plus 2pt minus 2pt}{0pt plus 2pt minus 2pt}
\titlespacing{\paragraph}{  0pt}{  0.5\baselineskip}{  1em}
\newcommand{\appendixref}[1]{\hyperref[#1]{Appendix \ref{#1}}}
\definecolor{dark-red}{rgb}{0.4,0.15,0.15}
\definecolor{dark-blue}{rgb}{0.15,0.15,0.75}
\definecolor{medium-blue}{rgb}{0,0,0.5}
\usetikzlibrary{shapes,arrows,automata,fit}
\tikzstyle{info}=[circle,thick,draw=black,fill=black!25,minimum size=4mm]
\tikzstyle{uninfo}=[circle,thick,draw=black,fill=white,minimum size=4mm]
\tikzstyle{inforecog}=[circle,line width=1mm,draw=black!50,fill=black!25,minimum size=4mm]
\tikzstyle{uninforecog}=[circle,line width=1mm,draw=black!50,fill=white,minimum size=4mm]
\tikzstyle{traded}=[draw, line width=1mm]
\tikzstyle{recog}=[draw=black!50, line width=1mm]
\sectionfont{\color{DarkRed}}
\subsectionfont{\color{DarkRed}}
\subsubsectionfont{\color{DarkRed}}
\newcommandx{\nageeb}[2][1=]{\todo[linecolor=blue,backgroundcolor=blue!25,bordercolor=blue,#1]{#2}}
\newcommandx{\andreas}[2][1=]{\todo[linecolor=black,backgroundcolor=black!25,bordercolor=black,#1]{#2}}
\newcommandx{\navin}[2][1=]{\todo[linecolor=red,backgroundcolor=red!25,bordercolor=red,#1]{#2}}
\begin{document}


\begin{titlepage}
\title{Endogenous Barriers to Learning}
\date{May 20th 2025}
\author{Olivier Compte\thanks{Affiliation: Paris School of Economics, 48 Boulevard
Jourdan, 75014 Paris, and Ecole des Ponts Paris Tech \href{mailto:olivier.compte@gmail.com}{\color{dark-blue}olivier.compte@gmail.com}}\thanks{These ideas where
partly developped working on blotto games with Alessandra Casella. I thank
her, as well as Evan Friedman, Philippe Jehiel, the editor and two anonymous referees for their very useful comments.}}
\maketitle
\begin{abstract}
\noindent
Building on the idea that lack of experience is a source of errors but that
experience should reduce them, we model agents' behavior using a stochastic
choice model (logit quantal response), leaving endogenous the accuracy of their choices. In
some games, higher accuracy leads to unstable logit-response
dynamics. Starting from the lowest possible accuracy, we define the barrier to learning as the maximum accuracy
which keeps the logit-response dynamic stable (for all lower accuracies). This defines a limit quantal response equilibrium. We apply the concept to
centipede, travelers' dilemma, and 11-20 money-request games and to
first-price and all-pay auctions, and discuss the role of strategy
restrictions in reducing or amplifying barriers to learning.
\textbf{Keywords:}  Learning, Bounded Rationality, Stochastic choice
\textbf{JEL Classification Codes:} C72, D83, D90
\end{abstract}
\thispagestyle{empty}
\end{titlepage}

\setstretch{1.1}

\clearpage
\setcounter{tocdepth}{2} \thispagestyle{empty} \clearpage
\setcounter{page}{1}

\onehalfspacing

\section{Introduction}

Playing a Nash Equilibrium (NE) requires that each player comes to play a best
response to others' behavior. One original motivation for studying Quantal
Response Equilibria (QRE) rather than exact equilibria is that learning to
play a best response may be hard to accomplish, or require enough experience.
As \citet{mckelvey95} put it in their seminal work,

\textquotedblleft\textit{as a player gains experience playing a particular
game and makes repeated observations about the actual payoffs received from
various action choices, he/she can be expected to make more precise estimates
of the expected payoffs from different strategies.}"

In other words, when a player faces a stationary environment, we should expect
that the precision of her estimates improves with accumulated experience, and
that her Quantal or logit response eventually converges to a best response.
This paper pursues this line of thought, analyzing games in which the
strategic interaction may raise obstacles to learning from experience: we
observe by example that as the accuracy in payoff evaluations rises, the
logit-response dynamics may become unstable, thus potentially fueling
strategic uncertainty and hurting in return the accuracy of payoff estimates.
We propose a notion of limit QRE where the precision of payoffs evaluations is
obtained endogenously: starting from the lowest precision, we gradually
increase precision up to the limit level where the logit-response dynamic
ceases to be stable.

We illustrate this notion with several well-studied games where it is
well-known that participants deviate from the NE solution (the centipede game
(\citet{rosenthal81}), the traveller's dilemma (\citet{basu94}) and the 11-20
money request game (\citet{arad12})), and we evaluate how well our solution
performs. We also analyze auction games with a continuum of types and actions.
In these games, the strategy set is a space of functions (from an ex ante
perspective). This richness of the strategy set is an obstacle to learning, as
the performance evaluation of each feasible strategy is clearly out of reach.
On such strategy sets, the Quantal response is not even well-defined, unless
one adopts an interim perspective, defining quantal response conditional on
each type, or agent-QRE (\citet{mckelvey98}, \citet{goeree02}). A possible
path consists in exogenously restricting the set of strategies that players
actively consider or explore, i.e., strategies on which players are assumed to
accumulate experience and gather performance-related data. We consider a
family of linear bidding strategies and illustrate that all-pay auctions are
more subject to unstable dynamics than first-price auctions, and that richer
strategy sets do not necessarily improve stability, even when the strategy set
includes the Bayesian solution.

Formally, we start from the classic logit choice model, parameterized by a
real number $\beta\geq0$, with higher $\beta$ meaning higher precision. For
any game and any distribution $p$ over action profiles, we define the
\textit{logit-response function} $\phi_{\beta}(p)$ that maps the set of
mixed-action profiles to itself. The fixed points of this map are Quantal
Response Equilibria (QRE).


Having defined the logit-response function $\phi_{\beta}$, we consider its
long-run properties and examine the iterations $\phi_{\beta}^{(n)}=\phi
_{\beta}\circ\phi_{\beta}^{(n-1)}$, i.e., the logit-response dynamics. Our
main hypothesis is that, if, for a given $\beta$, a QRE $p_{\beta}$ is locally
stable under logit-response dynamics,\footnote{That is, if starting from a
small neighborhood of $p_{\beta}$, the\ iterations $\phi_{\beta}^{(n)}(p)$
converge to $p_{\beta}$.} then there is scope for improving the quality of the
player's response, i.e., scope for increasing the precision $\beta$ by a fixed
small increment ($\nu$).

We use this hypothesis to define by induction a sequence of (locally stable)
QRE ($\beta_{k},p_{k})$. We start from the lowest precision\textit{ }%
$\beta_{1}=0$\textit{ }and uniform distribution $p_{1}$. Then for $k>1$, we
define $\beta_{k}=\beta_{k-1}+\nu$ and $p_{k}=\phi_{\beta_{k}}^{(\infty
)}(p_{k-1})$ up to the point where, possibly, for some finite $K,$
$\phi_{\beta_{K+1}}^{(\infty)}(p_{K})$ ceases to converge to a locally stable
QRE.
In the latter case, the induction process defines a limit QRE precision
$\beta^{\ast}=\beta_{K}$ and a limit QRE distribution $p^{\ast}=p_{K}$, or
more succinctly, a \textit{limit precision }and a \textit{limit distribution}%
.

\textbf{A game-specific prediction.} Our approach determines a barrier to
learning in the sense that it defines a maximum level of precision
$\beta^{\ast}$ below which convergence is guaranteed.
In particular, we do not think of $\beta^{\ast}$ as an exogenous
characteristic of the agents which would apply across all games, as is
frequently done in experimental work (for the sake of not having too many
degrees of freedom in picking game-specific noise). Rather, we think of it as
an endogenous level of precision with which alternative options are compared,
for the game under consideration\textit{.} Some games have stable dynamics at
all levels of precision, and for these games, we expect that agents play it
with a very high $\beta^{\ast}$ (i.e., learn to play a Nash equilibrium).
Other games may be subject to unstable dynamics even for low $\beta$'s, and
for these games we expect agents to play it with a low $\beta^{\ast}$. So we
do not expect the same $\beta^{\ast}$ to apply across all games. The method
allows us to discriminate between games and assess the degree to which
learning is difficult.


\textbf{A robust limit distribution.} The limit distribution $p^{\ast}$ is
robust in several ways. First, scaling up all payoffs does not affect
convergence properties of the logit function $\phi_{\beta}$, so rescaling
payoffs does not affect $p^{\ast}$ (in contrast\textit{, }the limit precision
$\beta^{\ast}$ can be hard to interpret because it depends on how payoffs are scaled).

In the applications we consider, we also examine the robustness to
coarsening/refining the strategy space. For example, considering a centipede
game where the cake rises exponentially over time, we investigate the effect
of the number of dates at which players can exit, or whether the game allows
for simultaneous exits. We find that the limit distribution is essentially
unaffected by these timing considerations. Similar robustness is obtained for
the traveler's dilemma and the 11-20 game, where we modify the set of claims
that each player can make.


We also check the robustness of $p^{\ast}$ to exogenous errors in how players
select alternatives.
Specifically, we introduce errors assuming that when a player targets an
alternative, she trembles and implements it with noise, selecting
\textquotedblleft nearby" alternatives with positive probability.\footnote{Our
trembles are in the spirit of \citet{nash53}, \citet{carlsson91} and the
global game approach (\citet{carlsson93}). Trembles are generally used as a
selection device, with \textit{trembles vanishing.} Here we are interested in
\textquotedblleft trembles" that need not be of arbitrarily small magnitude.
Errors could also come from an exogenous bound on precision in evaluating
expected gains, stemming from upper limit on sampling, for example.} This
defines a \textit{game over targets}, to which we can apply our limit~QRE
notion. We obtain a new limit precision $\widehat{\beta}^{\ast}$ and a limit
distribution over targets, which in turn, given the trembles, generates a
distribution $\widehat{p}^{\ast}$ over the alternatives.
We shall see in examples that the limit precision typically increases (i.e.,
$\widehat{\beta}^{\ast}>\beta^{\ast})$. However we also find that, so long as
trembles are not too large,\footnote{With large exogenous trembles, the
logit-response dynamics of the game over targets may become stable at all
$\beta$. In that case, the outcome is fully driven by the exogenous noise
structure. See \citet{compte18}
\href{http://www.parisschoolofeconomics.com/compte-olivier/Chapter19Unraveling.pdf}{\color{dark-blue}Chapter~19}%
, for examples.} the induced distribution over alternatives is mostly
unchanged (i.e., $\widehat{p}^{\ast}\simeq p^{\ast}$). In other words, the
inclusion of trembles, which can be seen as an exogenous device to limit
learning (as then players, by assumption, can never play the exact best
response), does not constitute an additional barrier to learning: exogenous
and endogenous limitations to learning are substitutes.
%

\textbf{A hierarchical adaptation process. }Since the seminal work of
\citet{maynard72}, convergence to Nash has often been examined through the
lens of dynamic processes where adaptation is slow (as in replicator dynamics
with a large population) or eventually slow (as in learning models -- see
\citet{fudenberg09} for a review.) In these learning models, players
accumulate experience indefinitely, and experience is used at any point in
time to estimate the value of each available action under the presumption that
the environment is stationary, but that presumption may be erroneous in games,
as play may cycle.

In contrast, our model links the issue of building precise value estimate to
the issue of convergence to a stationary environment. Precision may rise, but
only insofar as play is stationary. Limit precision can then be viewed as
stemming from a \textit{hierarchical adaptation process}, whereby, for a given
$\beta$, behavior adapts fast (i.e., the logit-response dynamics with fixed
$\beta$), while the precision parameter $\beta$ adapts slowly (upward -- so
long as behavior is stationary).

Keeping this hierarchical adaptation hypothesis, alternative models of
stochastic choice could be considered, for example based on the distribution
of payoff differences between alternatives (rather than expected payoff
differences), with errors stemming from limits on sample size, of expected
size $\beta$ (in the spirit of \citet{osborne98} or \citet{danenberg22}).
Stochastic choice could also be belief-based, with players subject to noisy
beliefs (\citet{friedman05}, \citet{friedman22}). Alternative assumptions
regarding behavioral adaptation could also be made, of the replicator dynamic
kind for example. This would in general raise the stability of the behavioral
adaptation process, leading to higher limit precision.

Whatever assumptions one makes regarding stochastic choice and behavioral
adaptation, our hierarchical adaptation process offers a simple tool to
discriminate between games (in terms of the difficulty in learning to play an
equilibrium) and make a single (possibly stochastic if limit precision is
finite) behavioral prediction.

\textbf{General implications for empirical work.} From an applied perspective,
we see several virtues to our approach. First, the logit parameter is often
used as a free instrument to fit the data, with a common logit $\beta$
calibrated to the panel of games considered. We see no reason for the common
logit restriction: if the data contains games for which learning to play a NE
is easy, we should expect a higher $\beta$ for these games. Our game-specific
prediction allows this. Furthermore, it involves no calibration of a free
parameter, as $\beta$ is endogenized for each game.

Second, our limit-distribution predictions are not sensitive to a rescaling of
all payoff magnitudes. This lack of sensitivity is consistent with the data
reported in \citet{mckelvey00} and it cannot be explained with a fixed logit
parameter. The robustness of our limit-distribution prediction to the
coarseness/fineness of the strategy set is also consistent with the data from
\citet{mckelvey92} and \citet{nagel98} (who examine the same centipede with
twice as many nodes).

Finally, when the limit precision is finite, choice is stochastic. So we
obtain behavioral dispersion for free, and this dispersion is unlikely to
disappear with more experienced players. In most of the games we study here,
this behavioral dispersion is large and consistent with the data, without
having to introduce additional frictions or exogenous heterogeneity
(heterogenous types, subjective beliefs or cognitive levels, as in
\citet{rogers09} -- or more generally the level-k literature, see
\citet{crawford2013}). This does not mean that there are no additional
frictions or heterogeneity: so long as exogenous sources are not too large,
the robustness of our prediction to trembles implies that identifying the
exact source of dispersion, i.e., whether it comes from exogenous frictions or
endogenous barriers to learning, is a difficult task.

\textbf{Extension.} While we value endogenizing behavioral dispersion with
homogenous players, introducing cognitive heterogeneity may sometimes be
justified. Some games may be cognitively more demanding than others, or some
strategies may be more appealing or focal than others, and in the spirit of
level-k and its many variants, naive or focal play may persist for a fraction
of the players. This naive play may then persistently shape the responses of
the more sophisticated ones.

We extend our approach to accommodate the possibility of naive play,
endogenizing the response of sophisticated players. We illustrate this in
Section~\ref{1120game}. We consider a variant of the 11-20 request game (the
costless-iteration version proposed by \citet{arad12}), for which our limit
distribution does not fit well with the data when players are assumed
homogenous. We introduce some heterogeneity, assuming that players are either
non-strategic (choosing 20) or strategic (choosing strictly less than 20). In
the spirit of the reputation literature, we fix the proportion $\pi$ of
non-strategic players, and endogenize the behavior of the strategic players by
finding the limit distribution (given $\pi$), obtaining a much better fit with
the data. We believe that this method of assigning a subset of strategies to a
strategic type (and endogenizing the distribution over the subset) may prove
useful in other contexts.


\textbf{Related Literature.} The centipede game, the traveler's dilemma and
the 11-20 money request game have led to extensive experimental work. Let us
mention \citet{mckelvey92} and \citet{nagel98} for the centipede game (see
also the literature review in \citet{garciapola20}); \citet{capra99} for the
traveler's dilemma and \citet{goeree17} for the 11-20 game. These applications
offer good examples of departures from the Nash prediction. At the heart of
the explanation for departures from Nash equilibrium are frictions coming from
evolutionary pressures (\citet{rand12}), the presence of behavioral types
(\citet{mckelvey92}), cognitive heterogeneity and level-k thinking
(\citet{nagel95}), imperfect learning or various forms of stochastic choice
(\citet{mckelvey95}). These frictions produce strategic uncertainty adequately
calibrated to explain the data observed.
Our contribution is in endogenizing a minimal level of strategic uncertainty,
characterized by our limit distribution $p^{\ast},$ that must show up in these
games from learning/stability considerations only. Except for a variant of the
11-20 game, the limit distributions obtained are consistent with the data.
Obviously, independently of stability considerations, other relevant sources
of noise arising from limited data processing, computation errors, exogenous
trembling, payoff uncertainty, or misspecified beliefs may be relevant in
explaining the data. However, our analysis of the games defined over targets
(where we combine exogenous trembles and endogenized quantal response)
suggests some robustness of our analysis: while these exogenous sources
generally improve the stability of the logit-response dynamics, the resulting
combined strategic uncertainty remains unaffected by the magnitude of the
exogenous errors.

We also study first-price and all-pay auctions with continuum of types and
actions. \citet{anderson02} examines a family of games with a continuum of
actions (including a continuous version of the traveler's dilemma). Auctions
have been examined through the QRE lens in \citet{anderson98},
\citet{goeree02} and \citet{camerer16}. \citet{anderson98} study an all-pay
auction with no types. \citet{goeree02} consider a first-price auction with a
small number of types and study the agent-based version of QRE where
stochastic play is defined independently for each type. \citet{camerer16} also
study an agent-based version of QRE. In contrast, we adopt an ex ante
perspective. We define a grid of a priori given (linear) strategies and derive
the limit~QRE of the induced game. The method allows us
to address the issue of learning in complex Bayesian environments (see also
\citet{casella25}), and identify in this respect a sharp contrast between
first price and all-pay (where instability prevails when strategies are
linear).

Strategy restrictions have been used by \citet{compte01}, in auctions, and
more systematically in \citet{compte18}, to model moderately sophisticated
agents. These restrictions are reminiscent of \citet{simon55}'s notion of
\textit{considered set}. While decision theory explores the related notion of
consideration set (see \citet{manzini14} for example), viewing it as a subset
of alternatives randomly generated prior to choice, our motivation for
strategy restrictions is closer to that of \citet{simon55}, who views the
considered set as a key aspect of what agents learn about. Similarly, we view
restrictions as defining the range of strategies on which players accumulate experience.



Using learning to discriminate between games echoes the work of \citet{roth95}
and \citet{erev98}, who compare speeds of learning across games, for a given
reinforcement learning mechanism. For example, \citet{erev98} (page~887)
proposes to study \textquotedblleft\textit{learning in games using simple
general models}" and see \textquotedblleft\textit{how particular games and
economic environments influence the dynamic of learning}". Here, we see the
convergence properties of $\phi_{\beta}^{(\infty)}$ as a simple all-purpose
vehicle for studying barriers to learning, a vehicle that we apply across
different well-known games. Regarding the response function $\phi_{\beta}$
assumed, we acknowledge that a more gradual behavioral adaptation in the
spirit of replicator dynamics or those studied in \citet{hopkins99,hopkins02}
and \citet{turocy05}, for example, would improve the convergence properties of
the response function. This would naturally tilt the limit precision
$\beta^{\ast}$ towards higher levels and induce limit distributions $p^{\ast}$
that are closer to Nash equilibria (see the working paper version
\citet{compte23} for examples).

Our paper also relates to the question of Nash equilibrium selection
(\citet{harsanyi88}). We focus here on games with a unique equilibrium
outcome, asking whether that outcome is easily learned. In games with several
candidate equilibrium outcomes, our method can be used as a device that
selects the one that is more easily learned (if any): starting from low
precision and gradually raising precision so long as the logit-response
dynamic is stable, our process eventually selects a unique equilibrium outcome
if precision can be raised without bound -- otherwise it selects a
distribution that may fall short of an equilibrium. We follow this path in \citet{compte25}.

Finally, let us mention the rational inattention literature as another vehicle
for endogenizing precision, based on a trade-off between the costs and
benefits of cognitive effort (see for example \citet{matejka15}). Due to
cognitive costs, players are inattentive and act as if they were trembling
away from the exact best response, with tremble size driven by payoff
consequences. For fixed trembles, our analysis can be performed and bounds on
precision stemming from our stability considerations could be derived. We
leave for further research the analysis of a model that would, \textit{in
addition}, for a fixed cognitive load, incorporate private incentives to
reduce tremble size.

The paper is organized as follows. Section 2 provides some key definitions and
the theoretical contribution. Section 3 deals with applications. Section 4
summarizes the main findings and concludes.

\section{Definitions}

\subsection{Quantal response functions and equilibrium.}

Consider first a decision problem between alternatives $k\in A$. Call $u^{k}$
the expected payoff associated with $k$, $\overline{u}$ the maximum payoff and
$u=(u^{k})_{k\in A}$ the vector of expected payoffs. With fully accurate
estimates of consequences, the decision maker would choose an alternative
achieving $\overline{u}$. With less precise estimates or lesser experience,
other actions may be played with positive probability. Quantal Response
assumes that the frequency $p^{k}$ with which $k$ is played depends on the
vector of payoffs $u,$ according to a \textit{quantal response function}
parameterized by a precision parameter $\beta$. Under the classic
\textit{logit formulation}, this yields a probability vector
\[
p\equiv h_{\beta}(u)\text{ where }p^{k}\text{ is proportional to }\exp\beta
u^{k}\text{.}%
\]
In the \hyperref[app1]{Appendix}, we briefly explore another functional form
capturing a notion of satisficing.


Turning to games where each player $i$ chooses $k_{i}\in A_{i}$, we let
$p_{-i}\in\Delta(A_{-i})$ denote the distribution over strategies used by
other players, $u_{i}(p_{-i})$ the vector of expected payoffs associated with
each alternative $k_{i}\in A_{i}$. Finally, for any profile $p\in\times
_{i}\Delta(A_{i})$, we let $u(p)=(u_{i}(p_{-i}))_{i}$ denote the profile of
payoff vectors induced by $p$. A game is characterized by its value function
$u$.

A \textit{Quantal Response Equilibrium (QRE) }is then defined as the solution
of the system%
\[
p_{i}=h_{\beta}(u_{i}(p_{-i}))\text{ for all }i
\]
or, in vectorial form, as a solution to
\begin{equation}
p=\phi_{\beta}(p) \label{eq}%
\end{equation}
where $\phi_{\beta}\equiv(\phi_{\beta}^{i})_{i}$ and $\phi_{\beta}%
^{i}(p)\equiv h_{\beta}(u_{i}(p_{-i}))$. We shall refer to $\phi_{\beta}$ as
the logit response. We denote by $\Sigma_{\beta}$ the set of QRE at $\beta$,
by $\Sigma$ the set of all QRE and by $G$ the graph of $\Sigma$. At $\beta=0$,
the QRE has each player mixing uniformly of their alternatives. Let $p^{0}$
denote that strategy profile. We denote by $G^{0}$ the (generically unique)
branch of $G$ that starts at $(\beta,p)=(0,p^{0})$.

One interpretation of quantal response is that it corresponds to a static
short-cut for modelling players with limited experience: when learning is
incomplete, the assessment of which strategy is the best performing one is
subject to errors, and $\beta$ is an accuracy parameter meant to characterize
experience.

The general view is also that absent exogenous limitations on experience
accumulation, and so long as play converges and remains stationary, the
performance evaluations of each action should become more accurate, i.e.,
$\beta$ should rise, and behavior should eventually approximate Nash
Equilibrium play. This paper is about qualifying that view, suggesting that,
in some games, there may be endogenous limits to learning, that is, a largest
$\beta^{\ast}$ for which $\phi_{\beta}$ has good convergence properties for
all $\beta<\beta^{\ast}$.

\subsection{Logit-response dynamics, stability and limit QRE.}

For any $\beta$ and $p$, the logit-response dynamics is defined by induction
on $n$ by
\[
\phi_{\beta}^{(n)}(p)=\phi_{\beta}(\phi_{\beta}^{(n-1)}(p)).
\]
We refer to the sequence $Q_{\beta}(p)\equiv(p,...,\phi_{\beta}^{(n)}(p)...)$
as a path. A path need not converge, but if it does, that is, if the limit
$\phi_{\beta}^{(n)}(p)$ exists as $n\rightarrow\infty$, we denote the limit
\[
\overline{\phi}_{\beta}(p)\equiv\lim_{n\rightarrow\infty}\phi_{\beta}%
^{(n)}(p)
\]
Clearly$\ \overline{\phi}_{\beta}(p)$ solves (\ref{eq}) so it is a QRE.
Furthermore, we are interested in the local stability properties of these
dynamics. For any $p$ and $\varepsilon$, $B_{\varepsilon}(p)$ refers to an
$\varepsilon$-neighborhood of $p$.

\textbf{Definition 1:} \textit{We say that an equilibrium }$p\in\Sigma_{\beta
}$ is\textit{ locally }$\beta-$\textit{stable if there exists }$\varepsilon
>0$\textit{ and }$B_{\varepsilon}(p)$\textit{ such that all paths from
}$B_{\varepsilon}(p)$\textit{ converge to }$p$\textit{. We denote by }%
$\Sigma_{\beta}^{\ast}$ \textit{the set of such }QRE\textit{.}


Call $\overline{r}_{\beta,p}$ the largest norm of the eigenvalues of the
Jacobi matrix associated with $\phi_{\beta}(p)$. As is well-known, a
sufficient condition for local stability is that $\overline{r}_{\beta,p}<1$.
Our main hypothesis is that so long as $p\in\Sigma_{\beta}^{\ast}$, there are
evolutionary pressures towards more precise performance evaluations, i.e. a
(small) raise in $\beta$.
This hypothesis motivates the following definition, with increments set to a
fixed small scalar $\nu$.

\textbf{Definition 2:} Set $\nu>0$. \textit{Starting from }$\beta_{1}%
=0$\textit{ and }$\overline{p}_{1}\equiv p^{0}$\textit{, we construct by
induction on }$k$\textit{ the (longest) sequence (}$\beta_{1},\overline{p}%
_{1}),...(\beta_{k},\overline{p}_{k})...$\textit{ having the following three
properties: (i)~}$\beta_{k}=\beta_{k-1}+\nu$\textit{; (ii)~}$\overline{p}%
_{k}=\overline{\phi}_{\beta_{k}}(\overline{p}_{k-1})$\textit{ and
(iii)~}$\overline{p}_{k}\in\Sigma_{\beta_{k}}^{\ast}$\textit{. We call it an
evolutionary path. If the path is finite, say of length }$K$\textit{, we set
}$\beta_{\nu}^{\ast}=\beta_{K}$\textit{ and }$p_{\nu}^{\ast}=\overline{p}_{K}%
$\textit{ and refer to }$\beta_{\nu}^{\ast}$\textit{ and }$p_{\nu}^{\ast}%
$\textit{ as the limit (QRE) precision and limit (QRE) distribution
respectively.}

At $\beta_{1}=0$, choice probabilities are not responsive to payoff
differences, so trivially, the logit response function is globally stable
(i.e. converging to $\overline{p}_{1}=p^{0}$ from any initial $p)$. As $\beta$
rises, choice probabilities become sensitive to payoff differences, possibly
making the logit response dynamic \textit{locally unstable. }By construction,
the induction proposed in Definition 2 may stop when $\overline{\phi}%
_{\beta_{k}}(\overline{p}_{k-1})$ does not exist or when $\overline{p}_{k}$ is
not locally stable. Having set $(\beta_{1},\overline{p}_{1})$, the
evolutionary path is well-defined for any game and any $\nu>0$. If the
sequence is infinite, it selects a Nash equilibrium. If the sequence is finite
it selects a locally stable QRE.

We are interested in the limit precision and distribution for arbitrary small
$\nu$. In simulations, we set $\nu$ at 1\% of the maximum payoff. We check
numerically that the limit distribution is robust to further decrease in $\nu$.

Our formal statement below consists in linking the limit precision to
properties of the continuous branch of the graph $G^{0}$ (starting at
$\beta=0$). First we define the continuous selection $(\beta,p_{0}(\beta))$
defined from $\beta=0$ up to a possible turning point $\overline{\beta}^{0}$
(i.e., a point of the branch where $\beta$ would start decreasing -- this
happens when $G^{0}$ is S-shaped). Then we define:
\[
\overline{\beta}\equiv\sup\{\beta,\beta\leq\overline{\beta}^{0}\text{ and
}\overline{r}_{\beta^{\prime},p_{0}(\beta^{\prime})}<1\text{ for all }%
\beta^{\prime}\leq\beta\}
\]
Proposition 1 below shows that the evolutionary path must remain on $G^{0}$ up
to at least $\overline{\beta}$. Formally, we define $\underline{\beta}^{\ast
}=\underline{\lim}_{\nu\searrow0}\beta_{v}^{\ast}$ and $\overline{\beta}%
^{\ast}=\overline{\lim}_{\nu\searrow0}\beta_{v}^{\ast}$. We have:

\textbf{Proposition 1:} $\underline{\beta}^{\ast}\geq\overline{\beta}%
$\textit{. Furthermore if the game has a unique QRE for all }$\beta$\textit{,
}$p_{0}(\beta)$\textit{ is uniquely defined and }$\overline{\beta}^{\ast}%
\leq\inf\{\beta,\overline{r}_{\beta,p(\beta)}>1\}$\textit{.}

For games that have a unique QRE for all $\beta$, Proposition 1 implies that
the limit precision is essentially determined by $\overline{\beta}%
$.\footnote{For generic games, we expect $\overline{r}_{\beta,p(\beta)}$ to be
strictly increasing at $\overline{\beta}$, in which case $\overline{\beta
}=\inf\{\beta,\overline{r}_{\beta,p(\beta)}>1\}$, hence $\underline{\beta
}^{\ast}=$ $\overline{\beta}^{\ast}=\overline{\beta}$.} For games with
multiple QRE, the evolutionary path must remain on the continuous branch of
$G^{0}$ until $\overline{\beta}$ is reached. Then, the evolutionary path may
end there, or possibly jump to another locally stable continuous selection
$p(\beta)$ of the graph $G$. If there are several candidate paths, we cannot
exclude the possibility that the path selected by the logit response dynamic
(i.e. condition (ii)) depends on the size of $\nu$, though we suspect this
issue is non-generic. We leave this for further research.

\textbf{Comments.} 1. \textit{Thick barriers.} For some games, there may be an
intermediate range $I=(\beta^{\ast},\overline{\overline{\beta}})$ of values of
$\beta$ for which $\phi_{\beta}$ has unstable dynamics, while for all
$\beta<\beta^{\ast}$ and $\beta>\overline{\overline{\beta}}$, $\phi_{\beta}$
has locally stable dynamics. In this case, the game has a locally stable Nash
equilibrium (and all QRE for high $\beta$ are locally stable). Yet, this Nash
equilibria may be hard to learn if the quality of expected payoff estimates
only \textit{gradually} improves from $\beta=0$. The size of $I$ can be
interpreted as the thickness of the barrier to learning.

2. \textit{Best-response cycles.}\textbf{ }For any $p$, denote by
$\overline{a}(p)$ the (set of) action(s) that get(s) the largest weight. To
get an idea of the extent of instability when $\beta>\beta^{\ast}$, we
consider the path $Q_{\beta}(p^{\ast})$ induced by $p^{\ast}$ and construct
the sequence of best-performing actions along that path, i.e., the sequence
$(\overline{a}^{(n)})_{n}$ where $\overline{a}^{(n)}=\overline{a}(\phi_{\beta
}^{(n)}(p^{\ast}))$. We refer to it as the best-response cycle.

3.\textit{ Mixed strategy equilibria.} We shall consider games where limit
precision is bounded despite the existence of a pure strict Nash equilibrium.
In some of the applications we consider (11-20 request game and all-pay
auctions), there are no pure strategy equilibria. For such games, generically,
limit precision is necessarily bounded. The reason is that for high enough
precision, QRE are locally unstable under logit-response dynamics (See the
Appendix). This bound on precision implies that the limit distribution cannot
converge to Nash play (except for non-generic games where uniform mixing is a
Nash equilibrium -- hence a QRE at all precision levels).


\subsection{Trembling and games over targets}

We interpret the limit precision $\beta^{\ast}$ as an endogenous barrier to
learning in the sense that it provides an upperbound on the accuracy with
which the performance of alternatives are evaluated, resulting in a limit
distribution $p^{\ast}$ that falls short of a Nash equilibrium.

In many economic environments, it is conceivable that other sources of errors
are relevant, for example reflecting the players' misperceptions of the
environment. Such errors may also contribute to departure from Nash play (of
the game without misperceptions). We wish to understand the robustness of the
limit distribution $p^{\ast}$ to these exogenous errors, and check whether and
how these exogenous barriers to learning compete with the endogenous barrier
we propose.

Formally we define a \textit{game over targets} $\kappa_{i}\in K_{i}$. We
assume that when a player targets $\kappa_{i}$, she actually implements it
with noise, selecting \textquotedblleft nearby" alternatives $k_{i}$ with
probability $\pi_{\kappa_{i}}^{q}(k_{i})$, where $q$ is a parameter
characterizing dispersion ($q=0$ means no errors).

The shape of the distributions $\pi_{\kappa_{i}}^{q}$ and the notion of
\textquotedblleft nearby strategies" may depend on the application considered.
It may require a natural topology on the strategy space, or it may reflect a
player's strategic thinking about the problem. In applications where there is
a well-defined distance $|k_{i}-\kappa_{i}|$ between the target $\kappa_{i}$
and the alternative $k_{i}$, we define $\pi_{\kappa_{i}}^{q}$ as the
probability distribution satisfying%
\begin{equation}
\pi_{\kappa_{i}}^{q}(k_{i})=q^{|k_{i}-\kappa_{i}|}\pi_{\kappa_{i}}^{q}%
(\kappa_{i})\text{ for all }k_{i}\text{,} \label{dyn1}%
\end{equation}
where $q\in\lbrack0,1)$ implies an exponentially decaying weight on
alternatives away from the target.

Given an exogenous noise structure characterized by $q$ and a profile of
targets $\kappa=(\kappa_{i})_{i}$, one can compute the induced distribution
over alternatives and the induced expected values%
\[
\widehat{u}_{i}^{q}(\kappa)=E_{k}[u_{i}(k)|\kappa,q]
\]
This defines, for each $q$, a game over targets.\footnote{Note that the value
$\widehat{u}_{i}^{q}(\kappa)$ is computed ex ante, taking into account the
tremble that $i$ herself is subject to. Player $i$ is implicitly gathering
information about the effect of selecting a particular target, and this
evaluation thus takes into account one's trembling.} When $q=0$, we are back
to the original game.

When studying the limit~QRE of the game defined over targets, we obtain for
each $q$ a limit precision $\widehat{\beta}_{q}^{\ast}$ and a limit
distribution over targets, which in turn, given the trembles, generates a
distribution $\widehat{p}_{q}^{\ast}$ over the alternatives. In other words,
the distribution $\widehat{p}_{q}^{\ast}$ is partially driven by the exogenous
trembles (given the targets chosen) and partially driven by the
value-estimation errors (characterized by $\widehat{\beta}_{q}^{\ast}$)
associated with the expected performance of targets. We shall see in examples
that trembles typically raise $\beta^{\ast}$ (i.e., $\widehat{\beta}_{q}%
^{\ast}>\beta^{\ast})$, but leave the induced distribution over alternatives
mostly unchanged (i.e., $\widehat{p}_{q}^{\ast}\simeq p^{\ast}$), so long as
trembles are not too large.


\subsection{An example}

To conclude this section, we provide a simple example of a symmetric game with
a unique pure strategy equilibrium, for which the limit~QRE does not converge
to it. The game has four strategies, with payoff structure as reported below,
with $\theta<1$. \begin{table}[h]
\caption{A four-action game}%
\centering
\scalebox{0.95}{
\begin{tabular}
[c]{c|cccc}
& 1 & 2 & 3 & 4\\\hline
1 & 0,0 & 0,2 & 2,0 & $\theta$,0\\
2 & 2,0 & 0,0 & 0,2 & 0,1\\
3 & 0,2 & 2,0 & 0,0 & 0,2\\
4 & 0,$\theta$ & 0,0 & 2,0 & 1,1
\end{tabular}
}\end{table}One should think of these payoffs as expected gains, not easily
observable by players. Otherwise, it may be too obvious that action 1 is
weakly dominated, and that as a consequence, iterated elimination of weakly
dominated strategies leads to action 4.

The three first strategies have a Rock-Scissors-Paper structure. For any
$\theta<1$, (4,4) is a pure Nash equilibrium of the game, and since strategy 1
is weakly dominated, for $\beta$ large enough, there is also a unique (stable)
quantal response equilibrium, which puts most weight on action $4$.

We observe that for any $\theta\geq0.5$, the limit QRE has bounded precision.
Intuitively, when $\theta$ is high enough and $\beta$ is large enough (but not
too large), action 1 gets sufficiently high weight that action 2 becomes a
best response, fueling instability. We report below limit distributions and
precision for various values of $\theta$. When the limit precision is bounded,
we also report the precision $\overline{\overline{\beta}}$ above which the
logit-response dynamics starts being stable again.

\begin{table}[h]
\caption{limit distribution and precision}%
\centering
\scalebox{0.8}{
\begin{tabular}{c|cccccc}
$\theta$ & $p^*_1$ & $p^*_2$ & $p^*_3$ & $p^*_4$ & $\beta^*$ & $\overline{\overline{\beta}}$\\\hline
0.49 & 0 & 0 & 0 & 1 &  $\infty$ &-- \\
0.5 & 0.23 & 0.19 & 0.15 & 0.43 & 3.02&3.24 \\
0.7 & 0.26 & 0.22 & 0.18 & 0.34 & 2.62& 7.24 \\
0.9 & 0.28 & 0.23 & 0.19 & 0.3 & 2.42 &29\\
\end{tabular}  }\end{table}

For precision levels $\beta\in(\beta^{\ast},\overline{\overline{\beta}})$, the
logit dynamic is unstable. If $\beta$ is above but close enough to
$\beta^{\ast}$, the best-performing strategy along the logit dynamic path
remains action 4. For larger $\beta$'s within $(\beta^{\ast},\overline
{\overline{\beta}})$, the best-performing strategy cycles as follows:
$4,2,3,4,2,3,4...$ Furthermore, the barrier to learning becomes thicker as
$\theta$ rises.

\section{Applications\label{Section3}}

\subsection{Centipede games\label{centipede}}

We consider two-player centipede games. Each player chooses an exit time
$t_{i}\in T_{i}$ where $T_{i}$ is a finite subset of $\mathcal{R}_{+}$. We let
$\tau=\min(t_{1},t_{2})$. The size of the pie to be shared is a function of
$\tau$, we denote it $S(\tau)$. The player who exits strictly before the other
gets a share $a(\tau)>1/2$, while the other gets the rest. We assume equal
sharing when $t_{1}=t_{2}$:%
\[
u_{i}(t_{i},t_{j})=%
\begin{cases}
a(\tau)S(\tau) & \text{ if }t_{i}<t_{j}\\
(1-a(\tau))S(\tau) & \text{ if }t_{i}>t_{j}\\
\frac{S(\tau)}{2} & \text{ otherwise}%
\end{cases}
\]

We examine various parametric specifications of $S(\tau)$ and $a(\tau)$. In
each case, we examine the robustness of predictions to timing (by varying the
sets $T_{i}$) and to trembles. We also compare the predictions with
experimental data.

\subsubsection{Linearly increasing pie\ and constant
sharing\label{sectionlinear}}

\begin{figure}[h]
\centering
\begin{minipage}{0.45\textwidth}
\centering
\includegraphics[scale=0.65]{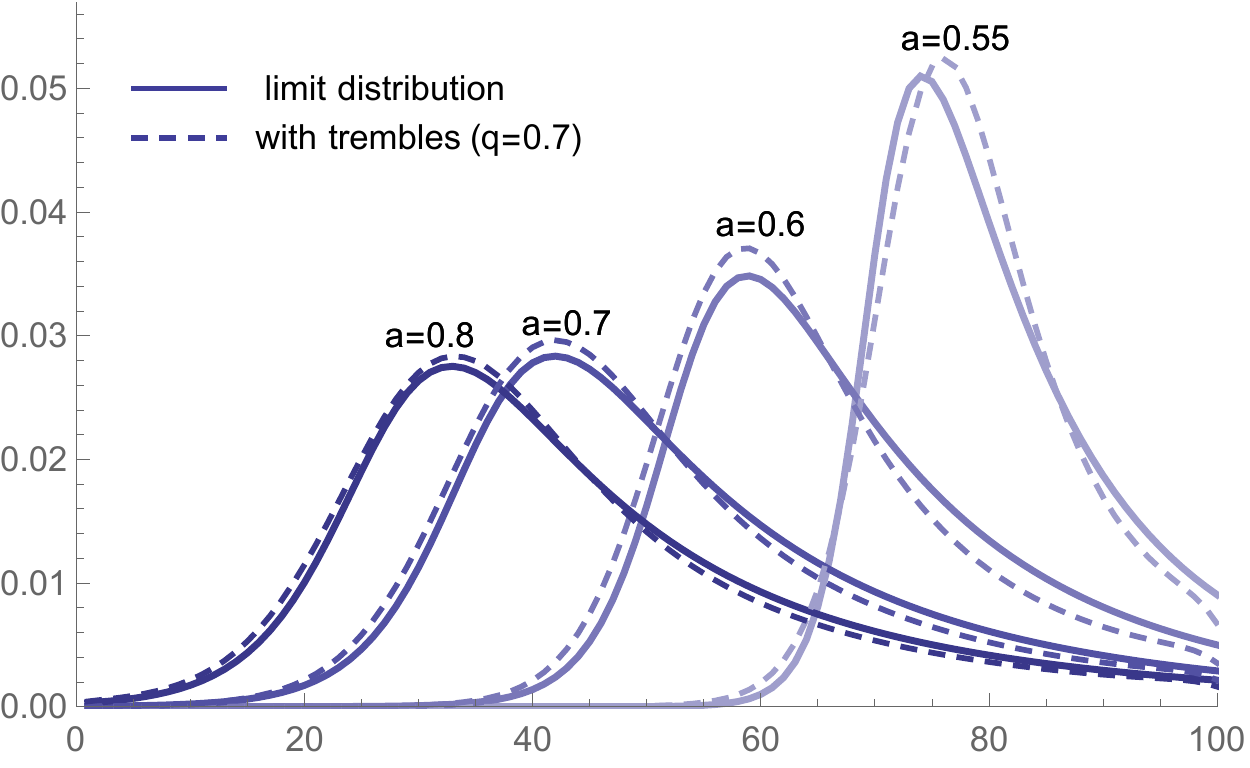}
\subcaption{Limit distributions}
\label{fig1}
\end{minipage}\hfill\begin{minipage}{0.45\textwidth}
\centering
	 \includegraphics[scale=0.6]{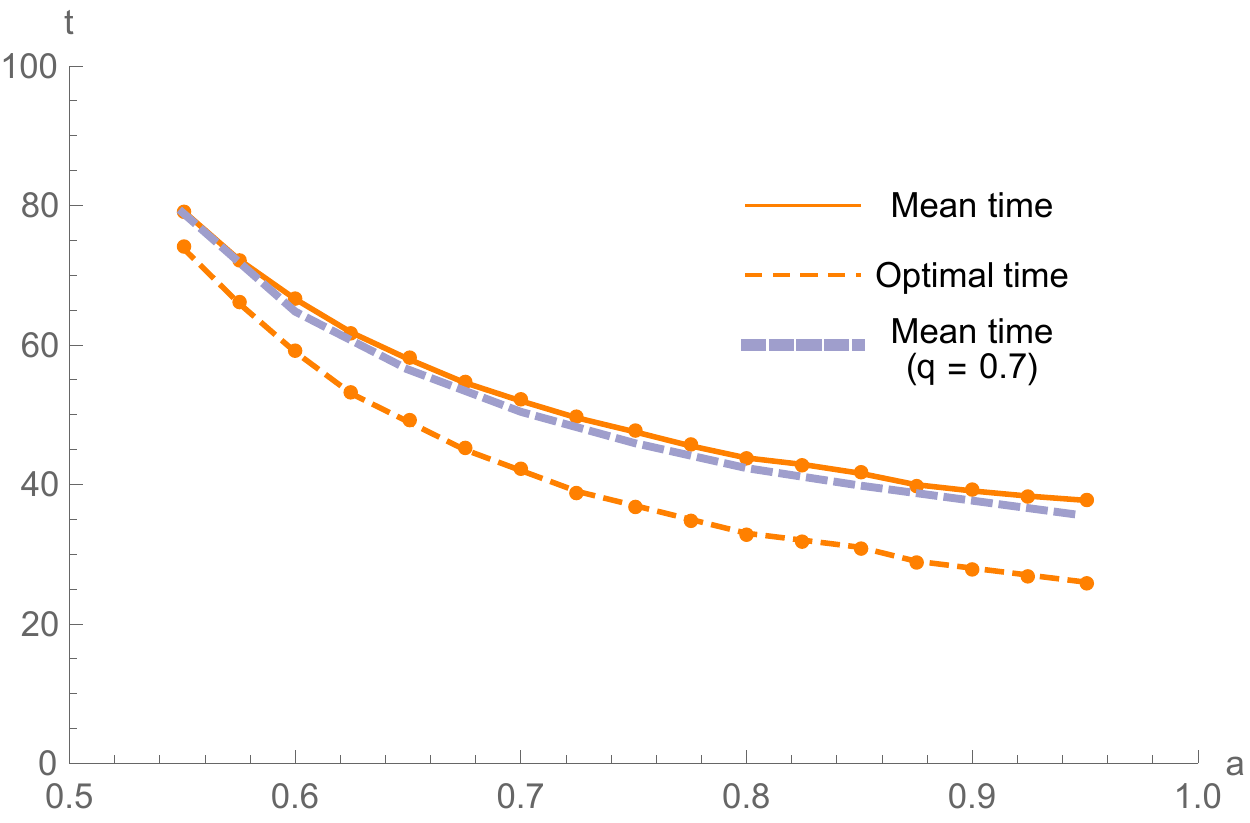}
\subcaption{\label{fig2} Mean and optimal times}
\end{minipage}
\caption{Centipede game. Linearly increasing pie}%
\end{figure}

We set $T_{1}=T_{2}=\{1,...,\overline{\tau}\}$, with $\overline{\tau}$ large
($\overline{\tau}=100$), $S(\tau)=\tau$ and $a(\tau)=a$, and we examine the
effect of changing $a$ on limit distributions, for $a>1/2$. Unless the other
player's $t$ is low, each player has an incentive to slightly undercut his
opponent, so there is an equilibrium force towards low values of $t$. This
force is more pronounced for larger values of $a$, and with sufficiently large
$\beta$, the Quantal Response equilibrium would naturally concentrate most
weight on low values of $t$. Nevertheless, there is an endogenous barrier
$\beta_{a}^{\ast}$ which limits this unravelling. Figure~\ref{fig1} plots the
limit distributions obtained for different values of $a$. The mode corresponds
to the most profitable strategy given the limit distribution. As one expects,
a higher value of $a$ generates more competitive pressures, hence a
distribution that shifts to lower $t$'s.

We also report (Figure~\ref{fig2}), as a function of $a$, the mean exit date
and best-performing strategy (i.e., the mode of the distribution) obtained
under the limit distribution.\footnote{That is, given the limit distribution
$p_{a}^{\ast}$ (which is defined for each $a$), we report $\overline{t}%
_{a}^{\ast}\equiv E_{p_{a}^{\ast}}t_{i}$ and $t_{a}^{\ast}=\arg\max_{t_{i}%
}u_{i}(t_{i},p_{a}^{\ast})$. We do that for $a\in\{0.55,0.6,...,0.95\}$.} The
mean exit date is larger because all high $t$'s yield similar gains, and this
generates a fat tail for high $t$'s, hence a distribution skewed to the right.
This right-skewness is responsible for the instability: whenever $\beta$
starts being high enough to be conducive to low $t$'s, the right-skewness
provides incentives to choose\ high $t$'s.

This instability can be illustrated by examining the \textit{best-response
cycles} for $\beta>\beta_{a}^{\ast}$ for say,\textit{ }$a=0.7$. Convergence
obtains up to $\beta_{a}^{\ast}=0.3$ and the limit distribution $p_{a}^{\ast}$
has a mode at $t_{a}^{\ast}=42$. Over the range $\beta\in I=[0.31,29.9]$, the
logit-response dynamics computed from $p_{a}^{\ast}$ cycles, and along that
sequence, the best-performing action cycles as well. Cycles have low amplitude
when $\beta$ is close to $\beta^{\ast}$, and the amplitude increases as
$\beta$ increases. For $\beta=0.5$, the best-response cycle is:%
\[
58,52,45,39,34,29,25,58,52...
\]
For $\beta=10$, the best-response cycle takes almost all values between 60
down to 3 before jumping back to $60$. And it takes $\beta\geq30$ to obtain a
dynamic where the best-performing action does not cycle (and converges to
$2$). In other words, over a large range of precision $\beta$, cycling
prevails, suggesting a thick barrier to unravelling.

Regarding limit precision, we observe that $\beta_{a}^{\ast}$ decreases with
$a$. Intuitively, a higher $a$ undermines the stability of the logit response
dynamics, and $\beta^{\ast}$ has to decrease to restore stability. Furthermore
there is more unravelling at the limit QRE when $a$ rises, so payoffs are
lower. So both effects (lower gains and endogenously lower $\beta^{\ast}$) are
conducive to higher behavioral dispersion as $a$ rises, as Figure~\ref{fig1}
confirms. In the Appendix, we further illustrate the difference between
exogenous and endogenous precision.



Finally, we examine the robustness of the limit distribution to trembles. We
consider the game over target exit dates $\tau_{i}$, assuming that these
targets are implemented with noise according to (\ref{dyn1}), with a noise
magnitude parameterized by the exponential decay parameter $q$. For any $a$,
trembles systematically induce more stable dynamics (hence higher $\beta
_{a}^{\ast}$) -- i.e, $\widehat{\beta}_{a}^{\ast}>\beta_{a}^{\ast}$), hence a
more concentrated distribution over targets. So long as $q$ is not too large
($q\leq0.7$), the overall \textit{induced distribution over dates} remains
mostly unchanged (see Figure~\ref{fig1}, where distributions are drawn for
$q=0.7$). For large enough values of $q$, $\widehat{\beta}_{a}^{\ast}$ may
rise to $\infty$, in which case the outcome becomes fully driven by the
structure of the exogenous noise assumed, as in \citet{compte18}.

\subsubsection{\label{otherparametric}Other parametric formulations}

We turn to parametric formulations closer to those analyzed in experiments. We
keep our long-horizon assumption ($\overline{\tau}$ large), consider
exponentially growing cake size, and allow for non-stationary sharing rules
(as in \citet{fey96}).

\textbf{Exponentially growing cake size.} We use a constant sharing rule
$a>1/2$ and assume an exponentially growing cake, that is:%
\begin{equation}
S(t)=S_{0}\exp bt/\overline{\tau} \label{eqEXP}%
\end{equation}
We set $T_{1}=T_{2}=\{0,1,2,...,\overline{\tau}\}$ with $\overline{\tau}=100$
and first report limit distributions when $a=0.8$ and (i) $b=4.15$ or (ii)
$b=1.5$. Case (i) is chosen so that over the full duration of the game, the
size of the cake is multiplied by $64$ (which corresponds to the magnitude
considered in many experiments). Case (ii) implies that over the full duration
of the game, the size is multiplied by $4.5$.\footnote{Limit distributions do
not depend on the scaling factor $S_{0}$. To better represent the shape of
$S(t)$, we use a different scaling in each case ($S_{0}=1$ for $b=4.15$) and
$(S_{0}=10$ for $b=1.5$).} $\ $

\begin{figure}[h]
\centering
\begin{minipage}{0.45\textwidth}
\centering
\includegraphics[scale=0.6]{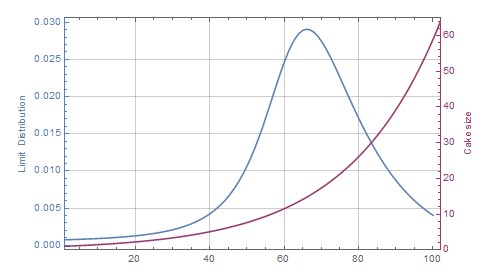}
\subcaption{$a=0.8$ and $b=4.15$}
\label{fig3}
\end{minipage}\hfill\begin{minipage}{0.45\textwidth}
\centering
\includegraphics[scale=0.6]{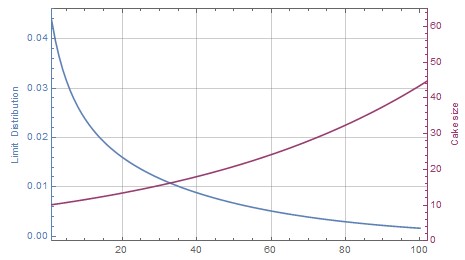}
\subcaption{\label{fig4} $a=0.8$ and $b=1.5$}
\end{minipage}
\caption{Exponentially increasing cake. Limit distributions}%
\end{figure}

The main difference with Section \ref{sectionlinear} is that in case (ii), the
limit distribution has a mode at the lowest date $t=1$. At a given $\beta$,
there are two opposing forces: one that favors unravelling, as for any date
$t$ chosen by the opponent, one prefers choosing a slightly smaller date;
another that, given the strategic uncertainty (fueled by the limited
precision), one favors the choice of higher dates. When the cake does not grow
fast enough, the latter force is weak, and $t=1$ becomes the optimal strategy.
Note that as before, unravelling cannot be complete at the limit distribution,
because whenever $\beta$ becomes just large enough to produce strong
unravelling, ex ante payoffs become small, and choosing higher dates generates
payoffs that are not substantially smaller than early dates, thus fueling
extensive dispersion in the dates chosen.

\begin{figure}[h]
\centering
\includegraphics[scale=0.6]{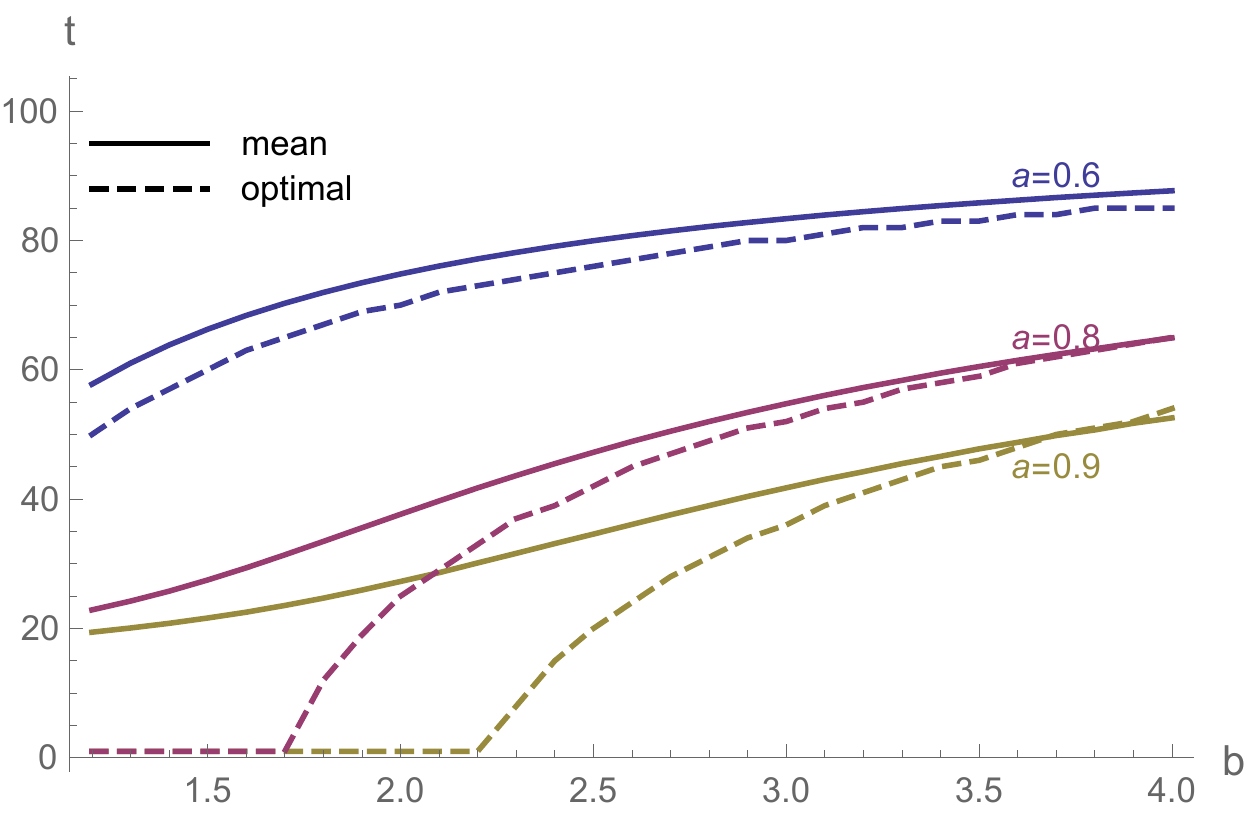}\caption{Exponentially
increasing cake. Mean and optimal times}%
\label{figEXP}%
\end{figure}

To conclude this Section, we examine comparative statics with respect to $a$
and $b$, computing the limit distribution in each case. For each distribution,
we report mean and optimal times (the optimal time corresponds to the mode of
the limit distribution). Figure~\ref{figEXP} illustrates that smaller $a$ and
higher $b$ both promote larger mean and optimal times. Furthermore, if $b$ is
small and $a$ not too close to $0.5$, the limit distribution has a mode at $1$.

\textbf{Constant size with non-stationary sharing rules. }Next, in the spirit
of \citet{fey96}, we consider a cake of constant size (i.e., $b=0$ -- so the
game is zero-sum), and we examine the effect of a sharing rule that becomes
more competitive over time:%
\[
a(t)=1-1/2\exp-\alpha t/100
\]
That is, at $t=0$, the cake is shared equally. After that, the individual who
exits before the other gets a larger share, and over time, this advantage from
exiting first becomes larger. Unlike in previous cases, there is no joint
interest for waiting. However, if the other waits, one prefers to wait too
(though slightly less). We report limit distributions for three values of
$\alpha$.

\begin{figure}[h]
\centering
\includegraphics[scale=0.6]{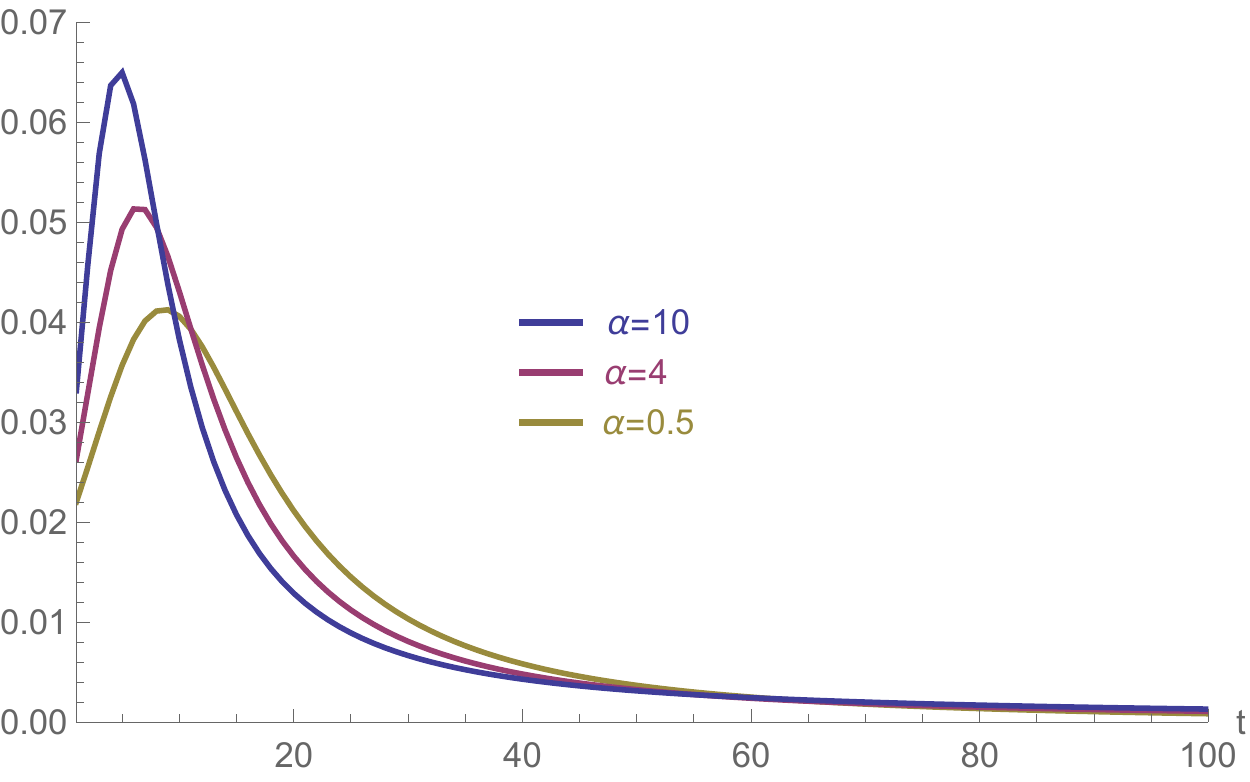}\caption{Constant-size
cake. Limit distributions}%
\label{figCONST}%
\end{figure}The limit distributions obtained illustrate that the pressure
towards unravelling is stronger when $\alpha$ is larger. This pressure does
not lead to a mode at $t=1$, because over some range of dates, $a(t)$ remains
sufficiently close to $1/2$. However for $\alpha$ sufficiently high
($\alpha>90)$, it would.

\subsubsection{Robustness to timing.}

We examine the robustness of the limit distribution to timing. First we vary
the number of exit opportunities, that is, we fix $n$ and set
\[
T_{1}=T_{2}=\{0,\rho,2\rho,..,n\rho\}\text{ with }\rho=\overline{\tau
}/n\text{.}%
\]
To fix ideas, we consider an exponentially increasing cake with a constant
sharing rule: in simulations we set $a=0.8$ and $b=4.15$. In other words, we
change the number of \textquotedblleft periods" in the sense that there are
now $1+n$ exit opportunities rather than $1+\overline{\tau}$, but we do not
change the broad payoff structure: the ratio between cake sizes at the last
and initial \textquotedblleft periods" is $\frac{S(\overline{\tau})}{S_{0}%
}=\exp b$ for all $n$ examined.

\begin{figure}[h]
\centering
\includegraphics[scale=0.6]{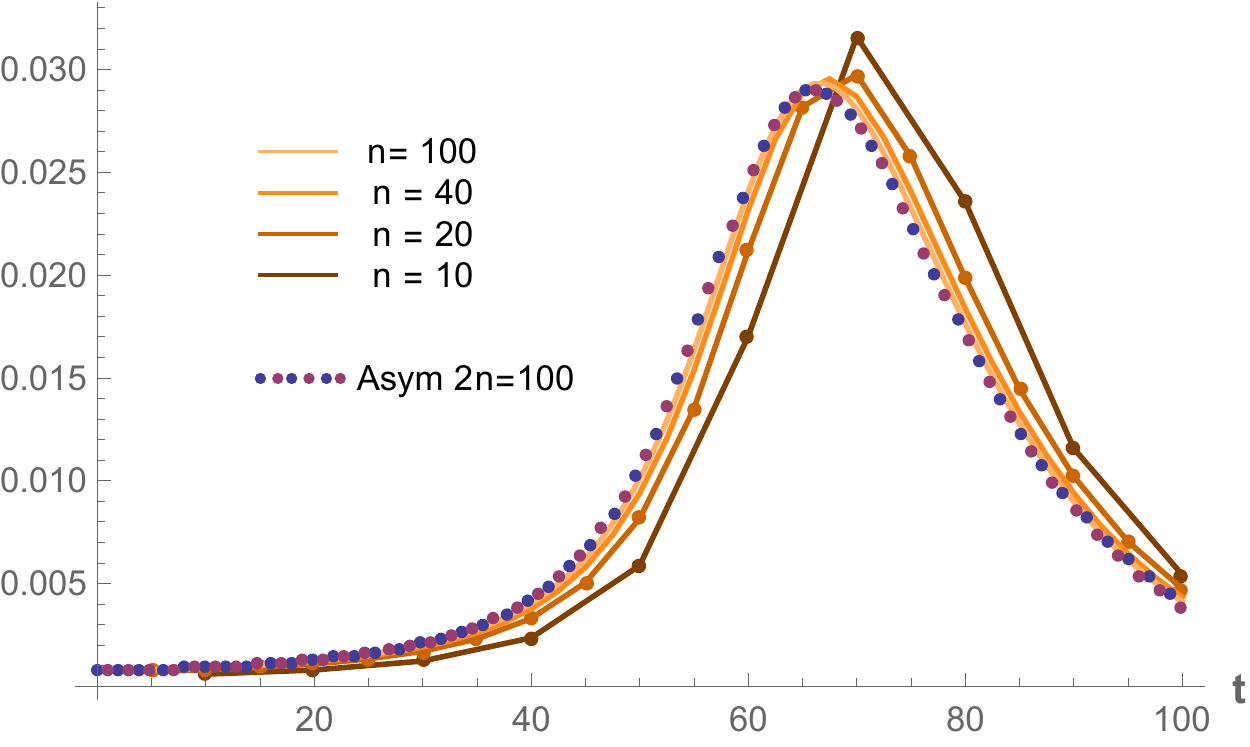}\caption{Limit distribution:
robustness to the number $n$ of exit opportunities}%
\label{figtiming}%
\end{figure}

Next we examine the more standard centipede where\textit{ players move in
alternation}, that is, we consider $2n$ exit opportunities ($n$ for each
player), and we set
\[
T_{1}=\{0,2\rho,4\rho,...(n-1)\rho\}\text{ and }T_{2}=\{\rho,3\rho
,..,n\rho\},
\]
with $\rho=\overline{\tau}/n$. Compared to the symmetric game examined above,
the main difference is that there cannot be date profiles where the cake is
shared equally. Figure \ref{figtiming} reports limit distributions for the
symmetric game, with $n=100,40,20$ and $10$. We also indicate on that figure
the limit distribution of the asymmetric game for $2n=100$.\footnote{Note that
to be able to compare distributions, we rescale the vertical axis. When there
are $1+n$ points in the distribution rather than $1+\overline{\tau}$, we
rescale the distribution by a constant factor $(1+n)/(1+\overline{\tau})$.}

Figure \ref{figtiming} shows a remarkably robust prediction for large $n$: the
limit distributions for $n=100$ and $n=40$ almost coincide. The distance to
the asymmetric timing case is also quite small. Intuitively, with many exit
opportunities, and given the strategic uncertainty at the limit distribution,
the chance of identical dates is negligible, so, at the limit distribution,
payoffs under both timing assumptions are almost identical.

\begin{figure}[h]
\centering
\begin{minipage}{0.45\textwidth}
\centering
\includegraphics[scale=0.6]{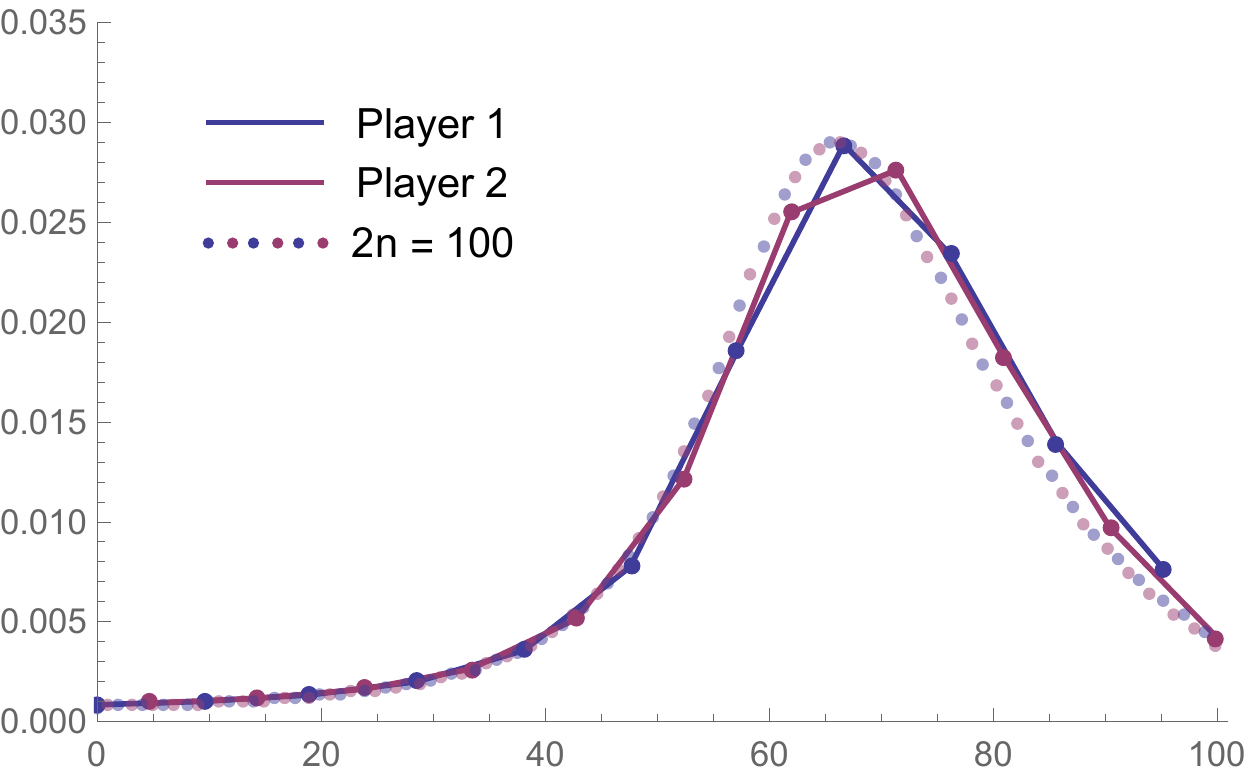}
\subcaption{\label{timingcase10}$2n = 20$}
\end{minipage}\hfill\begin{minipage}{0.45\textwidth}
\centering
	 \includegraphics[scale=0.6]{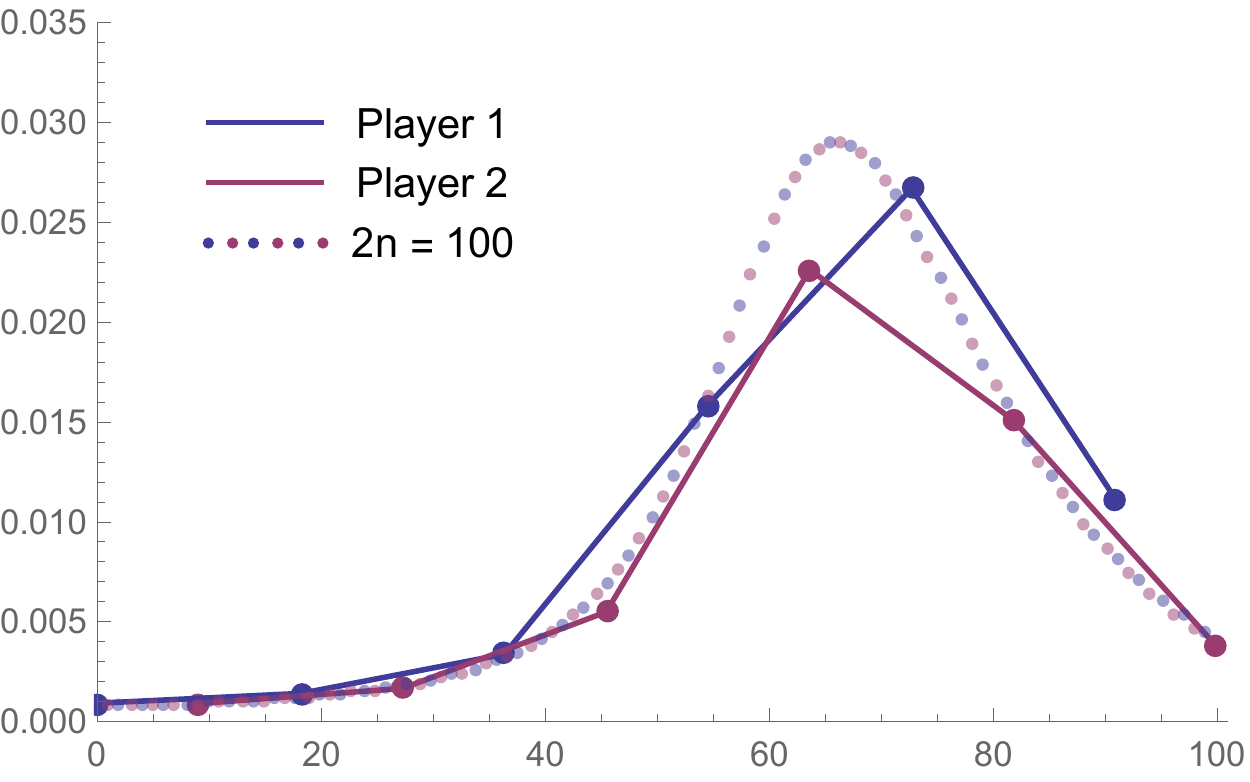}
\subcaption{\label{timingcase10}$2n = 10$}
\end{minipage}
\caption{Limit distributions. Alternating exit dates. Robustness to $n$}%
\label{alternatingdates}%
\end{figure}

As the number of exit opportunities is reduced, one observes a shift towards
higher dates. The reason is that with low $n$, the chance of players choosing
the same date becomes significant, and choosing the same high date yields a
high payoff (and is not far from being a Nash equilibrium). This shift towards
higher dates does not arise under the asymmetric timing assumption.
Figure~\ref{alternatingdates} reports the limit distributions obtained for
different values of $n$ when timing is asymmetric.

In the Appendix, we perform simulations for constant-size cakes.
Similar robustness to timing is obtained.

\subsubsection{Comparison with experimental data}

\begin{figure}[h]
\centering
\includegraphics[scale=0.6]{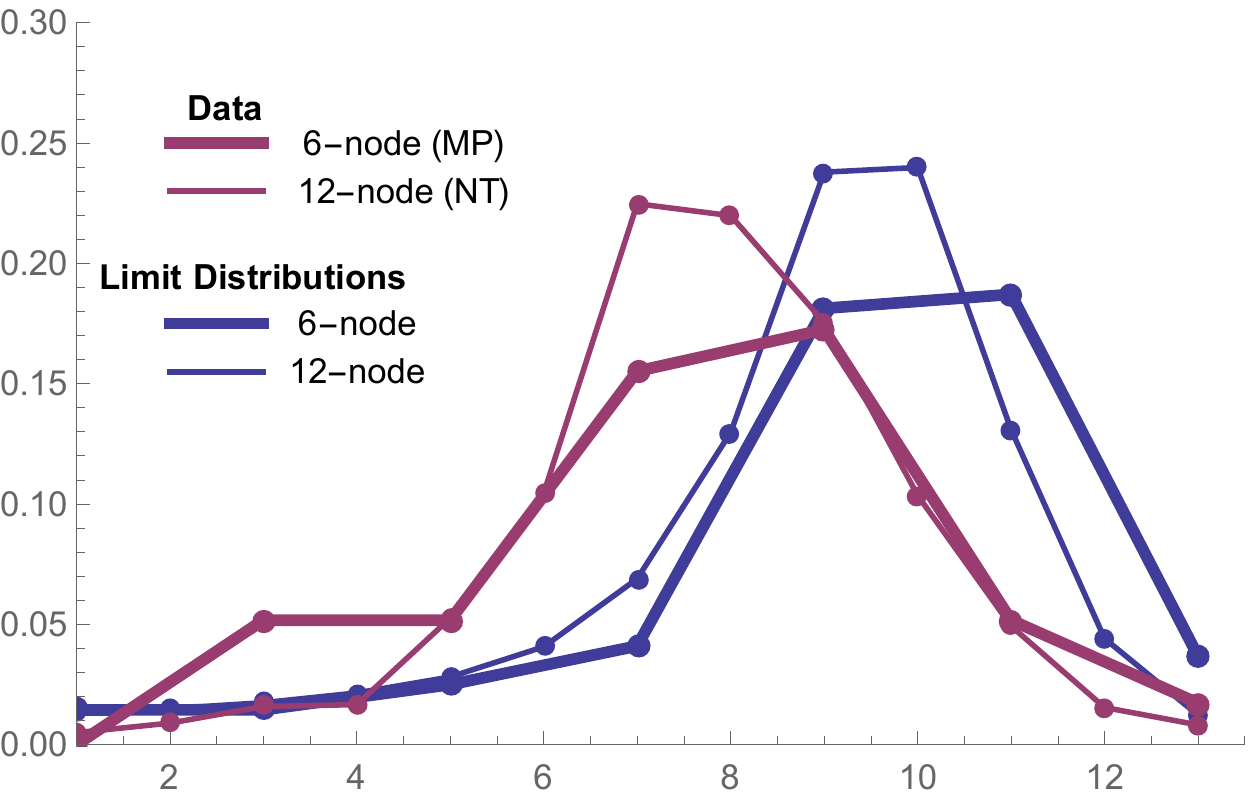}\caption{Data
versus limit distributions}%
\label{figMPNT}%
\end{figure}

We close our exploration of centipede games by considering games analyzed in
\citet{mckelvey92} (MP) and \citet{nagel98} (NT). These two games have a
payoff structure very close to $a=0.8$ and $b=4.15$, with a different number
of nodes ($2n=6$ for MP, $2n=12$ for NT). Exact payoffs are recalled in the
Appendix. For each game we obtain the limit distribution of dates chosen by
each player. We use these distributions to obtain the distribution over
terminal nodes. Figure \ref{figMPNT} reports the data and limit distributions
over terminal nodes.

Figure \ref{figMPNT} shows a robustness to timing (both in terms of dispersion
and locus of mode). While the data and the limit distribution over nodes
exhibit similar dispersion, the data shows a mode at lower dates, compared to
the limit distribution.

To reconcile the simulation with the data, one could look for QRE with higher
$\beta$'s (i.e, higher than $\beta^{\ast}$). One would indeed obtain QRE with
lower modes, but the dispersion would be significantly reduced (see Appendix).
One could look for a limit equilibria where the last node gets no
weight.\footnote{Obviously dominated strategies get positive weight under
quantal response, and one might expect that players learn to avoid these.}
This would conflict with the data for this last node, but it would shift the
limit distribution to the left, closer to the observed data (See Appendix).

\begin{figure}[h]
\centering
\begin{minipage}{0.45\textwidth}
\centering
\includegraphics[scale=0.55]{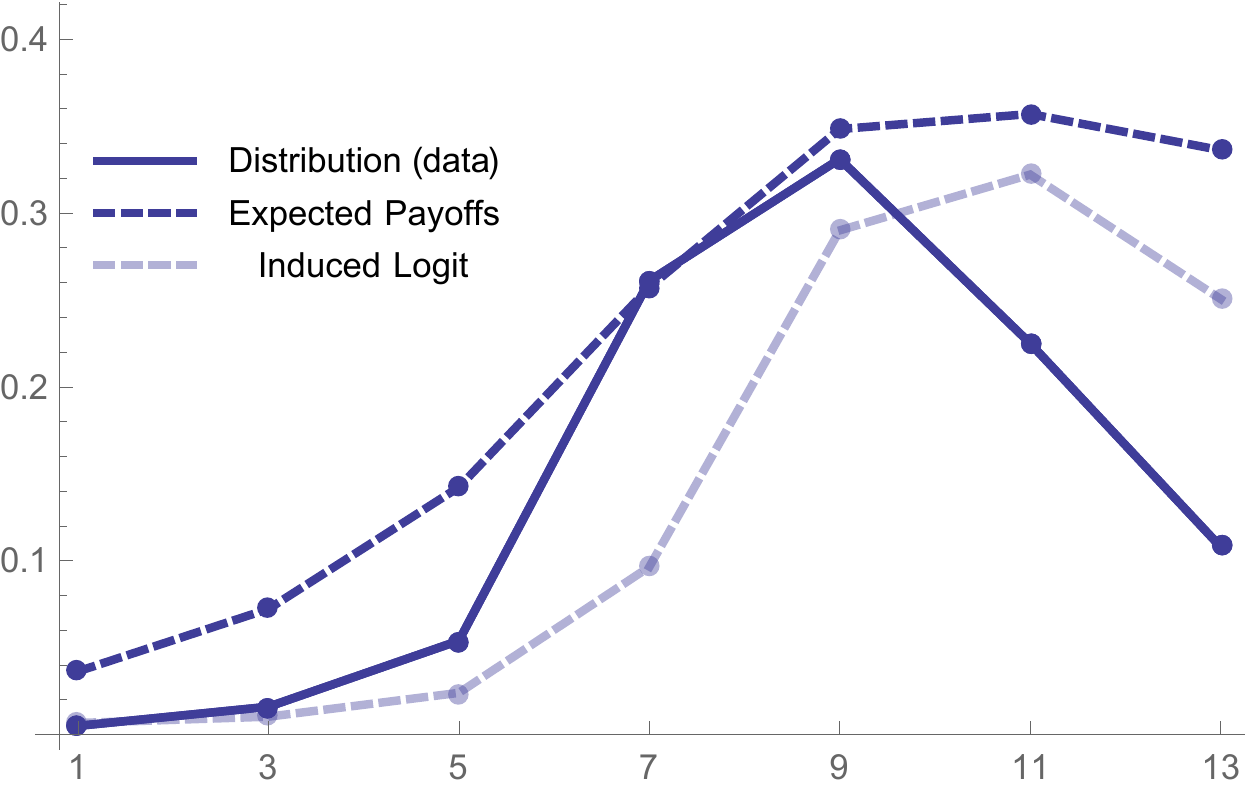}
\subcaption{\label{expectedpayoffs1}Player 1}
\end{minipage}\hfill\begin{minipage}{0.45\textwidth}
\centering
	 \includegraphics[scale=0.55]{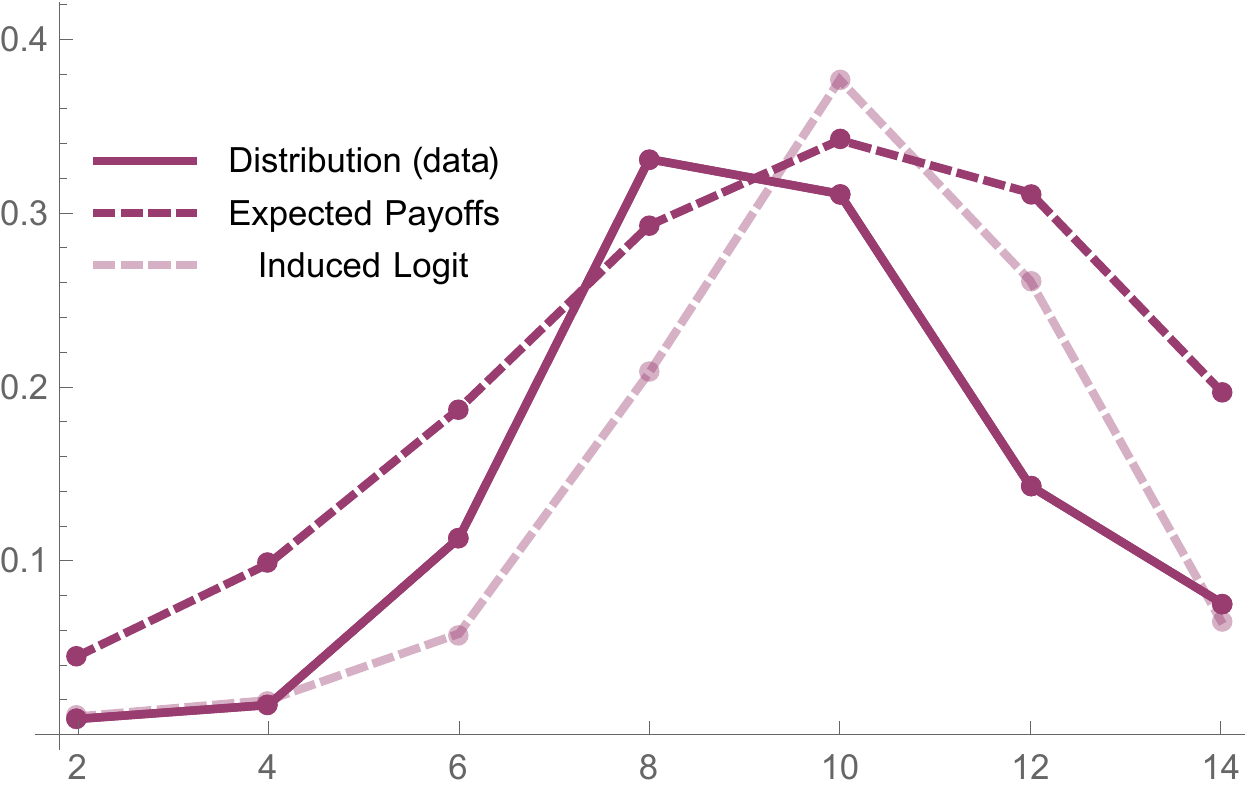}
\subcaption{\label{expectedpayoffs2}Player 2}
\end{minipage}
\caption{Expected payoffs against the empirical distribution}%
\label{expectedpayoffs}%
\end{figure}

Another way to interpret the discrepancy between the predictions and the data
is the following. We compute ex ante expected payoffs against the empirical
distribution and find that ex ante payoffs are not in line with the choices
made: players exit too early. Figure \ref{expectedpayoffs} reports these
expected payoffs for \citet{nagel98} (the Appendix reports a similar
\textit{early-exit bias} for other data sets). We also indicate the shape of
the logit distribution induced by these expected payoffs (where the logit
parameter $\beta$ is chosen to generate a dispersion over choices comparable
to that of the data). 

This suggests that if players' choices are truly driven by expected payoff
comparisons, we should expect with more experience, players' choices would
shift to higher dates, in the direction of our limit distribution. It could
also be that player's choices are not driven by expected payoff comparisons.
For example, it could be that when stakes become large, risk aversion or
regret matters.\footnote{If players maximize expected utility with
$u(x)=x^{c}$, with $c<1$, then this is as if they were maximizing expected
payoffs in a game where $b$ is reduced by a factor $c$, which generates more
unravelling.}

\subsection{The Traveler's dilemma}

\begin{figure}[h]
\centering
\begin{minipage}{0.45\textwidth}
\centering
\includegraphics[scale=0.65]{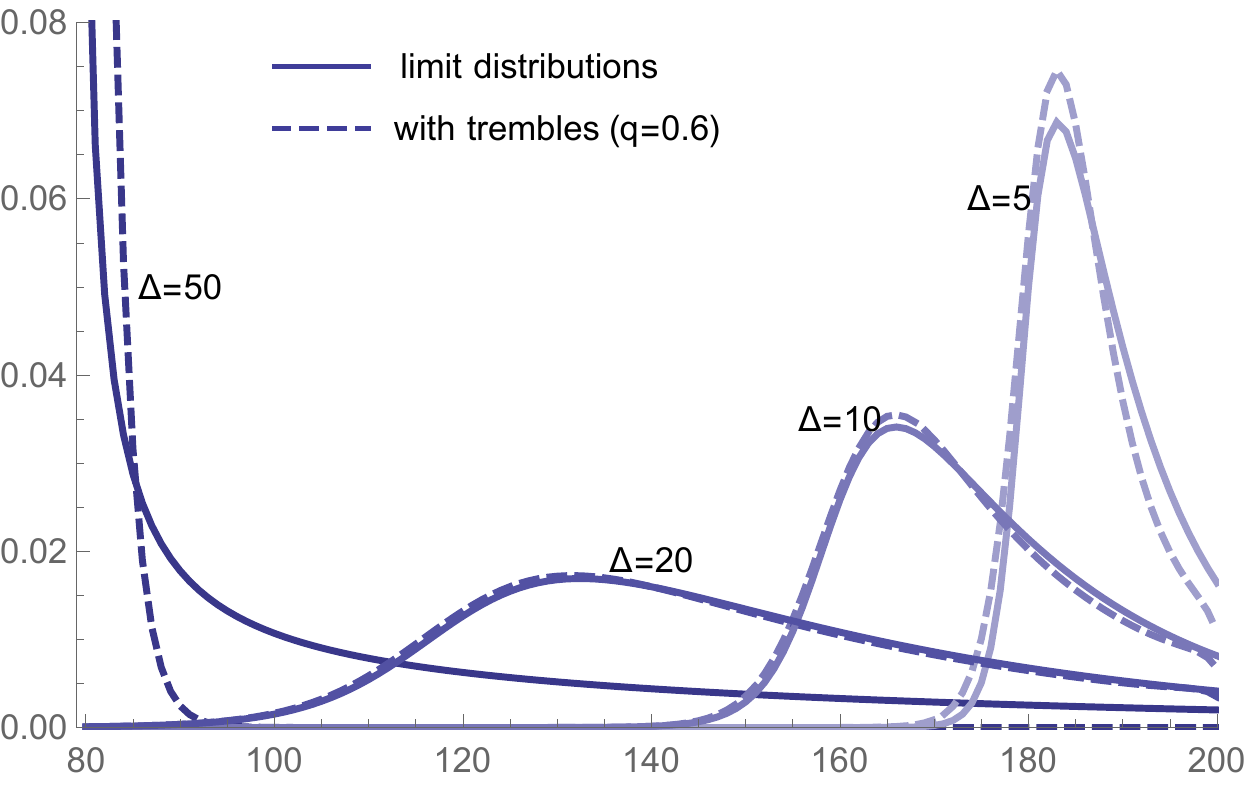}
\subcaption{\label{figtrav1}Limit distributions}
\end{minipage}\hfill\begin{minipage}{0.45\textwidth}
\centering
	 \includegraphics[scale=0.6]{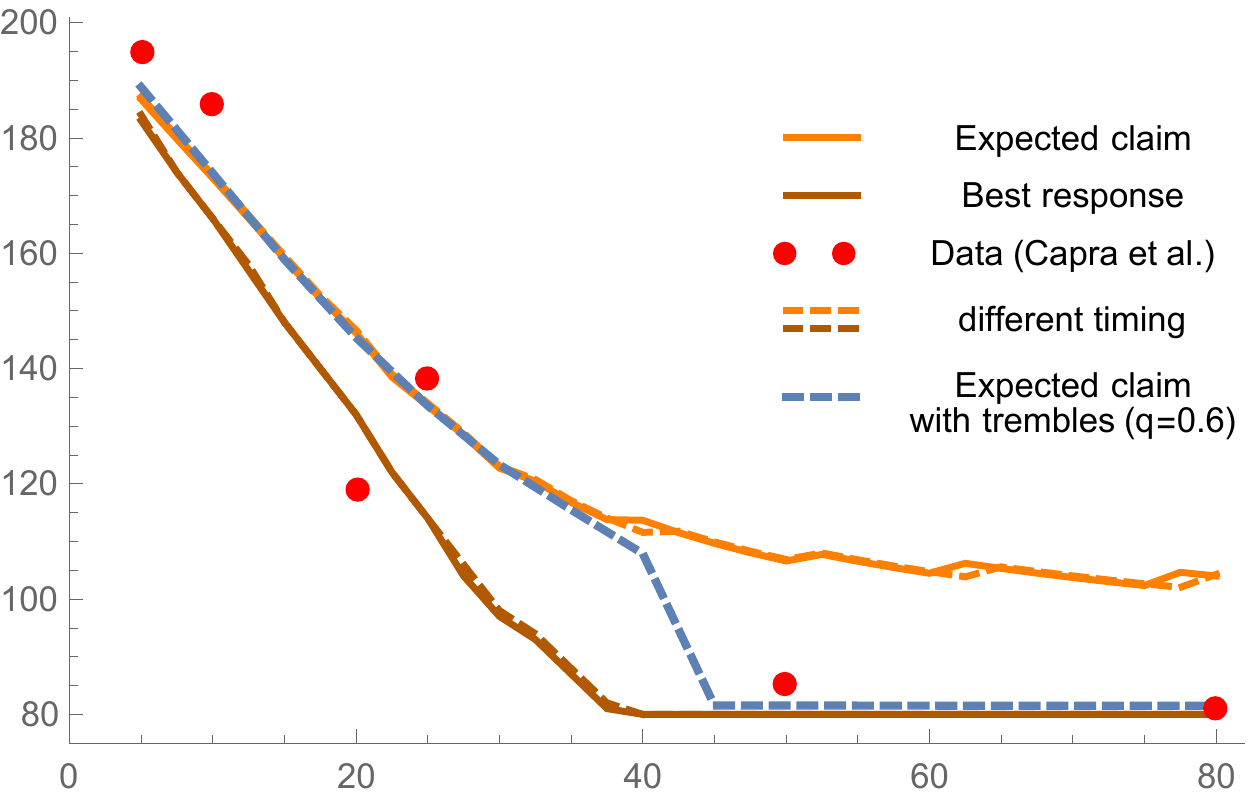}
\subcaption{\label{figtrav2}Expected and optimal times}
\end{minipage}
\caption{Traveler's dilemma}%
\label{trav}%
\end{figure}

We next consider the traveler's dilemma, where two players report a claim
$t_{i}\in\mathcal{T}\equiv\{80,..,T\}$, with $T=200$ as in \citet{basu94}.
When reports are $t_{i}$ and $t_{j}$, they each get $\tau=\min(t_{1},t_{2})$,
but the player making a strictly lower report gets a bonus $\Delta$, while the
other gets a penalty $\Delta$. Formally:%
\[
u_{i}(t_{i},t_{j})=%
\begin{cases}
\tau+\Delta & \text{ if }t_{i}<t_{j}\\
\tau-\Delta & \text{ if }t_{i}>t_{j}\\
\tau & \text{ otherwise}%
\end{cases}
\]
This game has a structure similar to the centipede game examined in
Section~\ref{sectionlinear}: the pie grows linearly with $\tau$, and the early
(late) mover is rewarded (penalized). The difference is that the magnitude of
the reward and penalty are constant (rather than proportional to $\tau$).
Figure~\ref{figtrav1} reports limit distributions for different values of
$\Delta.$
Varying $\Delta$, we also report for each limit distribution, the expected and
optimal exit time.

Higher penalties generate more unravelling, and for $\Delta$ above $40$, the
optimal time has a mode at the lowest claim $(80)$. Figure~\ref{figtrav2} also
reports the data from \citet{capra99}. For small $\Delta,$ expected times are
broadly consistent with the data. For large $\Delta$ however we do not get
full unravelling at the limit distribution. The optimal date is $80$ but many
dates get positive weight. This fat tail generates an expected time above the
lowest claim, unlike what the data suggests. As we shall see below however,
for large $\Delta$, this prediction is not robust to small exogenous noise.

\textbf{Discussion.} \textit{Robustness to timing.} Figure~\ref{figtrav2}
indicates the limit distribution obtained in the game where players can only
report even numbers: as in the centipede game, the limit distribution is
robust to the number of claims/dates available. It is also robust to the
length $T$ of the game, in the following sense. Call $d_{i}$ the distance to
the maximum feasible claim (i.e., $d_{i}=T-t_{i}$). So long as $\Delta$ is not
too large (so that the limit distribution puts negligible weight on the lowest
claims), the limit distribution over $d_{i}$ does not depend on $T$%
.\footnote{This is because so long as $d_{j}<T$, utility differences
$u_{i}(d_{i},d_{j})-u_{i}(d_{i}^{\prime},d_{j})$ do not depend on $T$.} If one
interpret this centipede game as a finitely repeated prisoner's dilemma (see
\citet{compte23}),
$d_{i}$ can be interpreted as the length of the end game (as viewed from
player $i$), which determines when she starts defecting. At the limit
distribution, this length is thus increasing in $\Delta$ and independent of
the length the game.

\textit{Limit precision. }Regarding how $\beta^{\ast}$ varies with $\Delta$,
we find that over the range $\Delta\in(1,40)$, a higher $\Delta$ induces lower
$\beta^{\ast}$. For example, $\beta_{10}^{\ast}=0.31>\beta_{25}^{\ast}=0.12$.
The intuition is similar to that of Section~\ref{centipede}: a higher $\Delta$
undermines the stability of the logit-response dynamics, and a lower precision
permits to restore stability. The consequence is as before a larger behavioral
dispersion (as Figure~\ref{figtrav1}\ confirms).

\begin{figure}[h]
\centering
\begin{minipage}{0.45\textwidth}
\centering
\includegraphics[scale=0.55]{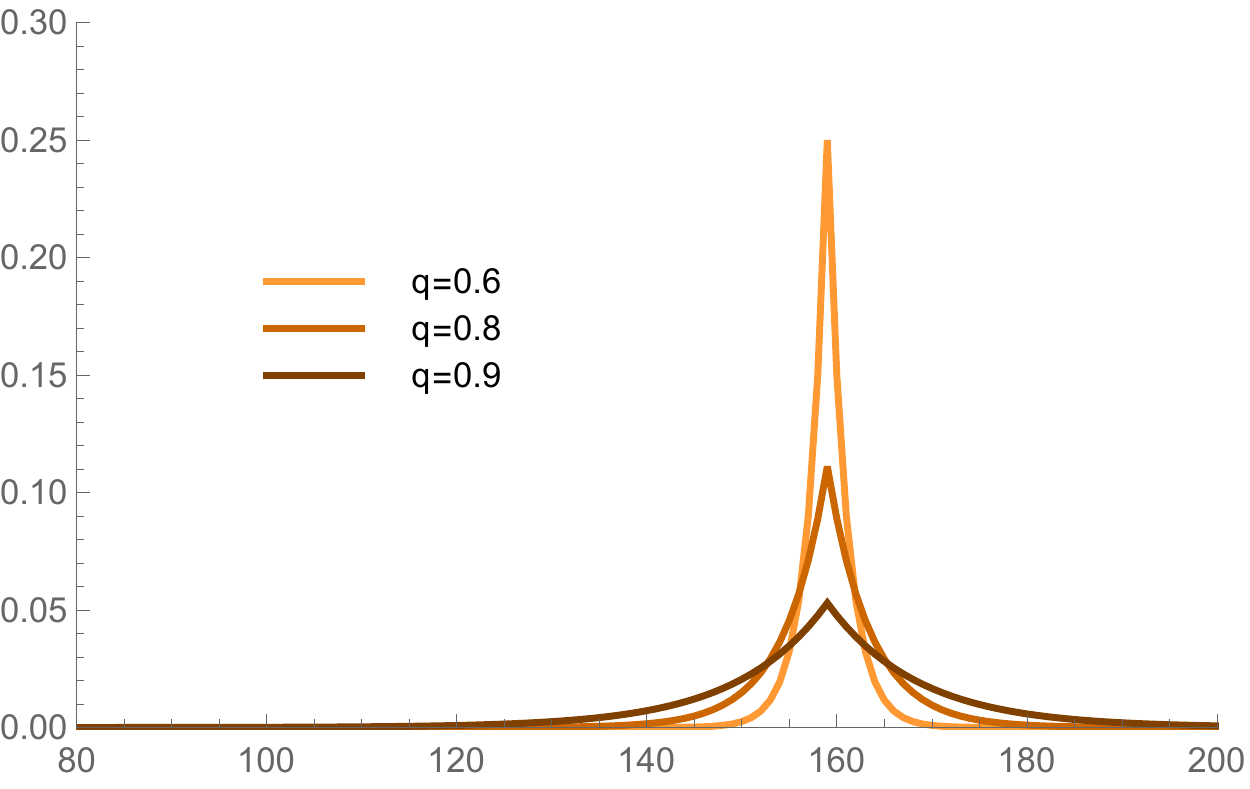}
\subcaption{\label{fignoise}Cond. distributions over claims given $\tau=160$}%
\end{minipage}\hfill\begin{minipage}{0.45\textwidth}
\centering
	\includegraphics[scale=0.55]{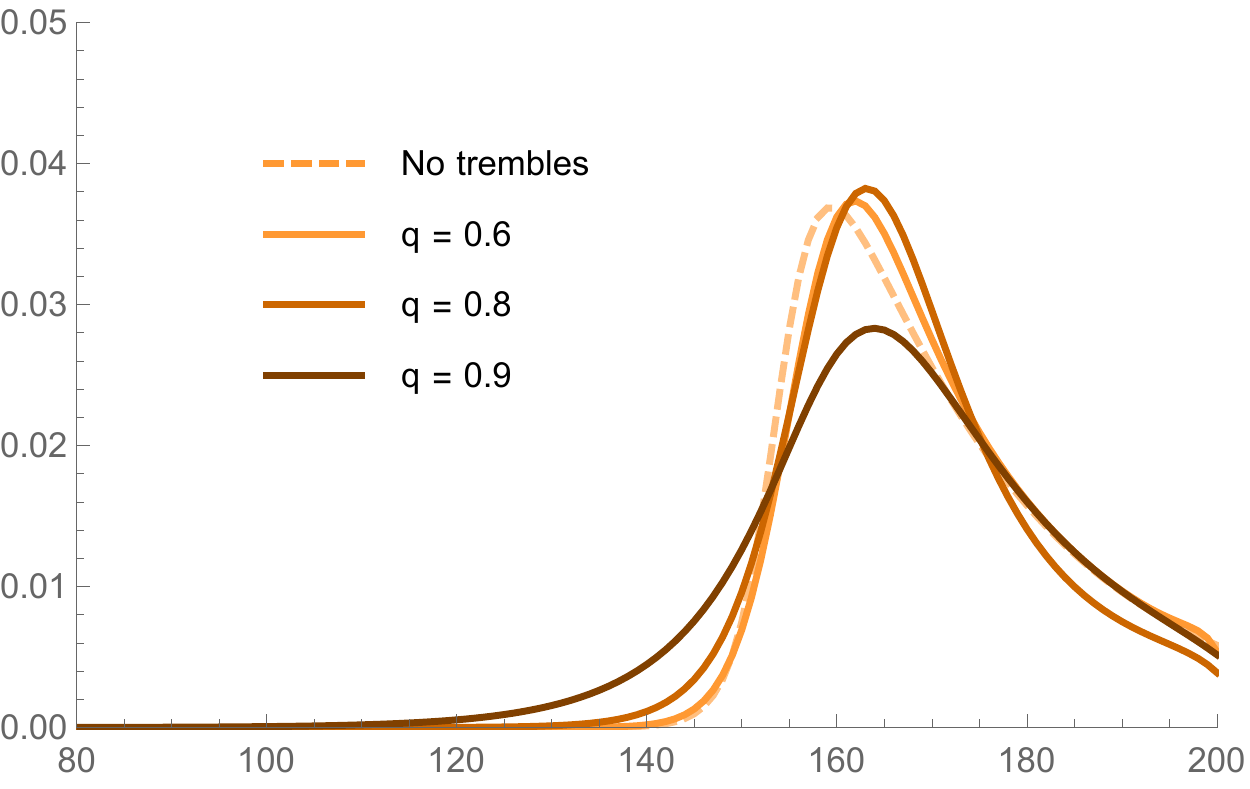}
\subcaption{\label{figlimitnoise} Induced distributions over claims when $\Delta=10$ }%
\end{minipage}
\caption{Game over targets}%
\end{figure}

\textit{Robustness to exogenous trembles. } We proceed as in the previous
section. We analyze the game over target (claims) for different values of the
noise magnitude $q$, and compute the induced distribution over claims.
Figure~\ref{fignoise} plots the distributions over claims for $q=0.6,0.8$ and
$0.9$, for a given target $\tau=160$.

Trembles again stabilize the logit-response dynamics over targets (i.e.,
$\widehat{\beta}^{\ast}>\beta^{\ast}$), but this increased stability has
different consequences depending on the magnitude of $\Delta$.

For $\Delta$ not too large, the consequence is as in the centipede game, with
only minor effect on the induced distribution over claims.
For $\Delta=10$ for example, Figure~\ref{figlimitnoise} provides these
distributions for different values of $q$. The distributions shift slightly to
the right: the modes of these distributions are respectively 160 ($q=0$), 162
($q=0.6)$ and 163 ($q=0.8$ and $q=0.9$). At $q=0.9$, the exogenous noise is
sufficiently large that the combined effect leads to a more dispersed
distribution. For $q=0.93$, not drawn in the figure, the exogenous noise is
strong enough that targeting $192$ is actually a NE of the game over targets,
and the limit distribution over targets converges to it (i.e., $\widehat{\beta
}_{10}^{\ast}$ tends to $\infty$).

For large values of $\Delta$ ($\Delta>40$) however, (even small) exogenous
uncertainty may be enough to increase stability of the best response dynamics
up to the point where there are no barriers to learning: in the game over
targets, $\widehat{\beta}^{\ast}$ tends to $\infty$, there is complete
unravelling and the equilibrium target is the lowest possible one ($\tau
^{\ast}=80$). Figure~\ref{figtrav2} reports expected claims associated with
each induced distribution ($q=0.6$ and different values of $\Delta$).
Intuitively, for large $\Delta$, say $\Delta=80$, there is a barrier to
learning at $\beta^{\ast}=0.04$, but, unlike in the centipede game, the
barrier is very thin: the best response dynamics is unstable for a tiny
interval $(0.04,0.07]$ and stable again for all $\beta>0.07$. With exogenous
trembles, stability improves, and there are no barriers to learning. For small
$\Delta$, the barrier to learning is thicker, and small trembles do not confer
enough additional stability to \textquotedblleft cross\textquotedblright\ the
barrier to learning (for $\Delta=20$ for example, the best response dynamics
is unstable for $\beta\in(0.15,0.42)$, and when $q=0.6$, $\widehat{\beta
}^{\ast}$ rises up to $0.16$).

\textit{Further comments}\textbf{.} \citet{compte18} (see
\href{http://www.parisschoolofeconomics.com/compte-olivier/Chapter19Unraveling.pdf}{\color{dark-blue}Chapter~19}%
) examines another target version of the traveler's dilemma, where the targets
$d_{i}=T-t_{i}$ are implemented with \textit{multiplicative noise}. Given this
noise structure, CP compute a Nash equilibrium of the game in targets,
observing that even small variance fosters high claims (i.e., high cooperation
level), and that higher stakes $\Delta$ lowers the equilibrium target and the
equilibrium dispersion induced. This increased dispersion there was obtained
as a by-product of the multiplicative noise assumption. Here we simultaneously
endogenize the magnitude and shape of the noise, given the logit response
assumption, with a similar qualitative prediction regarding comparative
statics with respect to $\Delta$, for target levels played and dispersion.

It is obviously not novel that noise -- whether coming from imperfect
information, stubbornness, or payoff or game-length uncertainty -- helps
foster cooperation (\citet{kreps82}). Our contribution here is to suggest,
based on difficulties in learning to play the game, an endogenous limit to
precision (hence a limit to unraveling), given the logit response assumed.

\subsection{The 11-20 Money Request game\label{1120game}}

We review the game proposed by \citet{arad12}. In its basic version, each
player $i\in\{1,2\}$ chooses a claim $t_{i}\in\mathcal{T\equiv}\{11,...,20\}$
and obtains it. In addition, player $i$ obtains a bonus equal to $20$ whenever
$t_{i}=t_{j}-1$. The payoffs are thus summarized by%
\[
u_{i}(t_{i},t_{j})=%
\begin{cases}
t_{i}+20 & \text{ if }t_{i}=t_{j}-1\\
t_{i} & \text{ otherwise}%
\end{cases}
\]
Like in previous games, each player has an incentive to undercut the other
player. But there are two notable differences. Undercutting does not reduce
the payoff that the other side can secure, as each can secure 20 by choosing
$t_{i}=20$. So standard equilibrium analysis cannot give rise to full
unravelling to the lowest feasible claim and the Nash equilibrium must be in
mixed strategy. Second, reducing one's claim (which is costly) delivers a
bonus \textit{only in the event where one undercuts by just one unit}. Since
behavior is likely to incorporate some randomness, undercutting is potentially
risky, and even more so for lower claim levels (as one forgoes sure gains),
unless these lower claims are played with higher probability.

\begin{figure}[h]
\centering
\includegraphics[scale=0.65]{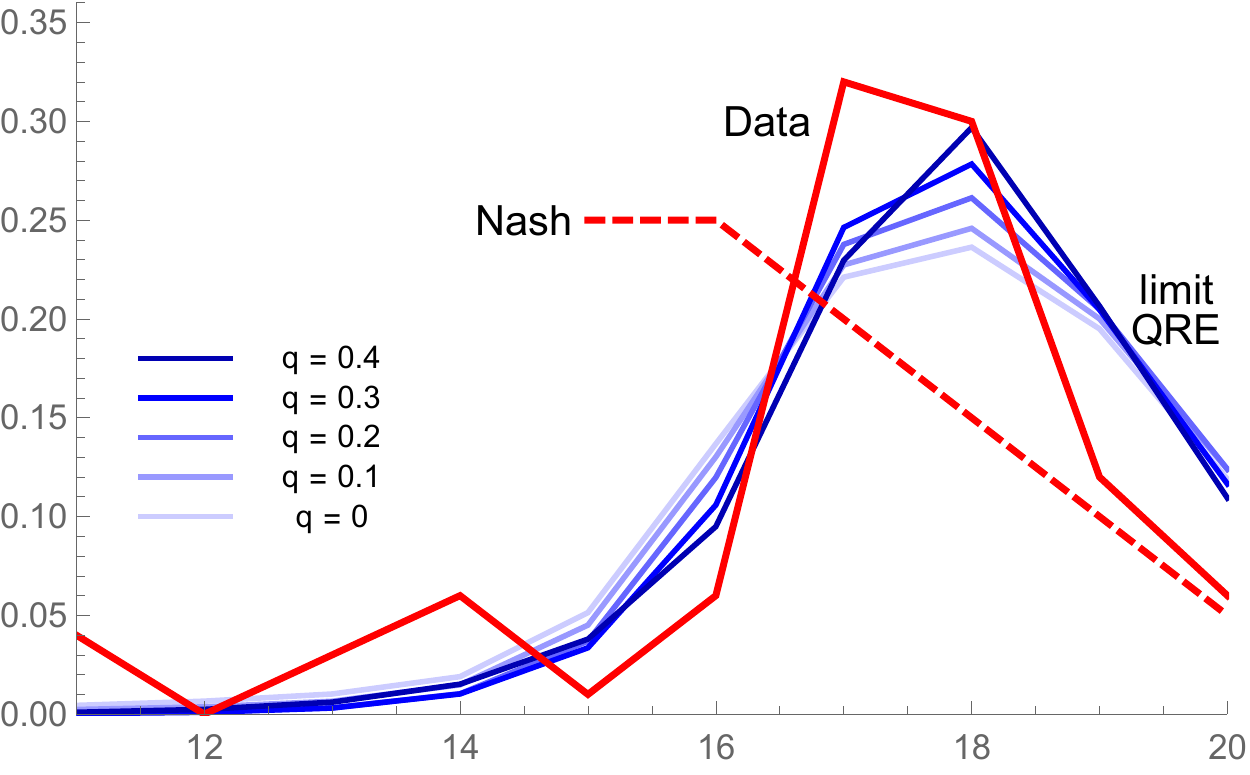}\caption{The 11-20 game}%
\label{fig6}%
\end{figure}

This explains why, in the Nash equilibrium of this game, among the claims $t$
chosen with positive probability $p(t)$, lower claims are chosen with higher
probability in equilibrium (i.e., $p(t)$ is downward sloping). As
\citet{arad12} show, the equilibrium outcome conflicts both with data and
intuition (see Figure~\ref{fig6}, Nash distribution plotted in dashed red,
data in red), and the authors suggest that some form of level-$k$ thinking may
be at work.

We examine this game through the limit~QRE lens. As in the previous Section,
we consider the game over targets for different level of noise. For each $q$,
we compute the limit distribution of the game over targets, and the induced
distribution over claims, given the trembles.
We report below these limit~QRE distributions over claims for different values
of $q$ (darker curves for higher values of $q\ $\ where $q$'s are multiples of
$0.1$ from $0$ to $0.4$).

The exogenous trembles improve the stability of the best response
functions $\phi_{\beta}$, with the consequence that the limit precision rises
with $q$, which yields significantly more concentrated limit distributions
over \textit{targets} as $q$ rises. However, the induced distribution over
claims shows only slightly higher concentration (on the mode at 18). In all
cases, the range 17-19 gets a weight approximately equal to 70\%.

\textbf{Discussion. }We investigate below the robustness of our predictions to
the presence of stubborn or unsophisticated types that always choose the naive
claim $t=20$. We also examine other payoff structures defined in
\citet{arad12} and find that limit QRE is a good fit when strategic types
choose among strategic claims only, i.e. claims strictly below $20$. We
revisit the classic level-k explanation in light of this observation.

\begin{figure}[h]
\centering
\includegraphics[scale=0.65]{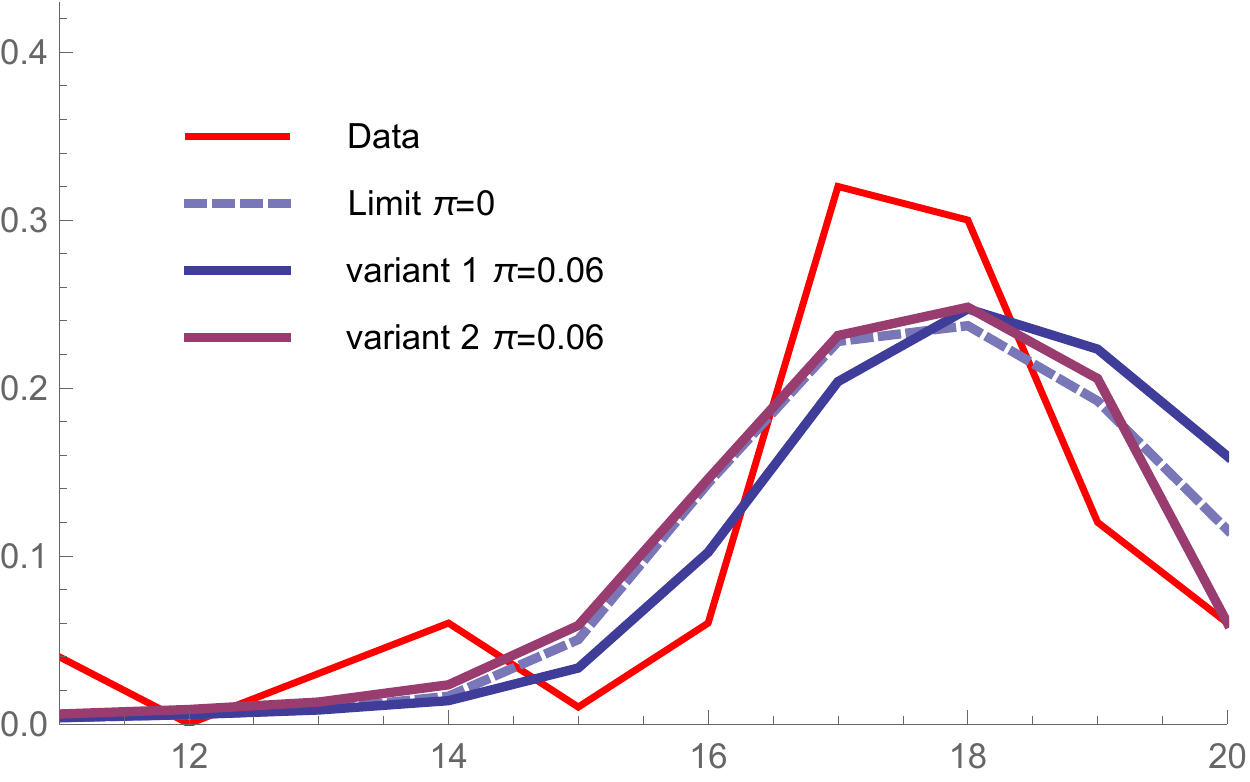}\caption{The 11-20 game with
stubborn types}%
\label{figvariant}%
\end{figure}

\textit{Robustness to stubborn types.} Given the particular role played by the
naive strategy $t=20$, we investigate how the presence of a fraction $\pi$ of
stubborn types playing $t=20$ affects the overall distribution over claims.
For a given $\pi\geq0$, the payoffs associated with $t=(t_{i},t_{j})$ is thus
\[
u_{i}^{\pi}(t)=(1-\pi)u_{i}(t)+\pi u_{i}(t_{i},20),
\]
which defines a new game. We consider two variants. In variant 1, the
strategic types optimize over all possible claims, including the naive claim
$20$. In variant 2, the strategic types focus on \textquotedblleft%
\textit{strategic claims}\textquotedblright, i.e., $t\in\{11,...,19\}$.
Variant 2 thus considers strategic players who understand or believe that they
are better off being strategic (we will check that indeed this is the case),
so they ignore the possibility of behaving non-strategically: their logit
response will put $0$ weight on $t=20$. Figure~\ref{figvariant} reports the
limit distributions over claims for the game without types ($\pi=0)$, and the
limit distribution over claims for each variant, for $\pi=0.06$. We choose
$\pi=0.06$ so that for variant 2, the weight on 20 coincides with the data.

Figure~\ref{figvariant} shows a robustness of the limit
distribution, with a small drift towards higher claims. Under variant 1, the
presence of stubborn types provides strategic players with incentives to lower
the weight on 20 and increase the weight on 19. Overall however, when both
types are combined, the weight on 20 increases. Under variant 2, the weight on
20 is constrained to be $\pi=0.06$, and the effect on other weights is negligible.

\textit{Other payoff structures.} \citet{arad12} examine two other payoff
structures. In the first one (the \textit{cycle version}), a player also wins
the bonus if she sets $t=20$ and the other chooses $11$. This seemingly adds
to the appeal of playing $20$, to the extent that the opponent plays 11 with
positive probability.

In the second one (the \textit{costless-iterations} version), a player wins 20
if he claims $20$. If he claims $t_{i}<20$, he obtains
\[
u_{i}(t_{i},t_{j})=37\text{ if }t_{j}=t_{i}+1\text{ and }u_{i}(t_{i}%
,t_{j})=17\text{ otherwise.}%
\]
In the latter version, there is a cost of 3 of choosing a claim below 20, but
all claims below 20 involve the same cost. In both versions, \citet{arad12}
obtain an empirical distribution that puts significantly more weight on $t=20$
(13 and 15\% respectively) and $t=19$ (47 and 40\% respectively). The weights
on $t=19$ are surprisingly large. The best response to the empirical
distribution points toward lower claims, so it is plausible with more
experience, players would learn to spread claims over a larger range. Still,
the later version is interesting because it suggests that in previous
versions, the larger cost associated with choosing lower claims may not be the
reason for players putting large weights on high claims.

We report the data from \citet{arad12} and limit distributions under both
versions and both variants. In each case, we calibrate $\pi$ so that the
weight on 20 is consistent with the data: we use $\pi=0.06$ for variant 1 and
$\pi=0.15$ or $0.13$ for variant $2$.

\begin{figure}[h]
\centering
\begin{minipage}{0.45\textwidth}
\centering
\includegraphics[scale=0.55]{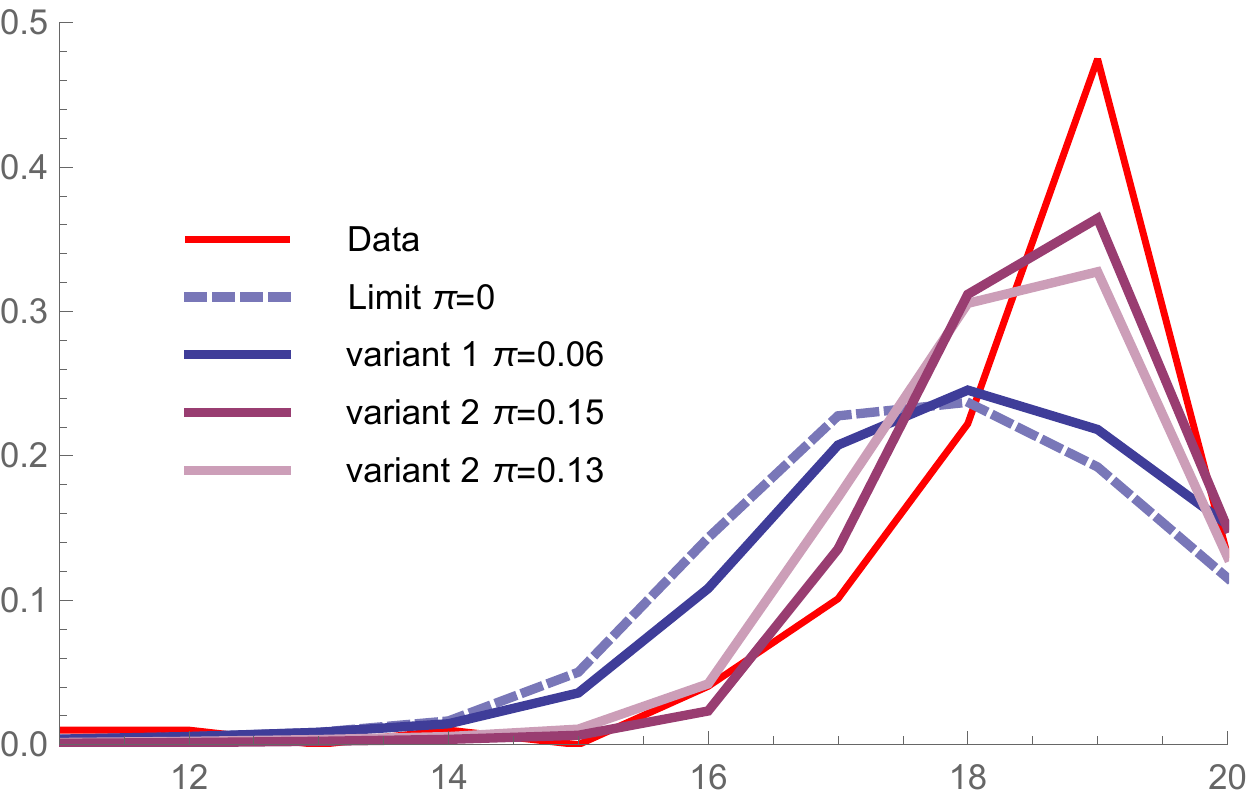}
\subcaption{\label{figcycleversion}The cycle version}%
\end{minipage}\hfill\begin{minipage}{0.45\textwidth}
\centering
	\includegraphics[scale=0.55]{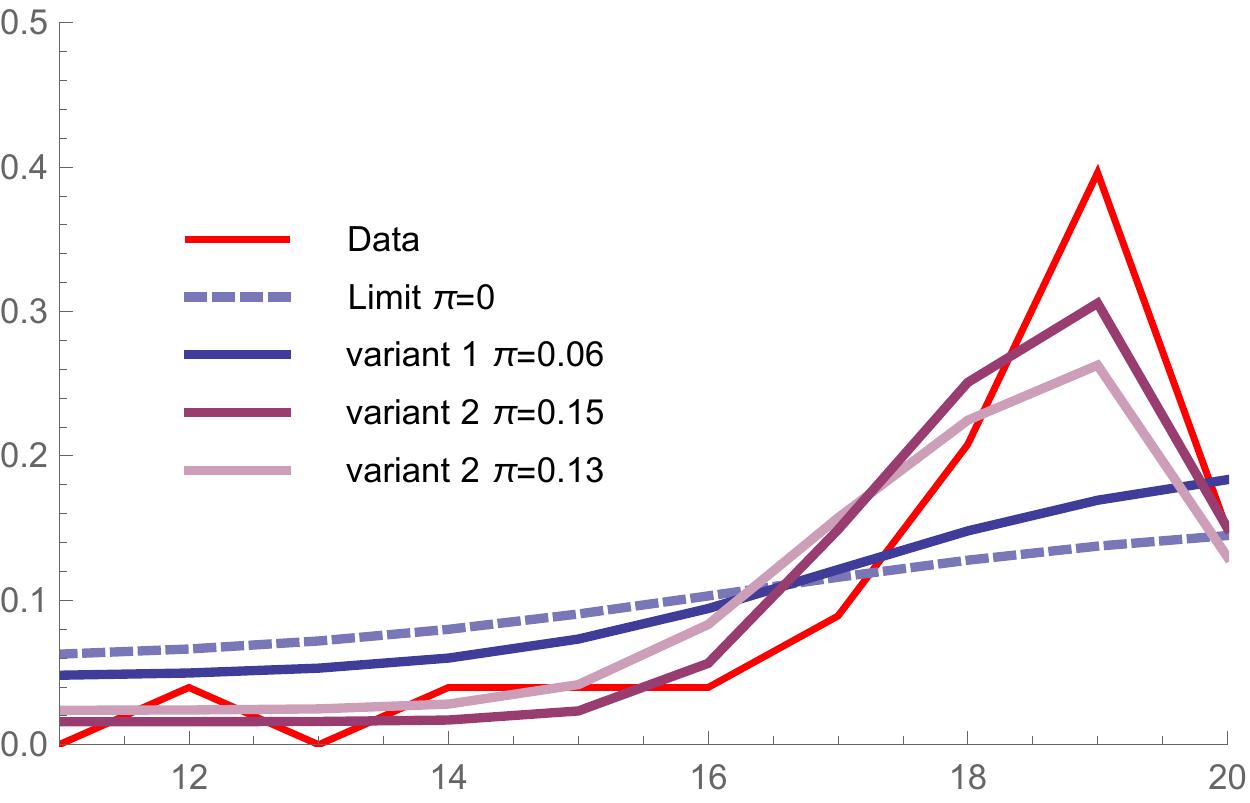}
\subcaption{\label{figcostless}The costless iteration version}%
\end{minipage}
\caption{The 11-20 game with different payoff structures}%
\end{figure}In the \textit{cycle version}, limit distributions are essentially
unchanged (for a fixed $\pi$) compared to the basic version, because the
weight on $11$ is negligible. For variant 2, there is a difference with
Figure~\ref{figvariant} because we used $\pi=0.06$: a larger $\pi$ has the
effect of shifting the distribution of play to the right, with a mode at 19
rather than 18.

In the \textit{costless-iteration} version, lower claims are less penalized so
they get more weight than in the cycle version. Conditional on $t\leq19$,
limit distributions are upward sloping, and a higher $\pi$ induces more
concentration on higher claims. A key difference between the two variants in
that in variant 1, the strategic player includes $t=20$ in her strategy set,
contributing to a large overall weight on $t=20$: the unconditional limit
distribution is upward sloping with a mode at 20. This sharply contrasts with
variant 2, which provides a much better fit with the data for $p$ large enough
(see Figure~\ref{figcostless}).

Another interesting observation about variant 2, in the costless-iterations
version, is that for all $\pi<0.11,$ the expected payoff obtained by strategic
players at the limit distribution is below 20. In other words, strategic
players only have incentives to remain strategic if $\pi\geq0.11$.

\textit{Discussion. }Level-k thinking may certainly be at work in structuring
how people with little experience play or think about the game. Here, the
cognitive levels are proxies for the strategies that players use (according to
the relationship $k=20-t$).
The observation that a large majority of claims lie in the 17-19 range could
be interpreted as a bound on sophistication.

For the basic and cycle versions of the game, our analysis of limit
distributions suggests an alternative explanation, as even with long
experience,
unravelling towards lower claims is unlikely: a large fraction of claims
(about 70\%) are in the 17-19 range, with a mode on 18.


For the costless-iterations version (and to some extent for the cycle
version), our limit distributions do not fit well with the data, unless
variant 2 is considered. Our analysis thus confirms \citet{arad12}'s insight
that a priori strategic thinking about the game may play a key role. Our
method suggests a parsimonious way to incorporate some basic strategic
thinking, partitioning types into non-strategic and strategic ones (as if
there way only two cognitive levels). This leaves partially exogenous the
fraction non-strategic types (partially because we check that strategic types
do not want to mimic non-strategic ones), but it endogenizes the behavior of
strategic ones. The limit distribution obtained can be interpreted as a
distribution over cognitive levels (as in level-k), but not necessarily so. It
just results from (imperfectly) learning which strategic claim is a better
one, with learning being performed by strategic agents only.

As a final observation, we consider a costless-iterations version of the 11-20
game where the strategy set is made finer: players report a claim between 110
and 200 (as if they were making claims in units of 10 cents). There is a bonus
from making a claim between \$0.10 and \$1 below the other, and the largest
bonus is obtained when one is just \$0.10 below.\ Specifically, player $i$
gets 20 when $t_{i}=200$. Otherwise, for any $t_{i}<200$, payoffs are:%
\[
u_{i}(t_{i},t_{j})=17+20(1-\alpha(t_{j}-t_{i}-1))\text{ if }t_{j}-10\leq
t_{i}<t_{j}\text{ and }u_{i}(t_{i},t_{j})=17\text{ otherwise.}%
\]
This game has a payoff structure similar to the previous one. We report in the
Appendix (see Figure~\ref{figARtimed}) the limit distribution obtained for
$\alpha=0.01$ and $0.1,$ and $\pi=0.1$. \ For variant 2 and $\alpha=0.01$, the
limit distribution shows several modes (at 190 and a smaller one at 183).
Strategic claims between 170 and 199 get 67\% of total weight. This weight is
consistent with that obtained for the coarse strategy space (11-20) version of
the game, where, for the same variant 2, claims 17-19 get 64\% of total
weight, illustrating once again the robustness of limit distributions to the
richness of the strategy space.

Regarding the classic level-k explanation, we note that strategic claims
between 170 and 185 get 30\% of total weight. Explaining this large weight
with level-k thinking would require an overly large number of cognitive levels
(at least 15 levels).

\subsection{Auctions}

Our objective in this Section is to analyze a complex game in which the set of
a priori feasible strategies is so rich that it precludes players from
evaluating the performance of all of them. For such games, it is plausible
that players would restrict attention to a subset of strategies. We apply our
limit QRE approach to the restricted strategy set.

Specifically, we consider an auction where each participant has a private
value $v_{i}$ for the object, drawn independently from the same distribution
$f$. In simulations to come, we further assume that $f$ is a lognormal
distribution with variance $\sigma$. The strategy space is a space of
functions, and in principle, all bid functions $b_{i}(v_{i})$ should be
compared. In the spirit of \citet{compte01} and \citet{compte18}, we assume
that players are more limited in their ability to gather experience: they only
compare a limited (here finite) number of strategies, defined as follows. We
fix an increment $\delta>0$ and consider a finite family or grid
$\Gamma_{\delta}$ of linear shading strategies, which, only for simplicity, we
assume identical across players:%
\[
b_{i}^{\lambda_{i}}(v_{i})=\lambda_{i}v_{i}\text{ with }\lambda_{i}\in
\Gamma_{\delta}=\{\delta k_{i},k_{i}\in K\}\text{ and }K=\{k\in\mathcal{N}%
,\delta k_{i}<1\}.
\]
For any profile of strategies $\lambda=(\lambda_{i},\lambda_{-i})$, each value
vector realization $v=(v_{i})_{i}$ induces a vector of bids $b^{\lambda
}(v)=(b_{i}^{\lambda_{i}}(v_{i}))_{i}$, from which we compute, given the
auction format considered, the gain $u_{i}(v_{i},b^{\lambda}(v))$. Taking
expectations over value vector realizations, we compute the (ex ante) expected
gains:%
\[
U_{i}(\lambda)=E_{v}u_{i}(v_{i},b^{\lambda}(v))
\]
Given $U_{i}(\lambda)$ and the grid $\Gamma_{\delta}$ assumed, we adopt the
limit QRE methodology to solve this game, and examine two auction formats,
first-price and all-pay auctions.

\subsubsection{First-Price Auctions\textbf{\ }}

\begin{figure}[h]
\centering
\begin{minipage}{0.45\textwidth}
\centering
\includegraphics[scale=0.57]{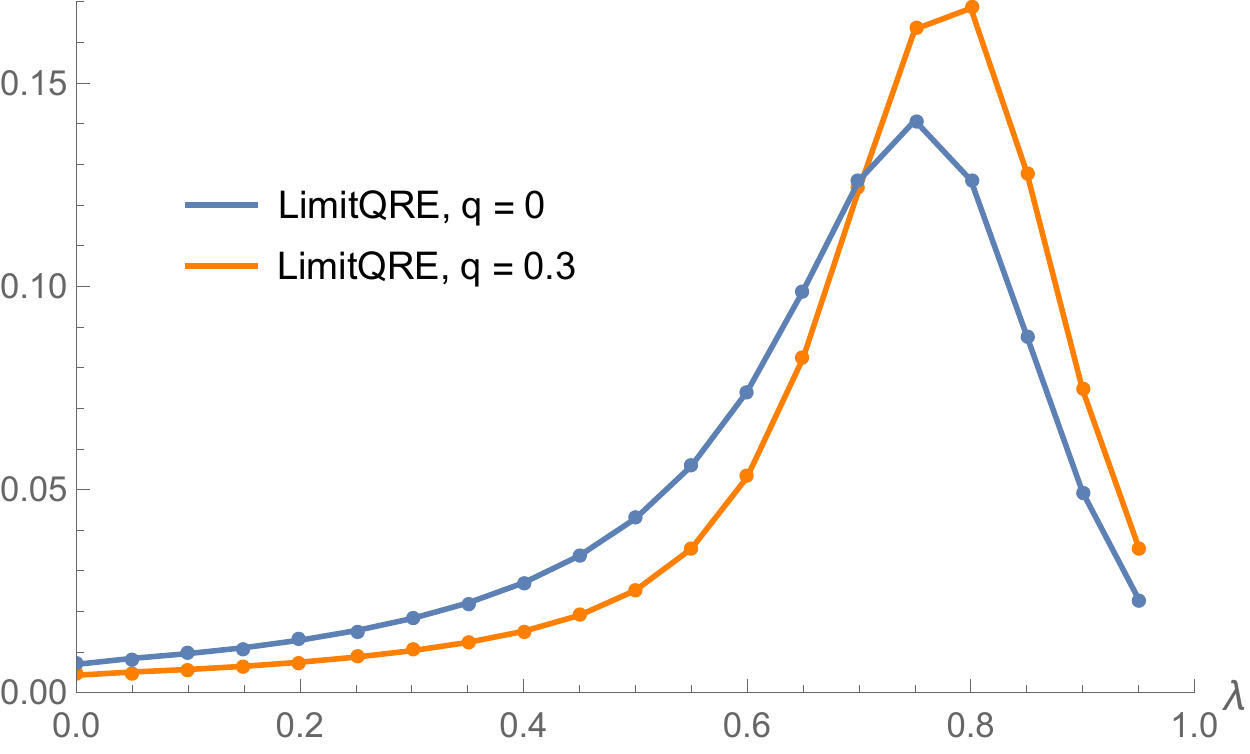}
\subcaption{\label{figauctionfirstpricelimitQRE}Limit distributions $\sigma=0.05$}%
\end{minipage}\hfill\begin{minipage}{0.45\textwidth}
\centering
	\includegraphics[scale=0.55]{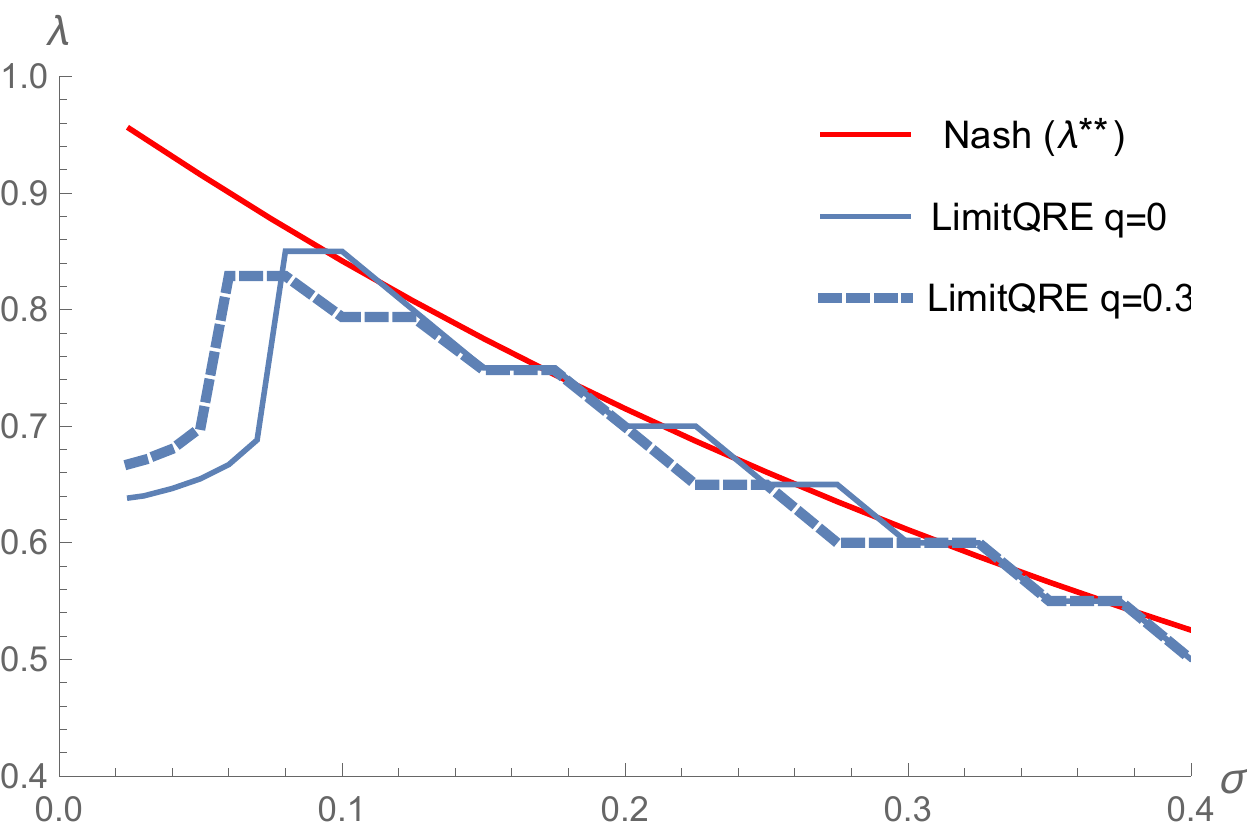}
\subcaption{\label{figauctionfirstpricecomparative}Comparative statics. Expected shading}%
\end{minipage}
\caption{Shading in first price auction.}%
\label{figfirstprice}%
\end{figure}

We focus on two players, choose $\delta=0.05$, and run simulations for
$\sigma\in\lbrack0.025,0.5]$. When values are not too concentrated (i.e.,
$\sigma\geq0.08)$, we find that limit precision is unbounded and the limit
distribution is a Nash equilibrium $\lambda_{\sigma}^{\ast}$ of the game
played over the restricted strategy set $\Gamma_{\delta}$. This Nash
equilibrium approximates the unique Nash equilibrium $\lambda_{\sigma}%
^{\ast\ast}$ of the game played over an arbitrarily fine grid $\delta\simeq0$,
which we plot in Figure~\ref{figauctionfirstpricecomparative}. As expected,
when $\sigma$ increases, competition weakens (i.e., $\lambda_{\sigma}%
^{\ast\ast}$ decreases).

When values are too concentrated (i.e., $\sigma\leq0.075$), limit precision is
bounded: when precision rises, competition is harsh and conducive to high
$\lambda$'s and thus gains close to 0; this makes low $\lambda^{\prime}s$ not
much worse than competitive ones, so (many) low shading factors pick up
significant weight (overall), which fuels instability: there is no reason to
behave very competitively when the other player bids low with a substantial probability.

Figure~\ref{figauctionfirstpricelimitQRE} reports the limit distribution for
$\sigma=0.05$. We also report the induced limit distribution over shading
factors for the game defined over target shading factors, with $q=0.3$.
Stability is improved with noise (limit precision increases and this is
pro-competitive), but at this noise level, the limit precision is still
bounded. The induced distribution over bids leads to slightly higher bids in
expectation. At a higher noise level ($q=0.35)$, limit precision becomes
infinite, and the induced distribution over bids is significantly higher,
though smaller than $\lambda_{0.05}^{\ast}$ because the noise is skewed to the
left for high bids.

Figure~\ref{figauctionfirstpricecomparative} reports expected shading factors
for the Nash outcome $\lambda_{\sigma}^{\ast\ast}$ and the limit~QRE
distribution over bids, for $q=0$ and $q=0.3$. For $\sigma<0.08$, bounded
limit precision fosters bid dispersion, which substantially decrease the
strength of competition, compared to the Nash outcome. For $\sigma=0.05$ for
example, the expected shading factor is equal to $0.65$ at the limit
distribution, while $\lambda_{\sigma}^{\ast\ast}=0.91.$ We suspect that a
similar drop in competition strength would obtain for larger $\sigma$ if we
raised the number of bidders.

In the Appendix, we report another example where the limit precision is
bounded despite the existence of a pure strategy Nash equilibrium. The example
involves dispersion uncertainty, with $\sigma$ taking two possible values
$\underline{\sigma}=0.05$ or $\overline{\sigma}=0.5$, with equal
probability.\ In equilibrium, the chance of a low dispersion realization
$\underline{\sigma}$ makes the Nash equilibrium rather competitive
($\lambda^{\ast}=0.8$). But the limit QRE stops short of this outcome, because
betting on the high dispersion realization $\overline{\sigma}$ yields gains
that are not significantly lower than the optimum, and this fuels instability.
The expected shading level in the limit QRE equilibrium is ($\lambda^{\ast
}=0.71$).


\subsubsection{All-pay Auctions.}

\begin{figure}[h]
\centering
\begin{minipage}{0.45\textwidth}
\centering
\includegraphics[scale=0.6]{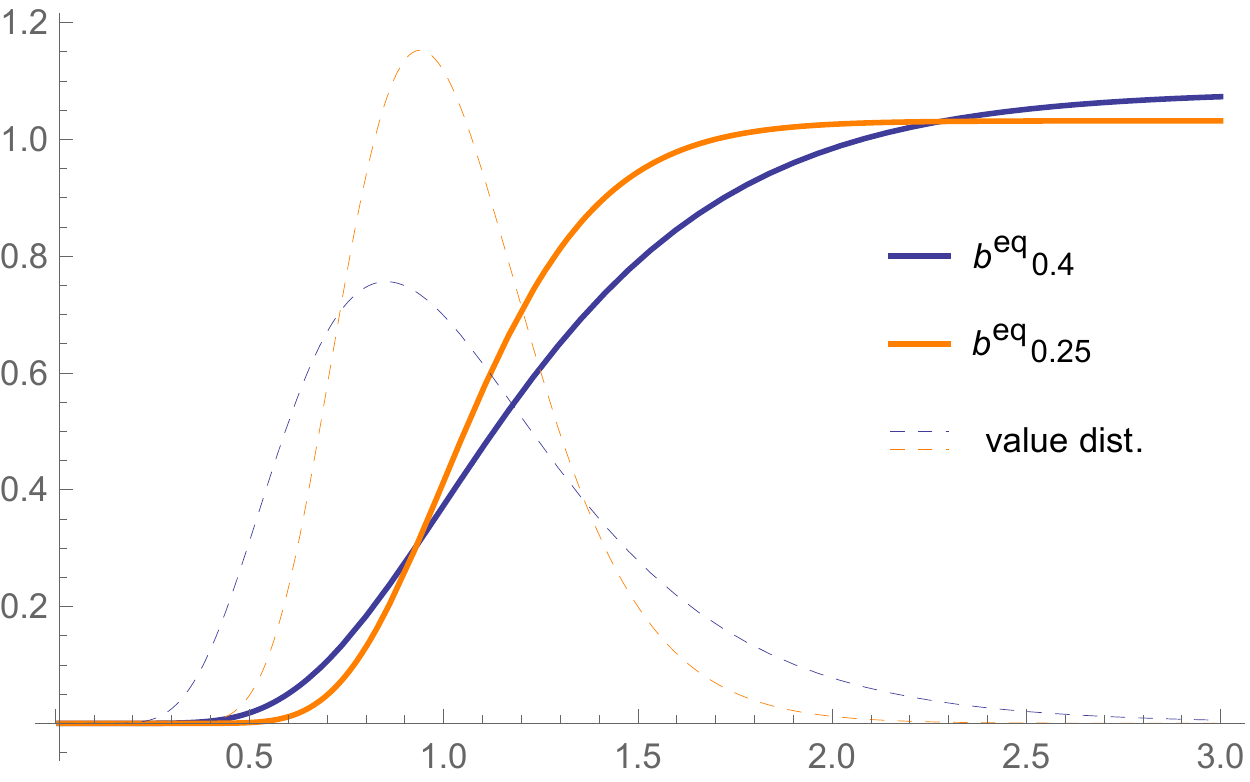}
\subcaption{\label{fig11b}Bayesian solutions for $\sigma=0.4$ and $\sigma=0.25$}
\end{minipage}\hfill\begin{minipage}{0.45\textwidth}
\centering
	 \includegraphics[scale=0.6]{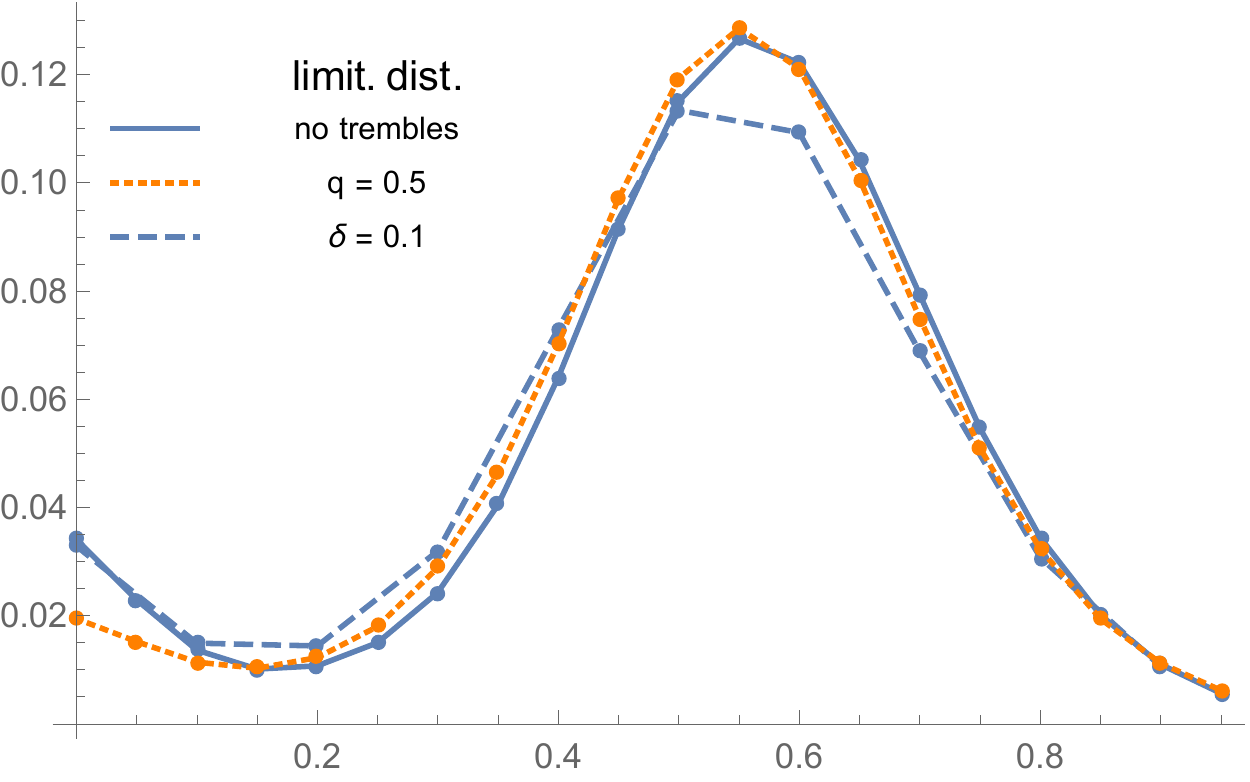}
\subcaption{\label{figallpay04}Limit distributions. $\sigma=0.4$}
\end{minipage}
\caption{All pay auction}%
\end{figure}

With 2 players, the standard Bayesian solution (where no strategy restrictions
are made) is a pure strategy equilibrium and the equilibrium bid can be
derived explicitly using revenue equivalence (See Appendix).
These equilibrium bid functions, denoted $b_{\sigma}^{eq},$ are $S$-shaped,
showing more steepness at the inflection point when $\sigma$ is lower (see
Figure~\ref{fig11b}).\footnote{We also indicate the distribution over values.
The corresponding dashed curves are scaled down vertically (by a factor 0.7).}
Under these Bayesian solutions, players behave as if they drew correct
inferences from getting a low draw of $v$: their chances of winning are slim
hence optimal bidding is close to 0; players are not spending unnecessary
resources on auctions they will most certainly lose.

With linear strategies, saving on low-value realizations cannot be done and
the strategic interaction may differ substantially. When $\sigma$ is larger
than 0.5, the game still has a pure Nash equilibrium $\lambda_{\sigma}^{\ast}%
$, and our limit QRE distribution converges to it. For lower $\sigma$, the
game has the structure of a Rock-Scissors-Paper game (akin to the 11-20 game
-- though with a different payoff structure): if one looks at the
best-response dynamics over pure strategies ($k_{1}$, $k_{2}$ best response to
$k_{1}$ etc...) when $\sigma=0.4$, the sequence obtained eventually cycles
over eight strategies:%
\[
k=0,1,3,6,9,11,12,13,0,1...
\]
Intuitively, when $\sigma$ is not too high, each player has an incentive to
overbid the other for a large range of shading factors: upward competitive
pressures are strong. But when $\lambda$ becomes too high, $\lambda=0$ becomes
a better option. This fuels a cycle of best-responses, and
there is thus no pure strategy equilibrium. Figure~\ref{figallpay04} plots the
limit distribution for $\sigma=0.4$. 

We also indicate the limit distribution obtained under a coarser grid of
linear strategies, where coefficients are multiple of $0.1$ as well as the
limit distribution obtained when players tremble (with $q=0.5$). These
distributions are similar to one another, showing some robustness of our
solution; and they have two modes, reflecting two plausible ways of playing
this game, safe ($\lambda=0$) or aggressive ($\lambda\simeq0.55$). The mode at
0 is the smaller one, but with more than two players, it would increase in magnitude.

\textbf{Including more sophisticated strategies. }
The restriction to linear strategies has a different impact for all-pay
auctions, compared to first price: it does not impair learning for first price
(unless $\sigma$ is very small), while it does for all-pay, giving the latter
auction the structure of a Rock-Scissors-Paper game with no pure strategy
equilibrium. It is thus no surprise that learning is difficult. We check below
whether the inclusion of $S-$shaped strategies, which are better suited for
all-pay, would favor their selection under limit~QRE and the convergence to a
pure NE when it exists. We find that it is not always the case.

Formally we consider strategy sets $\overline{\Gamma}=\Gamma_{\delta}%
\cup\{b_{S}\}$ that consists of the set of linear strategies $\Gamma_{\delta}$
plus an additional $S-$shaped strategy $b_{S}$. In all cases below, we choose
$b_{S}$ so that $b_{S}$ is the unique pure strategy equilibrium of the game
with strategy restriction $\overline{\Gamma}$. Unless otherwise specified, we
set $\delta=0.05$. We observe:

\begin{quote}
(i) If $\sigma=0.4$ and $b_{S}=b_{0.4}^{eq}$, the limit~QRE converges to
$b_{S}$.

(ii) If $\sigma=0.3$ and $b_{S}=b_{0.3}^{eq}$, the limit~QRE puts a weight on
$b_{S}$ equal to 10\%.

(iii) If $\sigma=0.4$, the limit~QRE puts a weight $68\%$ on $b_{S}$ if
$b_{S}=b_{0.3}^{eq}$, $37\%$ if $b_{S}=b_{0.25}^{eq}$

(iv) If $\sigma=0.3$, $b_{S}=b_{0.3}^{eq}$ and $\delta>0.26$ the limit~QRE
converges to $b_{S}$.
\end{quote}

Observation (i) shows that the inclusion of the Bayesian solution may allow
convergence to it. However convergence is not always guaranteed: by (ii),
convergence to $b_{S}$ does not obtain despite the fact that $b_{S}$ is a Nash
equilibrium of the restricted game. The reason is that many linear strategies
are good enough against $b_{S}$, so they get substantial weight, which creates
some instability.

Furthermore, observation (iii) shows that the inclusion of a strategy that is
only \textquotedblleft close to" the equilibrium may not be sufficient to
reach convergence, even if $b_{S}$ is a Nash equilibrium of the restricted
game. As $b_{S}$ goes away from $b_{0.4}^{eq}$, linear strategies are
suboptimal but some of them start performing well against $b_{S}$, and this
fuels instability. When $b_{S}=b_{\sigma}^{eq}$ with $\sigma<0.25$, linear
strategies are no longer suboptimal against $b_{S}$ -- so $b_{S}$ actually
ceases to be a Nash equilibrium in these cases.

Finally, (iv) indicates that a coarsening of the strategy set $\Gamma_{\delta
}$ may help the selection of the Bayesian solution. Instability stems from
outbidding the other player, using a higher linear coefficient. With a
coarsened strategy space, the best linear response cannot be well-tuned to the
other's behavior, and the instability is reduced.

\section{Summary and Conclusion}

We have proposed a general method for making predictions in games.
Technically, the method boils down to looking at the graph of the logit
quantal response equilibria, considering its continuous branch starting at the
lowest precision, and then iteratively raising precision along that branch so
long as at each iteration the logit-response dynamics converges and is locally
stable. The method selects a limit precision and a limit distribution.

Our motivation has been that so long as agents face a stationary environment,
more experience should increase the precision with which alternatives are
evaluated, and that, on the other hand, unstable dynamics put an upperbound on
precision. The method can be applied to any game and provides a game-specific
prediction. When limit precision is bounded, it means that even as experience
accumulates, play will not converge to a Nash equilibrium. There is a barrier
to learning that translates into an endogenously derived behavioral dispersion.

We have found our prediction to be robust in three ways. First, the limit
distribution does not depend on the scaling of payoffs. Second, when one adds
exogenous trembles on nearby strategies, limit precision rises, but, to the
extent that it remains bounded, the limit distribution that results from
endogenous stochastic choice and exogenous trembles is not notably affected.
Finally, considering games where the strategy space can be expressed as a
subset of an interval (spanning uniformly that interval), we found the limit
distribution barely sensitive to the discretization considered.

We have applied the method to well-known games for which departures from Nash
is well-documented: the centipede, the traveler's dilemma and the 11-20
request games. Our limit distributions are broadly consistent with the data,
except for a variant of the 11-20 request game. For that variant, the data
suggests a weight on 20 that cannot be explained with our method. However, we
also examine an extension where we distinguish between naive players who play
20 and strategic players who choose strictly below 20. Applying our method to
endogenize the behavior of strategic players, we obtain a good fit to the data.

We have also applied the method to first-price and all-pay auctions, assuming
that players compare a finite number of linear shading strategies. For
first-price, limit precision is unbounded, except for auctions where
dispersion of values is very small. For all-pay, for a large range of value
dispersion, the game has no pure strategy equilibrium, so naturally, limit
precision is bounded. However, limit precision may remain bounded even if we
include the Bayesian Nash solution in the strategy space. This failure to
converge to Nash even when the Nash solution is available suggests that
learning to play an equilibrium is harder for all-pay than first-price auctions.

The 11-20 request game and all-pay auctions with linear strategies have in
common that there are no pure strategy equilibria. For such games with no pure
strategy equilibria, limit precision is bounded, because for large enough
precision, the logit-response dynamics are necessarily unstable.

The centipede game, the traveler's dilemma, the first price auction where
dispersion of value is low, and the all-pay auction where the Bayesian
solution is added to the strategy space, all have in common that there exists
a pure strategy equilibrium, yet limit precision is bounded. In these four
games, competitive pressures lead to low equilibrium payoffs when precision is
high, but the loss to using alternative strategies is small. This implies
that, as precision rises, significant weight is put on these less competitive
alternatives, and this fuels instability, because slightly undercutting these
strategies (or slightly overbidding them in the case of auctions) becomes attractive.

\textbf{Further research}. We have identified some robustness of our
predictions to rescaling of payoffs and to the discretization of the strategy
space. It would be worth exploring more systematically the reality of this
robustness in experiments. We have modelled imperfect learning using a logit
choice model, with the logit parameter characterizing the extent of learning.
This is a crude modelling device, which implicitly assumes a particular shape
of estimation errors, around expected gains. In practice, the shape of errors
could be different, stem from matching or finite sampling from past data for
example, or depend on the action considered. It would be worth exploring
whether similar insights regarding limit precision could be derived under
alternative stochastic choice models. Finally, we note that we derive an
upperbound on precision from stability considerations only. In practice other
sources of errors may well be preponderant and drive behavior. Some strategies
may also appear focal to some players, shaping the responses of others. We
explored an example along this line when discussing a version of the 11-20
money request game, introducing a model with only two cognitive types,
strategic and non-strategic. The method may prove useful in other games, as a
way to remain parsimonious in the number of cognitive types assumed.

\begin{singlespace}
\addcontentsline{toc}{section}{References}
\bibliographystyle{chicago}
\bibliography{reflib}

\begin{thebibliography}{}

\bibitem[\protect\citeauthoryear{Anderson, Goeree, and Holt}{Anderson
  et~al.}{1998}]{anderson98}
Anderson, S.~P., J.~K. Goeree, and C.~A. Holt (1998).
\newblock Rent seeking with bounded rationality: An analysis of the all‐pay
  auction.
\newblock {\em Journal of Political Economy\/}~{\em 106\/}(4), 828--853.

\bibitem[\protect\citeauthoryear{Anderson, Goeree, and Holt}{Anderson
  et~al.}{2002}]{anderson02}
Anderson, S.~P., J.~K. Goeree, and C.~A. Holt (2002).
\newblock The logit equilibrium: A perspective on intuitive behavioral
  anomalies.
\newblock {\em Southern Economic Journal\/}~{\em 69\/}(1), 21--47.

\bibitem[\protect\citeauthoryear{Arad and Rubinstein}{Arad and
  Rubinstein}{2012}]{arad12}
Arad, A. and A.~Rubinstein (2012).
\newblock The 11-20 money request game: A level-k reasoning study.
\newblock {\em The American Economic Review\/}~{\em 102\/}(7), 3561--3573.

\bibitem[\protect\citeauthoryear{Basu}{Basu}{1994}]{basu94}
Basu, K. (1994).
\newblock The traveler's dilemma: Paradoxes of rationality in game theory.
\newblock {\em The American Economic Review\/}~{\em 84\/}(2), 391--395.

\bibitem[\protect\citeauthoryear{Camerer, Nunnari, and Palfrey}{Camerer
  et~al.}{2016}]{camerer16}
Camerer, C., S.~Nunnari, and T.~R. Palfrey (2016).
\newblock Quantal response and nonequilibrium beliefs explain overbidding in
  maximum-value auctions.
\newblock {\em Games and Economic Behavior\/}~{\em 98}, 243--263.

\bibitem[\protect\citeauthoryear{Capra, Goeree, Gomez, and Holt}{Capra
  et~al.}{1999}]{capra99}
Capra, C.~M., J.~K. Goeree, R.~Gomez, and C.~A. Holt (1999).
\newblock Anomalous behavior in a traveler's dilemma?
\newblock {\em The American Economic Review\/}~{\em 89\/}(3), 678--690.

\bibitem[\protect\citeauthoryear{Carlsson}{Carlsson}{1991}]{carlsson91}
Carlsson, H. (1991).
\newblock A bargaining model where parties make errors.
\newblock {\em Econometrica\/}~{\em 59\/}(5), 1487--1496.

\bibitem[\protect\citeauthoryear{Carlsson and van Damme}{Carlsson and van
  Damme}{1993}]{carlsson93}
Carlsson, H. and E.~van Damme (1993).
\newblock Global games and equilibrium selection.
\newblock {\em Econometrica\/}~{\em 61\/}(5), 989--1018.

\bibitem[\protect\citeauthoryear{Casella, Compte, and Si}{Casella
  et~al.}{2025}]{casella25}
Casella, A., O.~Compte, and S.~Si (2025).
\newblock Restricting strategy sets in complex games.

\bibitem[\protect\citeauthoryear{Compte}{Compte}{2001}]{compte01}
Compte, O. (2001).
\newblock The winner's curse with independent private values.

\bibitem[\protect\citeauthoryear{Compte}{Compte}{2023}]{compte23}
Compte, O. (2023).
\newblock Endogenous barriers to learning.
\newblock {\em Arxiv 2306.16904\/}.

\bibitem[\protect\citeauthoryear{Compte}{Compte}{2025}]{compte25}
Compte, O. (2025).
\newblock Learned collusion.

\bibitem[\protect\citeauthoryear{Compte and Postlewaite}{Compte and
  Postlewaite}{2018}]{compte18}
Compte, O. and A.~Postlewaite (2018).
\newblock {\em Ignorance and Uncertainty}.
\newblock Econometric Society Monographs. Cambridge University Press.

\bibitem[\protect\citeauthoryear{Crawford, Costa-Gomes, and Iriberri}{Crawford
  et~al.}{2013}]{crawford2013}
Crawford, V., M.~Costa-Gomes, and N.~Iriberri (2013).
\newblock Structural models of nonequilibrium strategic thinking: Theory,
  evidence, and applications.
\newblock {\em Journal of Economic Literature\/}~{\em 51\/}(1), 5–62.

\bibitem[\protect\citeauthoryear{Danenberg and Spiegler}{Danenberg and
  Spiegler}{2022}]{danenberg22}
Danenberg, T. and R.~Spiegler (2022).
\newblock Representative sampling equilibrium.

\bibitem[\protect\citeauthoryear{Erev and Roth}{Erev and Roth}{1998}]{erev98}
Erev, I. and A.~E. Roth (1998).
\newblock Predicting how people play games: Reinforcement learning in
  experimental games with unique, mixed strategy equilibria.
\newblock {\em The American Economic Review\/}~{\em 88\/}(4), 848--881.

\bibitem[\protect\citeauthoryear{Fey, McKelvey, and Palfrey}{Fey
  et~al.}{1996}]{fey96}
Fey, M., R.~D. McKelvey, and T.~R. Palfrey (1996).
\newblock An experimental study of constant-sum centipede games.
\newblock {\em International Journal of Game Theory\/}~{\em 25\/}(3),
  269‑287.

\bibitem[\protect\citeauthoryear{Friedman}{Friedman}{2022}]{friedman22}
Friedman, E. (2022, February).
\newblock Stochastic equilibria: Noise in actions or beliefs?
\newblock {\em American Economic Journal: Microeconomics\/}~{\em 14\/}(1),
  94–142.

\bibitem[\protect\citeauthoryear{Friedman and Mezzetti}{Friedman and
  Mezzetti}{2005}]{friedman05}
Friedman, J.~W. and C.~Mezzetti (2005).
\newblock Random belief equilibrium in normal form games.
\newblock {\em Games and Economic Behavior\/}~{\em 51\/}(2), 296--323.

\bibitem[\protect\citeauthoryear{Fudenberg and Levine}{Fudenberg and
  Levine}{2009}]{fudenberg09}
Fudenberg, D. and D.~K. Levine (2009).
\newblock Learning and equilibrium.
\newblock {\em Annual Review of Economics\/}~{\em 1}, 385--420.

\bibitem[\protect\citeauthoryear{García-Pola, Iriberri, and
  Kovářík}{García-Pola et~al.}{2020}]{garciapola20}
García-Pola, B., N.~Iriberri, and J.~Kovářík (2020).
\newblock Non-equilibrium play in centipede games.
\newblock {\em Games and Economic Behavior\/}~{\em 120}, 391--433.

\bibitem[\protect\citeauthoryear{Goeree, Holt, and Palfrey}{Goeree
  et~al.}{2002}]{goeree02}
Goeree, J.~K., C.~A. Holt, and T.~R. Palfrey (2002).
\newblock Quantal response equilibrium and overbidding in private-value
  auctions.
\newblock {\em Journal of Economic Theory\/}~{\em 104\/}(1), 247--272.

\bibitem[\protect\citeauthoryear{Goeree, Louis, and Zhang}{Goeree
  et~al.}{2017}]{goeree17}
Goeree, J.~K., P.~Louis, and J.~Zhang (2017).
\newblock Noisy introspection in the 11–20 game.
\newblock {\em The Economic Journal\/}~{\em 128\/}(611), 1509--1530.

\bibitem[\protect\citeauthoryear{Harsanyi and Selten}{Harsanyi and
  Selten}{1988}]{harsanyi88}
Harsanyi, J. and R.~Selten (1988).
\newblock {\em A general theory of equilibrium selection in games}.
\newblock MIT press.

\bibitem[\protect\citeauthoryear{Hopkins}{Hopkins}{1999}]{hopkins99}
Hopkins, E. (1999).
\newblock A note on best response dynamics.
\newblock {\em Games and Economic Behavior\/}~{\em 29\/}(1), 138--150.

\bibitem[\protect\citeauthoryear{Hopkins}{Hopkins}{2002}]{hopkins02}
Hopkins, E. (2002).
\newblock Two competing models of how people learn in games.
\newblock {\em Econometrica\/}~{\em 70\/}(6), 2141--2166.

\bibitem[\protect\citeauthoryear{Kawagoe and Takizawa}{Kawagoe and
  Takizawa}{2012}]{kawagoe12}
Kawagoe, T. and H.~Takizawa (2012).
\newblock Level-k analysis of experimental centipede games.
\newblock {\em Journal of Economic Behavior \& Organization\/}~{\em 82\/}(2),
  548--566.

\bibitem[\protect\citeauthoryear{Kreps, Milgrom, Roberts, and Wilson}{Kreps
  et~al.}{1982}]{kreps82}
Kreps, D., P.~Milgrom, J.~Roberts, and R.~Wilson (1982).
\newblock Rational cooperation in the finitely repeated prisoners' dilemma.
\newblock {\em Journal of Economic Theory\/}~{\em 27\/}(2), 245--252.

\bibitem[\protect\citeauthoryear{Levitt, List, and Sadoff}{Levitt
  et~al.}{2011}]{levitt11}
Levitt, S.~D., J.~A. List, and S.~E. Sadoff (2011).
\newblock Checkmate: Exploring backward induction among chess players.
\newblock {\em American Economic Review\/}~{\em 101\/}(2), 975--90.

\bibitem[\protect\citeauthoryear{Manzini and Mariotti}{Manzini and
  Mariotti}{2014}]{manzini14}
Manzini, P. and M.~Mariotti (2014).
\newblock Stochastic choice and consideration sets.
\newblock {\em Econometrica\/}~{\em 82\/}(3), 1153--1176.

\bibitem[\protect\citeauthoryear{Matějka and McKay}{Matějka and
  McKay}{2015}]{matejka15}
Matějka, F. and A.~McKay (2015).
\newblock Rational inattention to discrete choices: A new foundation for the
  multinomial logit model.
\newblock {\em The American Economic Review\/}~{\em 105\/}(1), 272--298.

\bibitem[\protect\citeauthoryear{Maynard~Smith}{Maynard~Smith}{1972}]{maynard72}
Maynard~Smith, J. (1972).
\newblock {\em On evolution}.
\newblock Edinburgh University Press.

\bibitem[\protect\citeauthoryear{McKelvey and Palfrey}{McKelvey and
  Palfrey}{1998}]{mckelvey98}
McKelvey, R. and T.~Palfrey (1998).
\newblock Quantal response equilibria for extensive form games.
\newblock {\em Experimental Economics\/}.

\bibitem[\protect\citeauthoryear{McKelvey and Palfrey}{McKelvey and
  Palfrey}{1992}]{mckelvey92}
McKelvey, R.~D. and T.~R. Palfrey (1992).
\newblock An experimental study of the centipede game.
\newblock {\em Econometrica\/}~{\em 60\/}(4), 803--836.

\bibitem[\protect\citeauthoryear{McKelvey and Palfrey}{McKelvey and
  Palfrey}{1995}]{mckelvey95}
McKelvey, R.~D. and T.~R. Palfrey (1995).
\newblock Quantal response equilibria for normal form games.
\newblock {\em Games and Economic Behavior\/}~{\em 10\/}(1), 6--38.

\bibitem[\protect\citeauthoryear{McKelvey, Palfrey, and Weber}{McKelvey
  et~al.}{2000}]{mckelvey00}
McKelvey, R.~D., T.~R. Palfrey, and R.~A. Weber (2000).
\newblock The effects of payoff magnitude and heterogeneity on behavior in 2×2
  games with unique mixed strategy equilibria.
\newblock {\em Journal of Economic Behavior and Organization\/}~{\em 42\/}(4),
  523--548.

\bibitem[\protect\citeauthoryear{Nagel}{Nagel}{1995}]{nagel95}
Nagel, R. (1995).
\newblock Unraveling in guessing games: an experimental study.
\newblock {\em American Economic Review\/}.

\bibitem[\protect\citeauthoryear{Nagel and Tang}{Nagel and
  Tang}{1998}]{nagel98}
Nagel, R. and F.~F. Tang (1998).
\newblock Experimental results on the centipede game in normal form: An
  investigation on learning.
\newblock {\em Journal of Mathematical Psychology\/}~{\em 42\/}(2), 356--384.

\bibitem[\protect\citeauthoryear{Nash}{Nash}{1953}]{nash53}
Nash, J. (1953).
\newblock Two-person cooperative games.
\newblock {\em Econometrica\/}~{\em 21\/}(1), 128--140.

\bibitem[\protect\citeauthoryear{Osborne and Rubinstein}{Osborne and
  Rubinstein}{1998}]{osborne98}
Osborne, M.~J. and A.~Rubinstein (1998).
\newblock Games with procedurally rational players.
\newblock {\em The American Economic Review\/}~{\em 88\/}(4), 834--847.

\bibitem[\protect\citeauthoryear{Palacios-Huerta and Volij}{Palacios-Huerta and
  Volij}{2009}]{palacios09}
Palacios-Huerta, I. and O.~Volij (2009).
\newblock Field centipedes.
\newblock {\em American Economic Review\/}~{\em 99\/}(4), 1619--35.

\bibitem[\protect\citeauthoryear{Rand and Nowak}{Rand and Nowak}{2012}]{rand12}
Rand, D.~G. and M.~A. Nowak (2012).
\newblock Evolutionary dynamics in finite populations can explain the full
  range of cooperative behaviors observed in the centipede game.
\newblock {\em Journal of Theoretical Biology\/}~{\em 300}, 212--221.

\bibitem[\protect\citeauthoryear{Rogers, Palfrey, and Camerer}{Rogers
  et~al.}{2009}]{rogers09}
Rogers, B.~W., T.~R. Palfrey, and C.~F. Camerer (2009).
\newblock Heterogeneous quantal response equilibrium and cognitive hierarchies.
\newblock {\em Journal of Economic Theory\/}~{\em 144\/}(4), 1440--1467.

\bibitem[\protect\citeauthoryear{Rosenthal}{Rosenthal}{1981}]{rosenthal81}
Rosenthal, R.~W. (1981).
\newblock Games of perfect information, predatory pricing and the chain-store
  paradox.
\newblock {\em Journal of Economic Theory\/}~{\em 25\/}(1), 92--100.

\bibitem[\protect\citeauthoryear{Roth and Erev}{Roth and Erev}{1995}]{roth95}
Roth, A.~E. and I.~Erev (1995).
\newblock Learning in extensive-form games: Experimental data and simple
  dynamic models in the intermediate term.
\newblock {\em Games and Economic Behavior\/}~{\em 8\/}(1), 164--212.

\bibitem[\protect\citeauthoryear{Simon}{Simon}{1955}]{simon55}
Simon, H.~A. (1955).
\newblock A behavioral model of rational choice.
\newblock {\em The Quarterly Journal of Economics\/}~{\em 69\/}(1), 99--118.

\bibitem[\protect\citeauthoryear{Turocy}{Turocy}{2005}]{turocy05}
Turocy, T.~L. (2005).
\newblock A dynamic homotopy interpretation of the logistic quantal response
  equilibrium correspondence.
\newblock {\em Games and Economic Behavior\/}~{\em 51\/}(2), 243--263.

\end{thebibliography}
\end{singlespace}

\appendix

\section*{Appendix}

\phantomsection\addcontentsline{toc}{section}{Appendix}

\subsection*{\label{proof}Proof of Proposition 1.}

Consider any $\widehat{\beta}<\overline{\beta}$. $\overline{r}_{\beta
,p(\beta)}$ is continuous, so it is bounded away from 1 on $[0,\widehat{\beta
}]$. In addition, over $[0,\overline{\beta}],$ the second derivatives of
$\phi_{\beta}(p)$ with respect to $p$ are uniformly bounded, so local
convergence to $p(\beta)$ obtains for a small enough ball, say of size
$\varepsilon_{0}$, \textit{uniformly over all }$\beta$\textit{'s in
}$[0,\widehat{\beta}]$. Next, since $p(\beta)$ is continuous over the compact
$[0,\overline{\beta}]$, we can choose $\nu_{0}$ so that for any $\nu\leq
\nu_{0}$ and any $\beta\in\lbrack\nu,\widehat{\beta}]$, $p(\beta-\nu)\in
B_{\varepsilon_{0}}(p(\beta))$. This ensures that condition (ii) is
automatically satisfied for any $\nu\leq\nu_{0}$ and any sequence $(\beta
_{k},p(\beta_{k}))_{k}$ for which condition (i) holds and $\beta_{k}%
\leq\widehat{\beta}$. So $\beta_{\nu}^{\ast}\geq\widehat{\beta}-\nu$ for all
$\nu\leq\nu_{0}$.

When the game has a unique QR equilibrium at all $\beta$, the continuous
selection $p(\beta)$ is uniquely defined and the evolutionary path necessarily
takes the form ($(\beta_{k},p(\beta_{k})))_{k}$. Let $\overline{\beta}%
^{\ast\ast}=\inf\{\beta,\overline{r}_{\beta,p(\beta)}>1\}$. If $\beta
_{k}>\overline{\beta}^{\ast\ast}$, then $\overline{r}_{\beta_{k},p(\beta_{k}%
)}>1$ so local stability fails, so we necessarily have $\beta_{\nu}^{\ast}%
\leq\overline{\beta}^{\ast\ast}$, hence $\overline{\beta}^{\ast}\leq
\overline{\beta}^{\ast\ast}$.


\subsection*{Centipede games}

\subsubsection*{Linearly increasing pie}

\begin{figure}[h]
\centering
\includegraphics[scale=0.7]{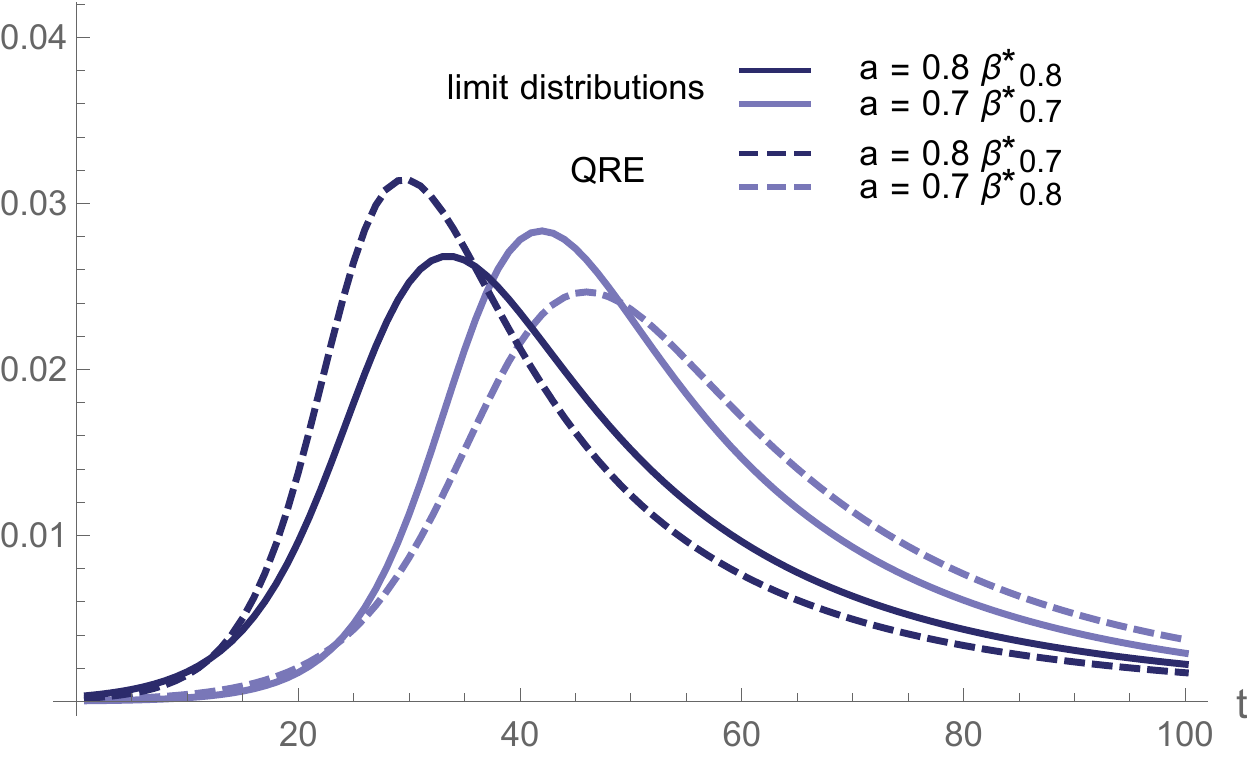}\caption{Endogenous
versus fixed $\beta$}%
\label{figcomparative}%
\end{figure}

\textbf{Effect of endogeneity of }$\beta$. When $a$ rises, incentives to
choose a lower date are stronger so, for a fixed $\beta$, there is more
unravelling. However there is a countervailing force because a higher $a$
induces more instability so $\beta_{a}^{\ast}$ decreases. Figure
\ref{figcomparative} illustrates this, comparing $a=0.7$ and $a=0.8$. We have
$\beta_{0.8}^{\ast}=0.23$ and $\beta_{0.7}^{\ast}=0.3$. We plot the limit
distributions, as well as the QRE outcomes when $a=0.8$ and $\beta=\beta
_{0.7}^{\ast}$, and when $a=0.7$ and $\beta=\beta_{0.8}^{\ast}$.
Figure~\ref{figcomparative} illustrates that the limit
distribution is less responsive to changes in the exogenous variable $a$ than
the QRE prediction with fixed $\beta$, in terms of both mean response and variance.


\textbf{An alternative to logit.} The logit formulation of QR assumes a
particular structure on errors. We wish to illustrate the properties of
limit~QRE with other formulations that implicitly incorporate a notion of
satisficing -- which may be relevant from a learning perspective.
Specifically, we fix $c\geq1$ and examine the stochastic choice model
\[
p_{k}\text{ is proportional to }\exp-(\beta(\overline{u}-u_{k}))^{c}%
\]
where $\overline{u}\equiv\max_{k}u_{k}$ is the maximum expected gain. For
$c=1$, this corresponds to logit choice. For $c$ large, there is sharp
decrease in weight on alternatives $k$ that are too far away from the best one
(i.e. for which $\beta(\overline{u}-u_{k})>1$) and a rather constant weight on
the others.

\begin{figure}[h]
\centering
\includegraphics[scale=0.55]{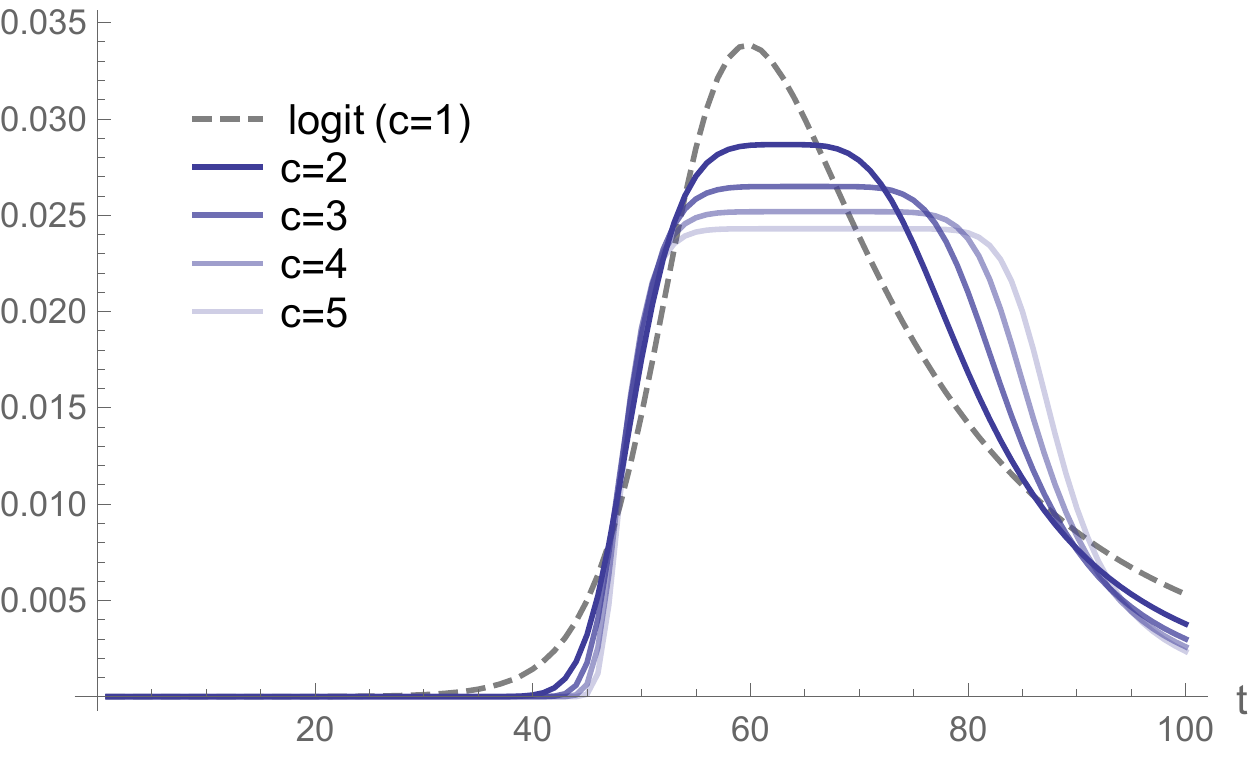}\caption{LimitQRE under an alternative
to logit}%
\label{fig15}%
\end{figure}

We illustrate the consequence of a higher $c$ in the centipede game examined
in Section~\ref{sectionlinear}, with linearly increasing pie, $T_{1}%
=T_{2}=\{1,...,\overline{\tau}\}$, with $\overline{\tau}=100$. Fixing $a=0.6$,
we compute the limit~QRE obtained for different value of $c$.
Figure~\ref{fig15} reports the limit distributions obtained. 

As one raises $c$, more strategies get comparable weight. For large $c$, the
limit QRE comes close to a set-concept that determines a range of \textit{good
enough strategies}, which are then all played with comparable weight, as if
the agent were only discriminating between good enough and poorly performing
strategies.\smallskip

\textbf{Robustness to payoff-rescaling. }If iterations of $\phi_{\beta}$
converge (resp. diverge), and if one multiplies all payoffs by a constant
$\alpha$, then iterations of $\phi_{\beta/\alpha}$ converge (resp. diverge)
for the rescaled game. The consequence is that the limit precision
$\beta^{\ast}$ depends on how payoffs are scaled while the limited
distribution does not. When only one player's payoffs are rescaled however,
the limit distribution may be affected.

\begin{figure}[h]
\centering
\includegraphics[scale=0.65]{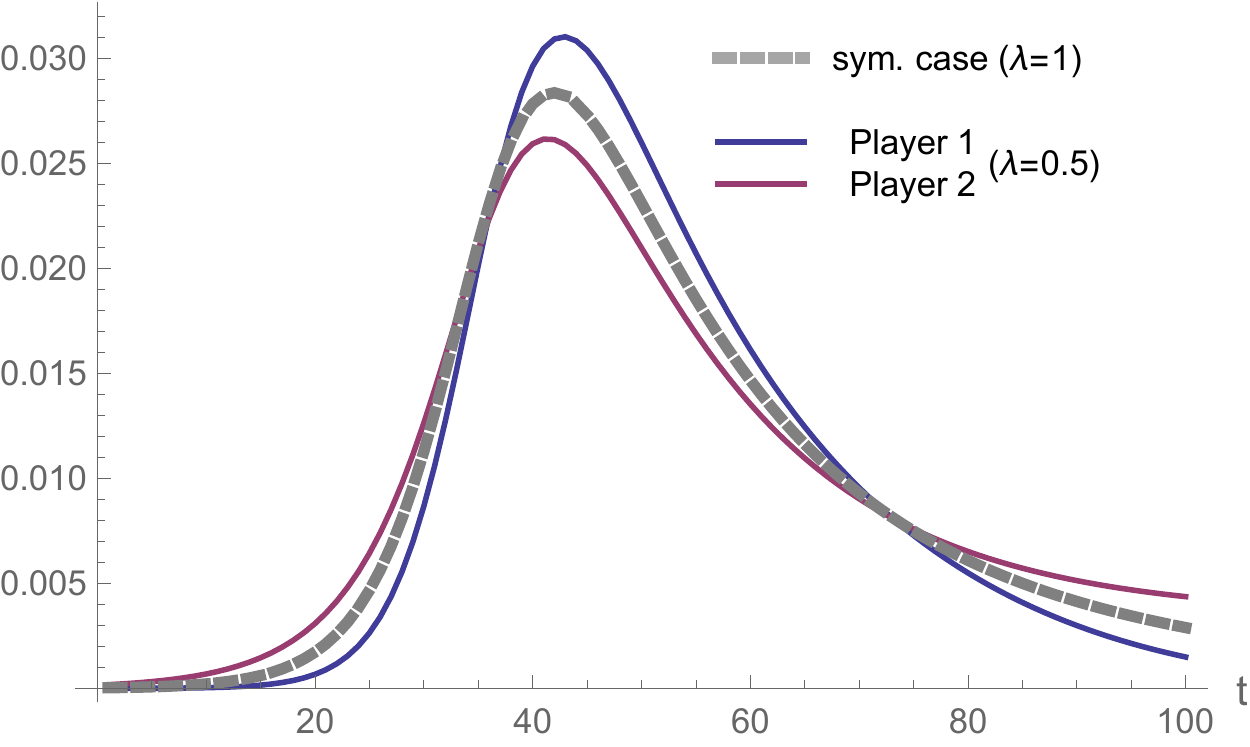}\caption{Limit distribution
under payoff-rescaling for player 2}%
\label{figrescaling}%
\end{figure}

We illustrate this with a rescaled version of the centipede game where only
player 2's payoffs are rescaled, by a factor $\lambda$. That is, player 2
obtains $u_{2}^{\lambda}(t_{1},t_{2})=\lambda u_{2}(t_{1},t_{2})$. Assuming
$a=0.7$, Figure~\ref{figrescaling} reports the limit distribution obtained in
the symmetric game where $\lambda=1$, and the limit distributions obtained in
the asymmetric game where $\lambda=1/2$.\ Distributions differ slightly, with
small shifts (upward for player 1 and downward for player 2), and more
dispersion for player 2. Overall however, expected exit times are very similar
(51.99 for player 1 and 51.9 for player 2, and 51.96 in the symmetric case).

\subsection*{\label{app1bis}Exponentially growing pie: comparison with
experimental data}

Figure \ref{figMPgame} reports the games studied in \citet{mckelvey92} and
\citet{nagel98}. \begin{figure}[h]
\centering
\begin{minipage}{0.45\textwidth}
\centering
\begin{tikzpicture}[font=\footnotesize,scale=0.45]
\tikzstyle{solid node}=[circle,draw,inner sep=1,fill=black];
\tikzstyle{hollow node}=[circle,draw,inner sep=1.2];
\node(0)[hollow node]{}
child[grow=down]{node[solid node]{}
}
child[grow=right]{node(1)[hollow node]{}
child[grow=down]{node[solid node]{}
}
child[grow=right]{node(2)[hollow node]{}
child[grow=down]{node[solid node]{}
}
child[grow=right]{node(3)[hollow node]{}
child[grow=down]{node[solid node]{}
}
child[grow=right]{node(4)[hollow node]{}
child[grow=down]{node[solid node]{}
}
child[grow=right]{node(5)[hollow node]{}
child[grow=down]{node[solid node]{}
}
child[grow=right]{node(6)[solid node]{}
}
}
}
}
}
};
\foreach \x in {0,2,4}
\node[above]at(\x){1};
\foreach \x in {1,3,5}
\node[above]at(\x){2};
\node[below]at(0-1){$\begin{array}
[c]{c}%
4\\
1
\end{array}$};
\node[below]at(1-1){$\begin{array}
[c]{c}%
2\\
8
\end{array}$};
\node[below]at(2-1){$\begin{array}
[c]{c}%
16\\
4
\end{array}$};
\node[below]at(3-1){$\begin{array}
[c]{c}
8\\
32
\end{array}$};
\node[below]at(4-1){
$\begin{array}
[c]{c}%
64\\
16
\end{array}$
};
\node[below]at(5-1){
$\begin{array}
[c]{c}%
32\\
128
\end{array}$
};
\node[right]at(6){
$\begin{array}
[c]{c}%
256\\
64
\end{array}$
};
\end{tikzpicture}
\subcaption{\label{figMPgame}6-node
(McKelvey-Palfrey)}
\end{minipage}\hfill\begin{minipage}{0.45\textwidth}
\centering
\begin{tikzpicture}[font=\footnotesize,scale=0.35]
\tikzstyle{solid node}=[circle,draw,inner sep=1.2,fill=black];
\tikzstyle{hollow node}=[circle,draw,inner sep=1.2];
\node(0)[hollow node]{}
child[grow=down]{node[solid node]{}}
child[grow=right]{node(1)[hollow node]{}
child[grow=down]{node[solid node]{}}
child[grow=right]{node(2)[hollow node]{}
child[grow=down]{node[solid node]{}}
child[grow=right]{node(3)[hollow node]{}
child[grow=down]{node[solid node]{}}
child[grow=right]{node(4)[hollow node]{}
child[grow=down]{node[solid node]{}}
child[grow=right]{node(5)[hollow node]{}
child[grow=down]{node[solid node]{}}
child[grow=right]{node(6)[hollow node]{}
child[grow=down]{node[solid node]{}}
child[grow=right]{node(7)[hollow node]{}
child[grow=down]{node[solid node]{}}
child[grow=right]{node(8)[hollow node]{}
child[grow=down]{node[solid node]{}}
child[grow=right]{node(9)[hollow node]{}
child[grow=down]{node[solid node]{}}
child[grow=right]{node(10)[hollow node]{}
child[grow=down]{node[solid node]{}}
child[grow=right]{node(11)[hollow node]{}
child[grow=down]{node[solid node]{}}
child[grow=right]{node(12)[hollow node]{}
}
}
}
}
}
}
}
}
}
}
}
};
\foreach \x in {0,2,4,6,8,10}
\node[above]at(\x){1};
\foreach \x in {1,3,5,7,9,11}
\node[above]at(\x){2};
\node[below]at(0-1){
$\begin{array}
[c]{c}%
4\\
1
\end{array}$};
\node[below]at(1-1){
$\begin{array}
[c]{c}%
2\\
5
\end{array}$};
\node[below]at(2-1){
$\begin{array}
[c]{c}%
8\\
2
\end{array}$
};
\node[below]at(3-1){
$\begin{array}
[c]{c}%
3\\
11
\end{array}$
};
\node[below]at(4-1){
$\begin{array}
[c]{c}%
16\\
4
\end{array}$};
\node[below]at(5-1){
$\begin{array}
[c]{c}%
6\\
22
\end{array}$
};
\node[below]at(6-1){
$\begin{array}
[c]{c}%
32\\
8
\end{array}$};
\node[below]at(7-1){$\begin{array}
[c]{c}%
11\\
45
\end{array}$
};
\node[below]at(8-1){$\begin{array}
[c]{c}%
64\\
16
\end{array}$
};
\node[below]at(9-1){$\begin{array}
[c]{c}%
22\\
90
\end{array}$
};
\node[below]at(10-1){
$\begin{array}
[c]{c}%
128\\
32
\end{array}$};
\node[below]at(11-1){
$\begin{array}
[c]{c}%
44\\
180
\end{array}$};
\node[right]at(12){
$\begin{array}
[c]{c}%
256\\
64
\end{array}$
};
\end{tikzpicture}
\subcaption{\label{figNTgame}12-node (Nagel-Tang)}
\end{minipage}
\caption{Centipedes with cake-size increasing exponentially}%
\end{figure}In the main text, we indicated the limit and empirical
distributions for both games. The raw data we use is the empirical
distribution over terminal nodes.\footnote{For the 6-node game, the
distribution is \{0, 3, 3, 9, 10, 3, 1\} out of 29 observations. For the
12-node game, we compute it from the ex ante distributions over choices
reported page 361 (Table I) of \citet{nagel98} for each player, i.e., \{0.005,
0.016, 0.054, 0.261, 0.331, 0.225, 0.108\} for player 1 and \{0.009, 0.017,
0.113, 0.331, 0.311, 0.143, 0.076\} for player 2.} Limit distributions have
similar dispersion as the empirical distribution, but they are shifted towards
higher dates. By raising $\beta$ above $\beta^{\ast}$, one obtains (unstable)
QR equilibria that induce more unravelling, with a mode that coincides with
data (see Figure \ref{figqrelargebeta}). However the prediction obtained is
too concentrated on the mode.

\begin{figure}[h]
\centering
\begin{minipage}{0.45\textwidth}
\centering
\includegraphics[scale=0.5]{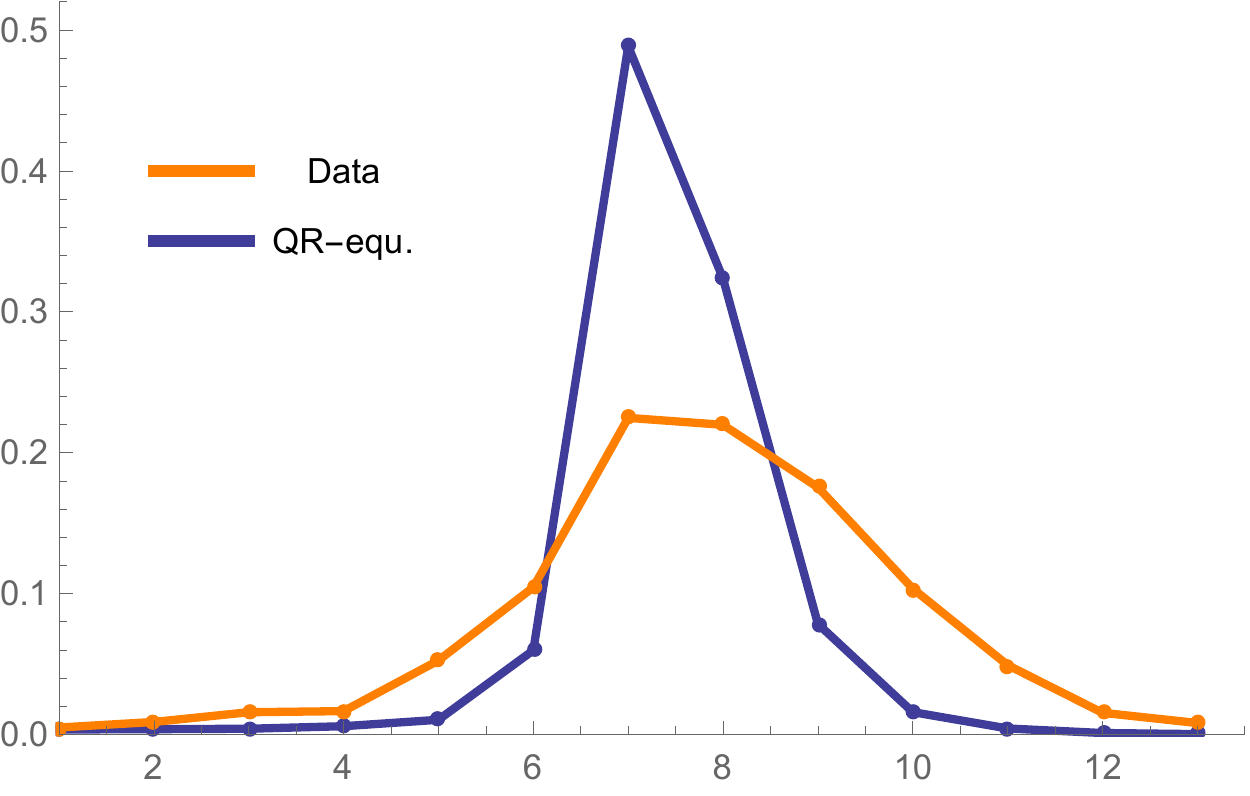}
\subcaption{\label{figntQRE} 12-node game}
\end{minipage}\hfill\begin{minipage}{0.45\textwidth}
\centering
	 \includegraphics[scale=0.5]{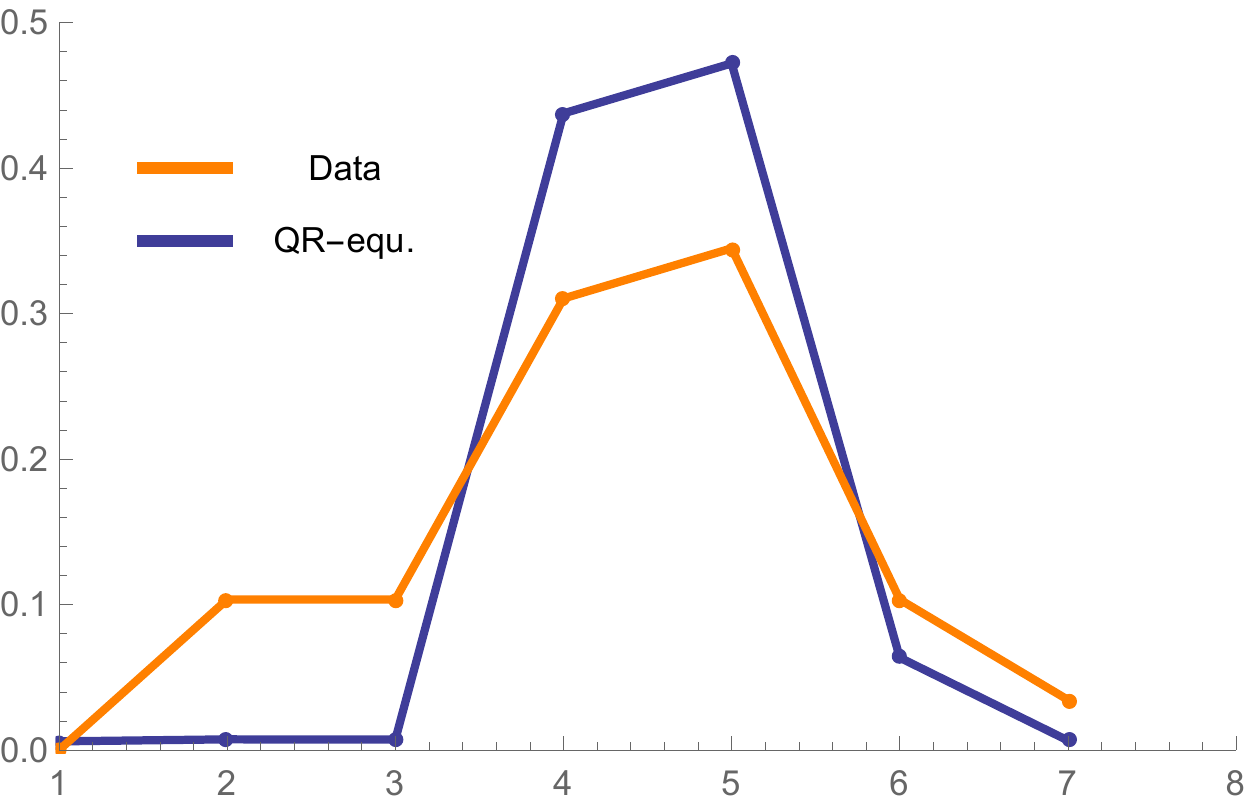}
\subcaption{\label{figmpQRE}6-node game}
\end{minipage}
\caption{QR equilibrium with larger $beta$}%
\label{figqrelargebeta}%
\end{figure}

Alternatively, one may force the last move of player 2 to be exit. In this
constrained strategy space, player 1 has $n$ moves, while player 2 has only
$n-1$ moves. We compute the limit distribution for that game, for each version
(MP and NT). The constraint, while not satisfied in the experimental data,
improves the fit for the rest of the distribution. \begin{figure}[h]
\centering
\begin{minipage}{0.45\textwidth}
\centering
\includegraphics[scale=0.5]{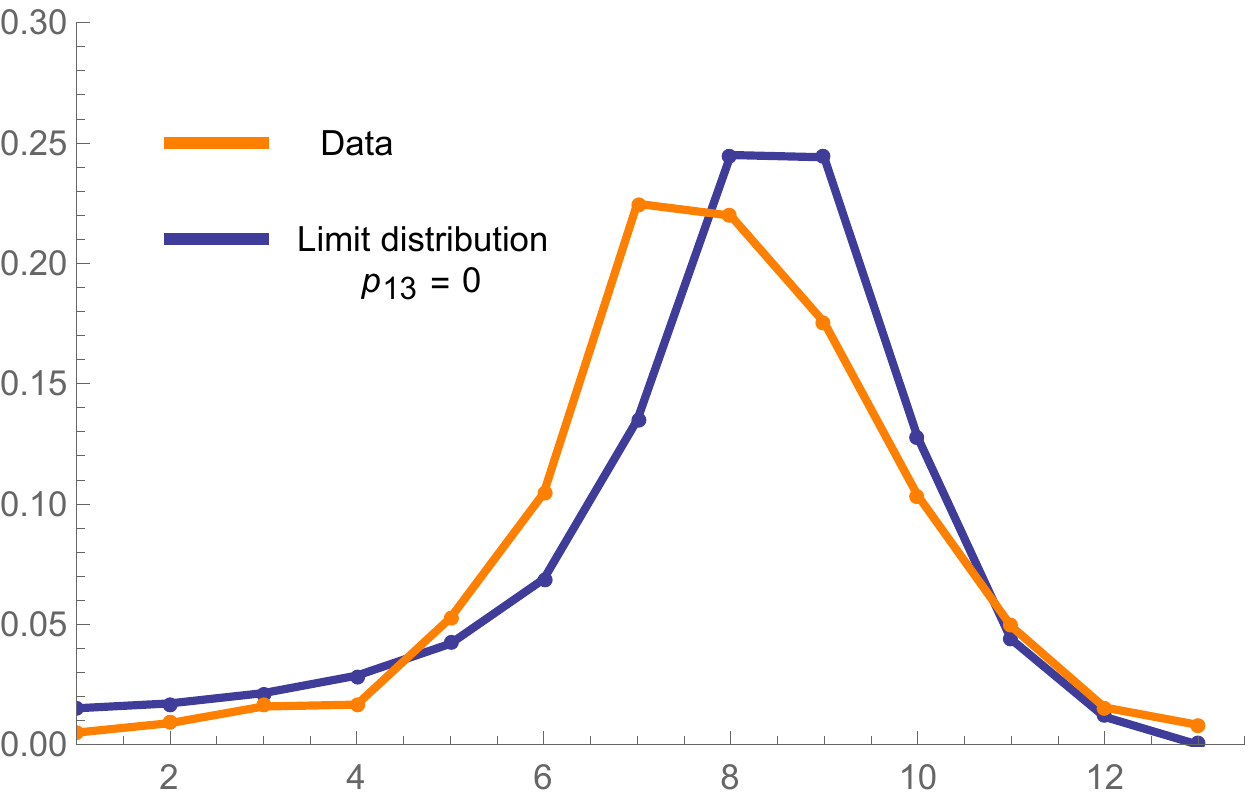}
\subcaption{\label{figntp13} 12-node game}
\end{minipage}\hfill\begin{minipage}{0.45\textwidth}
\centering
	 \includegraphics[scale=0.5]{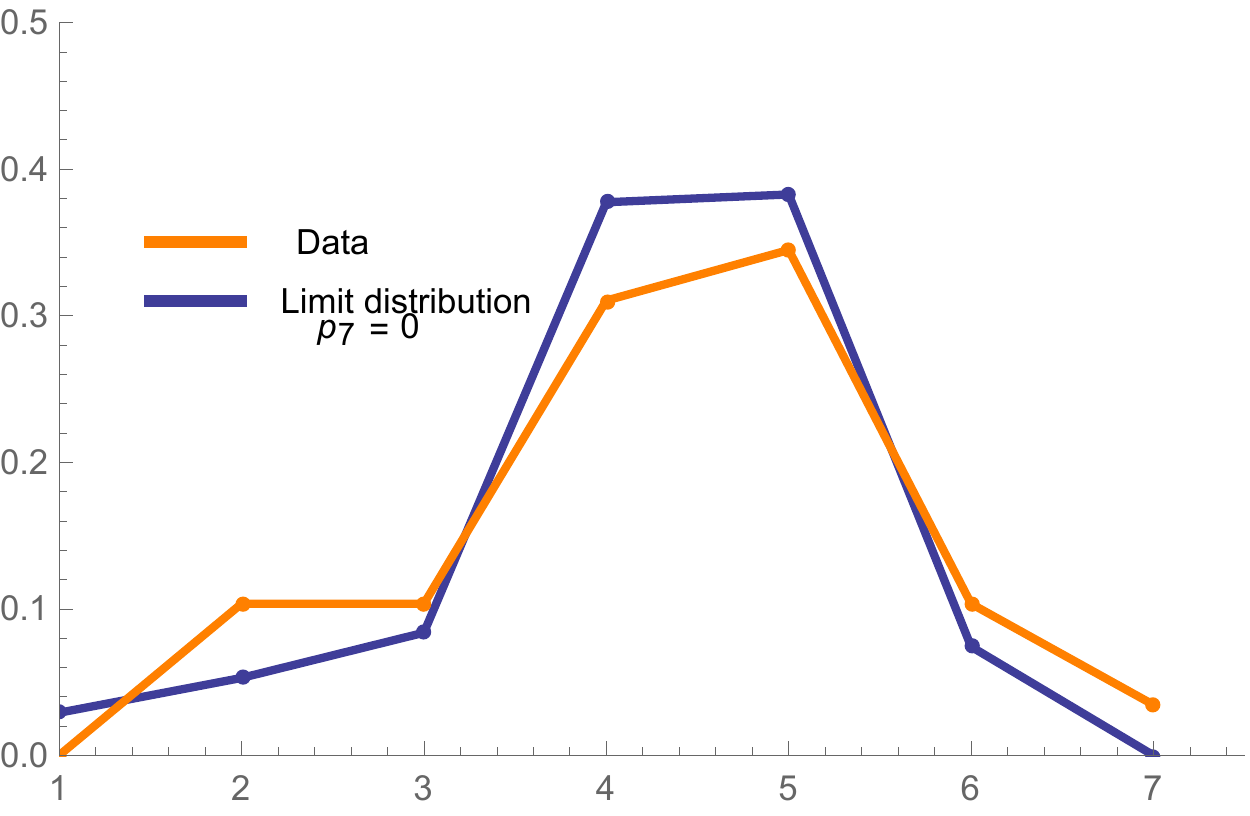}
\subcaption{\label{figmpp13}6-node game}
\end{minipage}
\caption{Limit distribution with last move constrained to exit}%
\label{figp13}%
\end{figure}

Finally, in the main text, given the empirical distributions over strategy
used by each player (in \citet{nagel98}), we computed the ex ante payoffs
associated with each exit time against the empirical distribution (see
Figure~\ref{expectedpayoffs}). We found that on average, given the dispersion
of exit times, subjects exit too early. The same exercise can be done for the
6-node game (MP).\footnote{We use the empirical distribution over nodes to
infer the ex ante probabilities of exit dates for each player.
\textquotedblleft no" corresponds to no exit throughout the game. We indicate
in bold the mode of the empiral distribution and the optimal exit date given
that distribution.} Player 1 exits at dates 1, 3, 5 or does not exit. Player 2
exits at dates 2, 4, 8 or does not exit. We use experimental data from
\citet{mckelvey92} (MP), \citet{kawagoe12} (KT), \citet{palacios09} (PVHs) and
\citet{levitt11} (SLS) and obtain again that subjects exit too early.

\begin{table}[h]
\caption{Expected gains against empirical distribution}%
\label{T1}%
\centering
\scalebox{0.9}{
\begin{tabular}
[c]{|cc||c|c|c|c||c|c|c|c|}\hline
& exit date & 1 & 3 & 5 & no & 2 & 4 & 6 & no\\\hline\hline
MP & \multicolumn{1}{|c||}{emp. dist.} & 0 & 11.5 & \textbf{63.2} & 25.3 & 10.4 &
35.1 & \textbf{40.9} & 13.6\\\hline
& \multicolumn{1}{|c||}{payoffs} & 40 & 145 & 379 & \textbf{510} & 80 & 287 &
\textbf{429} & 267\\\hline\hline
KT & \multicolumn{1}{|c||}{emp. dist.} & 2.3 & 2.3 & \textbf{62.7} & 32.6 & 2.3 &
7.1 & \textbf{69.7} & 20.9\\\hline
& \multicolumn{1}{|c||}{payoffs} & 40 & 156 & 585 & \textbf{764} & 78 & 306 &
\textbf{519} & 310\\\hline\hline
PVHs & \multicolumn{1}{|c||}{emp. dist.} & 7.5 & \textbf{41.8} & 40.6 & 10.1 &
16.2 & \textbf{59.1} & 24.6 & 0\\\hline
& \multicolumn{1}{|c||}{payoffs} & 40 & 137 & \textbf{208} & 129 & 74 & 179 &
\textbf{212} & 147\\\hline\hline
SLS & \multicolumn{1}{|c||}{emp. dist.} & 3.9 & 18.6 & \textbf{45.5} & 32 & 10.2 &
31.6 & \textbf{36.8} & 21.4\\\hline
& \multicolumn{1}{|c||}{payoffs} & 40 & 145 & 400 & \textbf{693} & 77 & 256 &
\textbf{490} & 285\\\hline
\end{tabular}
}\end{table}

\subsubsection*{Constant-size cakes, robustness to timing and experimental
data}

Following what we did for exponentially increasing cakes, we consider here the
effect of change in timing on the limit distribution. We also confront our
predictions to the data obtained in \citet{fey96} and \citet{kawagoe12} for
the constant-size centipede of Figure~\ref{payofffey}

\begin{figure}[h]
\centering
\begin{tikzpicture}[font=\footnotesize,scale=0.6]
\tikzstyle{solid node}=[circle,draw,inner sep=1.2,fill=black];
\tikzstyle{hollow node}=[circle,draw,inner sep=1.2];
\node(0)[hollow node]{}
child[grow=down]{node[solid node]{}
}
child[grow=right]{node(1)[hollow node]{}
child[grow=down]{node[solid node]{}
}
child[grow=right]{node(2)[hollow node]{}
child[grow=down]{node[solid node]{}
}
child[grow=right]{node(3)[hollow node]{}
child[grow=down]{node[solid node]{}
}
child[grow=right]{node(4)[hollow node]{}
child[grow=down]{node[solid node]{}
}
child[grow=right]{node(5)[hollow node]{}
child[grow=down]{node[solid node]{}
}
child[grow=right]{node(6)[solid node]{}
}
}
}
}
}
};
\foreach \x in {0,2,4}
\node[above]at(\x){1};
\foreach \x in {1,3,5}
\node[above]at(\x){2};
\node[below]at(0-1){$\begin{array}
[c]{c}%
16\\
16
\end{array}$};
\node[below]at(1-1){$\begin{array}
[c]{c}%
12\\
2
\end{array}$};
\node[below]at(2-1){$\begin{array}
[c]{c}%
23\\
9
\end{array}$};
\node[below]at(3-1){$\begin{array}
[c]{c}
6.8\\
25.2
\end{array}$};
\node[below]at(4-1){
$\begin{array}
[c]{c}%
26.9\\
5.1
\end{array}$
};
\node[below]at(5-1){
$\begin{array}
[c]{c}%
3.8\\
28.2
\end{array}$
};
\node[right]at(6){
$\begin{array}
[c]{c}%
29.2\\
2.8
\end{array}$
};
\end{tikzpicture}
\caption{6-node constant-size centipede (\citet{fey96})}%
\label{payofffey}%
\end{figure}

\begin{figure}[h]
\centering
\begin{minipage}{0.45\textwidth}
\centering
\includegraphics[scale=0.5]{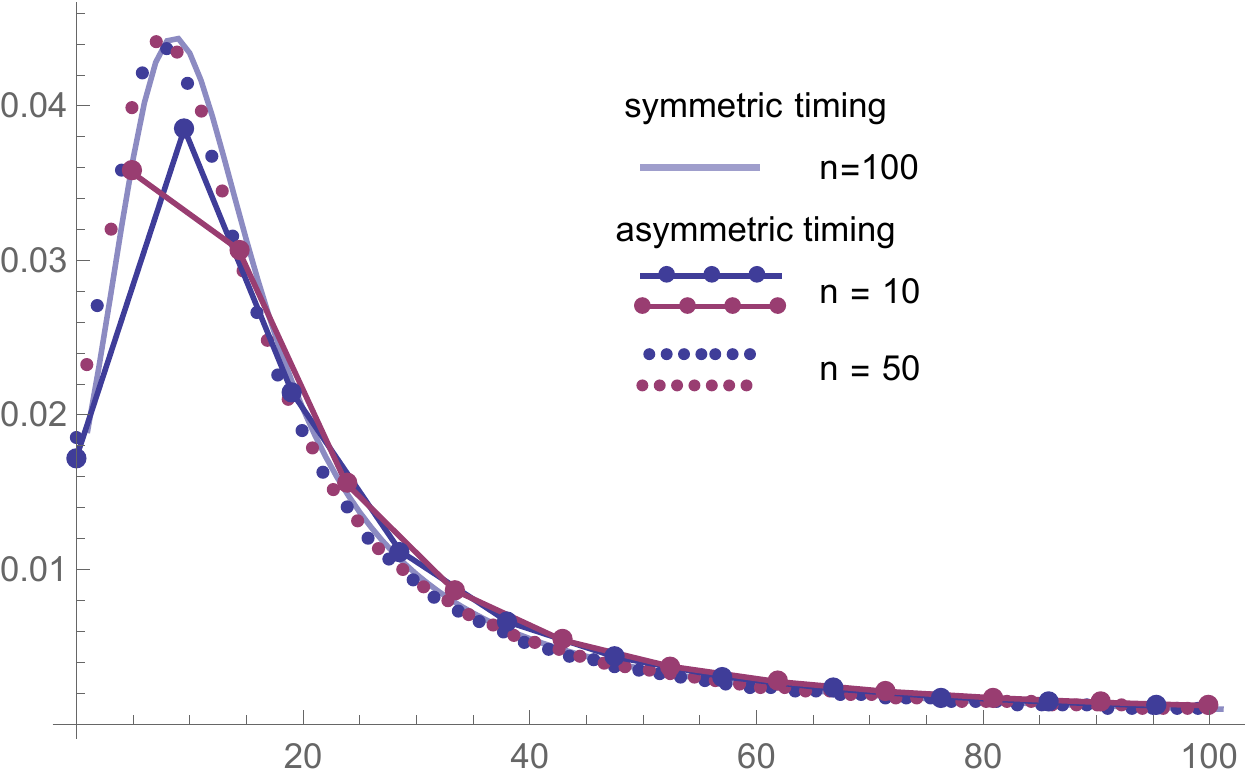}
\subcaption{\label{figfeytiming} Robustness to timing}
\end{minipage}\hfill\begin{minipage}{0.45\textwidth}
\centering
	 \includegraphics[scale=0.5]{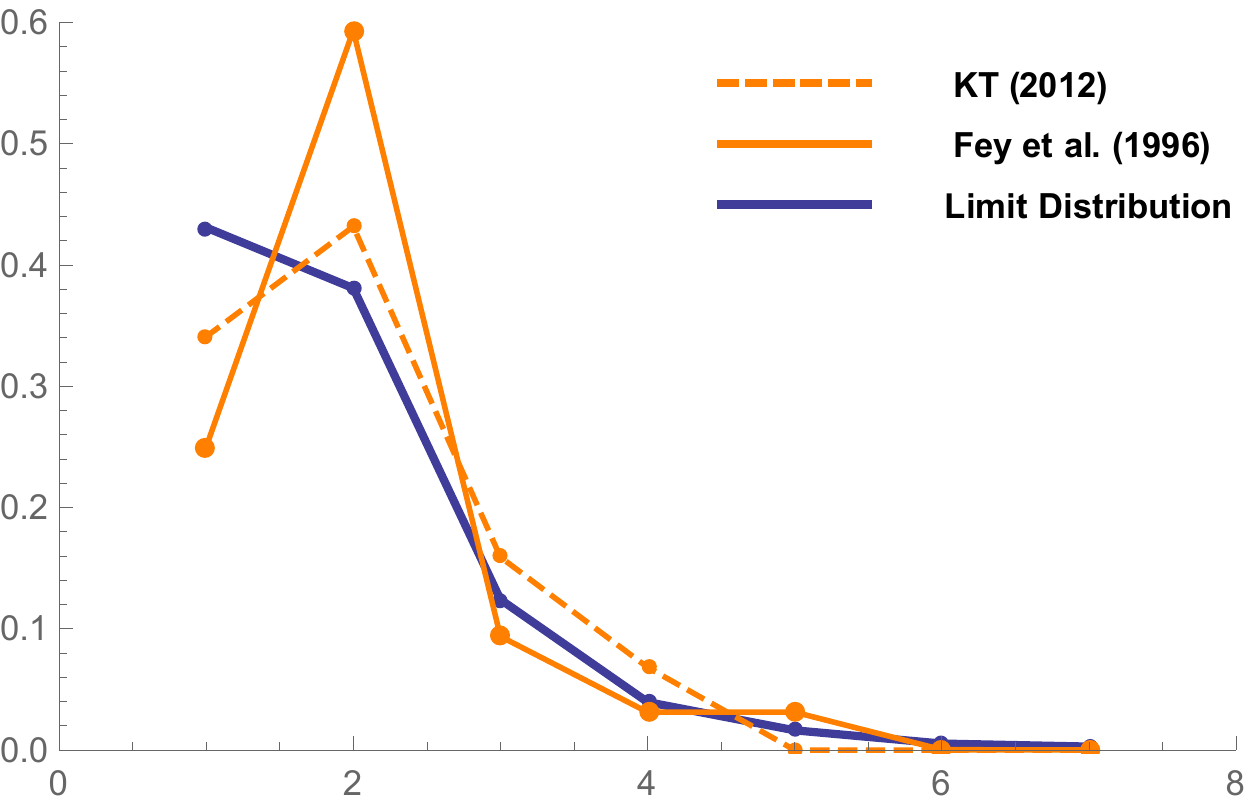}
\subcaption{\label{figdatafey}Data versus limit distribution}
\end{minipage}
\caption{Centipede with constant size cake}%
\label{figfeyappendix}%
\end{figure}

Following the steps of Section \ref{otherparametric}, we consider games where
players alternate in making decisions, and we vary the number of nodes,
without varying the payoff structure of the game. Formally, we consider $2n$
exit opportunities ($n$ for each player), and we set
\[
T_{1}=\{0,2\rho,4\rho,...(n-1)\rho\}\text{ and }T_{2}=\{\rho,3\rho
,..,n\rho\},
\]
with $\rho=\overline{\tau}/n$. We set the sharing-rule parameter so as it fits
with the experimental game, which corresponds to setting $\alpha=2$.
Figure~\ref{figfeytiming} reports limit distributions for each player for
$2n=100$ and $2n=20$ (player 1 is in Blue). We also indicate on that figure
the limit distribution of the symmetric game for $n=100$. 

Figure~\ref{figdatafey} reports the limit distribution and the data obtained
by the \citet{fey96} and \citet{kawagoe12}. The reason why the limit
distribution has a mode at $t=0$ (unlike what we report in Figure
\ref{figfeytiming}) is that when $n$ is small, incentives to undercut are
strong even for low dates. In both experiments, there is a mode on the second
date. Note however that given the data reported, player 1 has an incentive to
exit at the first date.\footnote{The distribution of exit dates are
respectively $(8,16,3,1,1,0)$ (out of 29) and $(15,19,7,3,0,0,0)$ (out of 44),
so waiting date 3 for player 1 yields strictly less than 16 in both cases.}
%

\subsubsection*{The 11-20 game. Costless iteration version with $t_{i}%
\in\{110,...,200\}$.}

We consider here the costless iteration version with a large strategy set.
Recall that player $i$ gets 20 when $t_{i}=200$. Otherwise, for any
$t_{i}<200$, payoffs are:%
\[
v_{i}(t_{i},t_{j})=17+20(1-\alpha(t_{j}-t_{i}-1))\text{ if }t_{j}-10\leq
t_{i}<t_{j}\text{ and }v_{i}(t_{i},t_{j})=17\text{ otherwise.}%
\]
Setting a weight $\pi=0.1$ on the non-strategic type, we report the limit
distributions over strategic claims (claims strictly below 200) for both
variants. Shapes are similar, though modes are more pronounced under variant 2.

\begin{figure}[h]
\centering
\begin{minipage}{0.45\textwidth}
\centering
\includegraphics[scale=0.65]{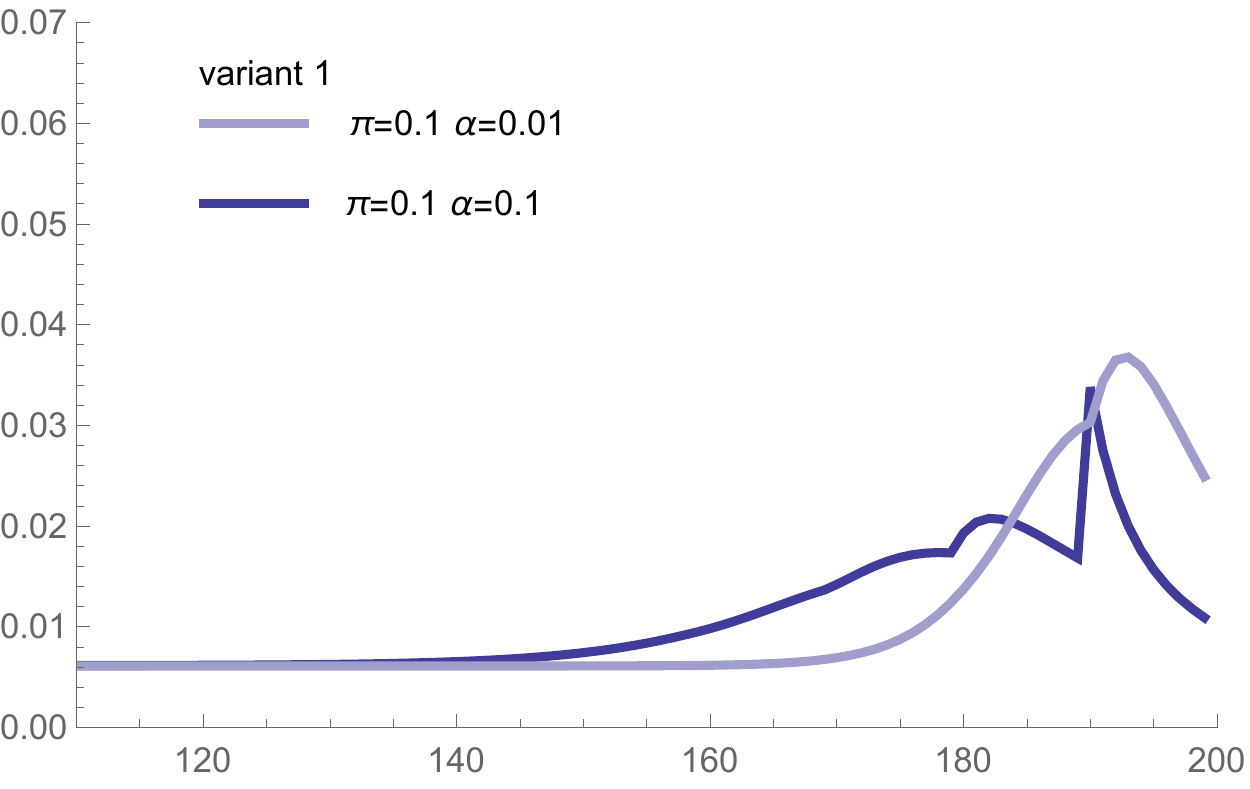}
\subcaption{\label{figcostlessLARGEvariant1}variant 1}%
\end{minipage}\hfill\begin{minipage}{0.45\textwidth}
\centering
	\includegraphics[scale=0.65]{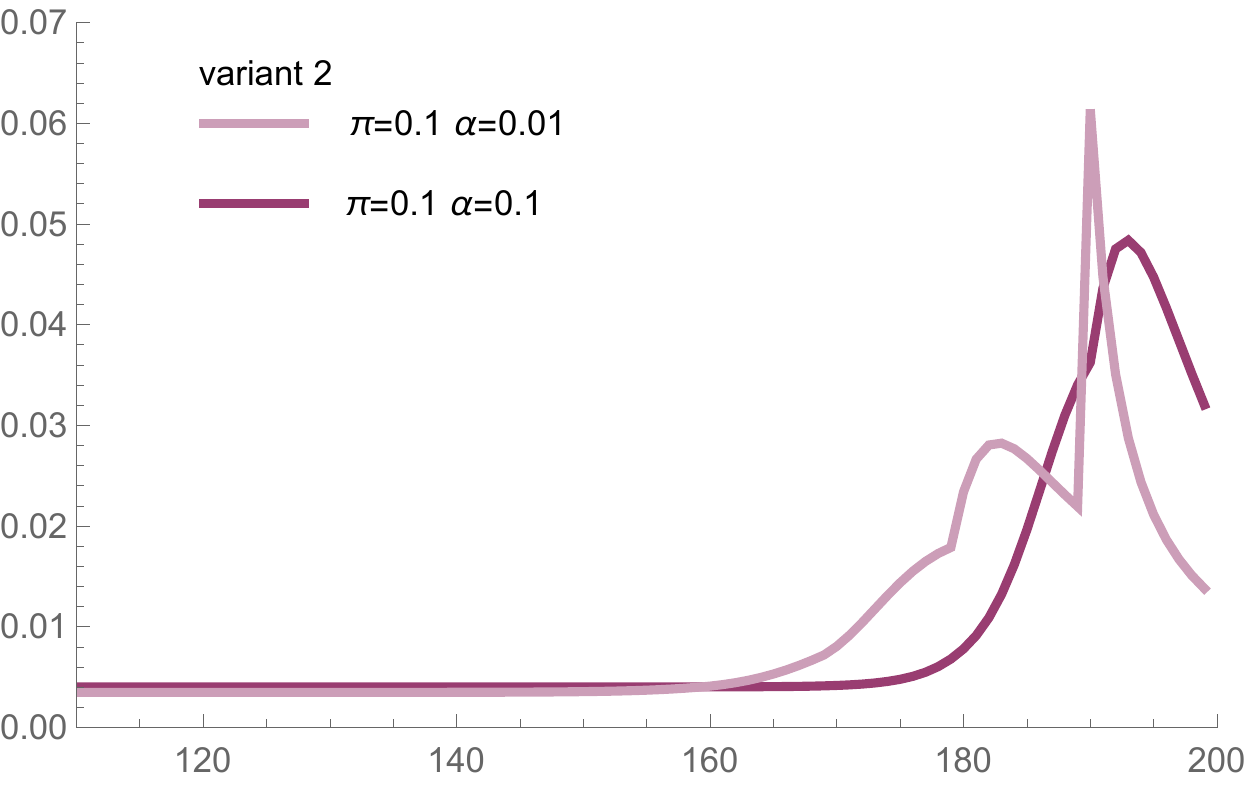}
\subcaption{\label{figcostlessLARGEvariant2}variant 2}%
\end{minipage}
\caption{The costless iteration version with large strategy sets}%
\label{figARtimed}%
\end{figure}

\subsubsection*{Auctions}

\paragraph{First price with dispersion uncertainty.}

\begin{figure}[h]
\centering
\begin{minipage}{0.45\textwidth}
\centering
	\includegraphics[scale=0.6]{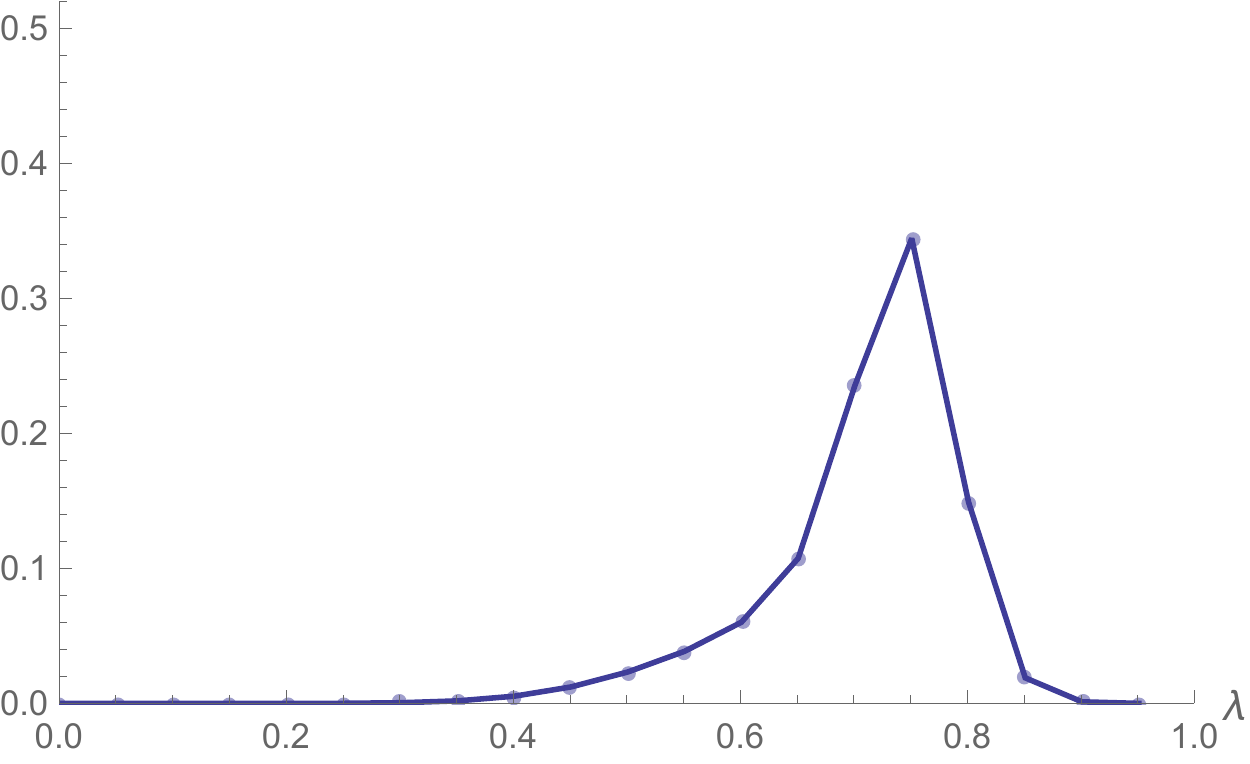}
\subcaption{\label{figdispa}Without trembles}%
\end{minipage}\hfill\begin{minipage}{0.45\textwidth}
\centering
	\includegraphics[scale=0.6]{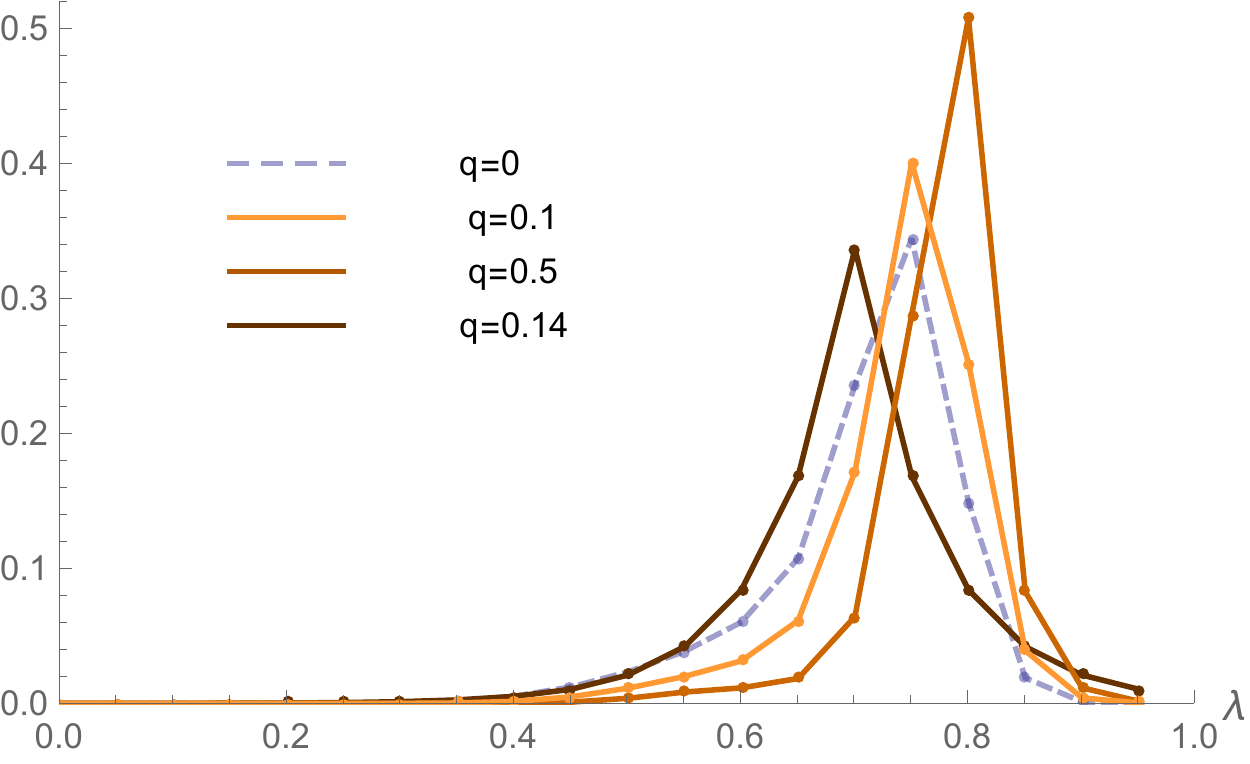}
\subcaption{\label{figdispc}With trembles}%
\end{minipage}
\caption{Limit distributions under dispersion uncertainty}%
\label{fig9}%
\end{figure}

We assume that the dispersion parameter is uncertain, say $\underline{\sigma}$
or $\overline{\sigma}$ with equal probability. We set $(\underline{\sigma
},\overline{\sigma})=(0.05,0.5)$. The game has a pure strategy equilibrium
($\lambda^{\ast}=0.8$) but the limit precision is bounded.
Figure~\ref{figdispa} reports the limit distribution obtained. Weights are
mostly spread over $\{0.6,0.65,...,0.8\}$, these strategies cumulating $86\%$
of the weights. This endogenous uncertainty weakens competition (compared to
the Nash outcome), with an expected shading level at the limit distribution
equal to $0.71(<0.8)$.

For the game defined over target shading factors $\lambda_{\kappa}$, we find
(as in our previous applications) that exogenous noise improves the stability
of the logit-response dynamics. This raises the limit precision, which is
pro-competitive. For $q\geq q^{\ast}(=0.15)$, limit precision is unbounded and
the limit~QRE of the game over targets converges to a Nash equilibrium of that
game (with $\lambda_{\kappa^{\ast}}=0.8$ for $q=q^{\ast}$). As $q$ increases
above $q^{\ast}$, noise is anti-competitive. Figure~\ref{figdispc} reports the
induced distribution over shading factors for $q=0.1,0.14$ and $0.5$,
illustrating the two effects above (pro-competitive due to an endogenous rise
in precision for $q=0.14$, and anti-competitive for large $q$).


\paragraph{All pay.}

\textit{Bayesian case.} By revenue equivalence, equilibrium bids of the
(unconstrained) Bayesian game are such that the utility $u(v)$ of type $v$
coincides with that obtained in a second price auction, so
$u(v)=vF(v)-b(v)=F(v)(v-E[v_{2}|v_{2}<v])$, where $v_{2}$ is the value of the
other player. This immediately gives the Bayesian solution $b^{eq}%
(v)=F(v)E[v_{2}|v_{2}<v])$

\textit{Coarse strategy space that includes the Bayesian solution}.
Table~\ref{tab3} reports expected gains obtained for the strategy set
$\overline{\Gamma}=\Gamma_{0.1}\cup\{b_{0.3}^{eq}\}$ when $\sigma=0.3$. By
inspection, one sees that $b_{0.3}^{eq}$ is the unique symmetric pure strategy
equilibrium of this game.

\begin{table}[h]
\caption{All pay. Payoff matrix. $\delta=0.1$, $\sigma=0.3$, $b_{S}%
=b_{0.3}^{eq}$}%
\label{tab3}%
\centering
\scalebox{0.8}{
\begin{tabular}{c|ccccccccccc}
$ k_1\backslash k _2$ & 0 & 1 & 2 & 3 & 4 & 5&6&7&8&9&$b_S$ \\\hline
0& 0.61 & 0. & 0. & 0. & 0. & 0. & 0. & 0. & 0. & 0. & 0. \\
1& 0.94 & 0.51 & -0.01 & -0.09 & -0.09 & -0.1 & -0.1 & -0.1 & -0.1 & -0.1 & 0.08 \\
2& 0.84 & 0.8 & 0.4 & 0.03 & -0.12 & -0.17 & -0.19 & -0.2 & -0.2 & -0.2 & 0.12 \\
3& 0.73 & 0.73 & 0.6 & 0.3 & 0.02 & -0.14 & -0.22 & -0.27 & -0.29 & -0.3 & 0.14 \\
4& 0.63 & 0.63 & 0.59 & 0.43 & 0.19 & -0.01 & -0.17 & -0.27 & -0.33 & -0.36 & 0.16 \\
5&0.52 & 0.52 & 0.51 & 0.44 & 0.28 & 0.09 & -0.08 & -0.22 & -0.32 & -0.39 & 0.16 \\
6&0.42 & 0.42 & 0.41 & 0.38 & 0.29 & 0.14 & -0.01 & -0.16 & -0.28 & -0.38 & 0.15 \\
7&0.31 & 0.31 & 0.31 & 0.3 & 0.25 & 0.15 & 0.02 & -0.11 & -0.24 & -0.35 & 0.12 \\
8&0.21 & 0.21 & 0.21 & 0.2 & 0.17 & 0.11 & 0.01 & -0.1 & -0.22 & -0.33 & 0.07 \\
9&0.1 & 0.1 & 0.1 & 0.1 & 0.09 & 0.05 & -0.01 & -0.11 & -0.21 & -0.32 & 0.01 \\
$b_S$& 0.61 & 0.51 & 0.4 & 0.3 & 0.19 & 0.09 & 0. & -0.1 & -0.18 & -0.25 & 0.18 \\
\end{tabular}
}\end{table}The limit distribution puts a weight only equal to 19\% on
$b_{0.3}^{eq}$, while strategies $\lambda=0.4,0.5,0.6$ and $0.7$ respectively
get weights $9\%,13\%$, $15\%$ and $12\%$.

\subsection*{Mixed strategy equilibria and local instability.}

We consider a general symmetric game defined over strategy set $\overline{S}$,
having a mixed strategy equilibrium $p^{\ast}$ with non-degenerate support
$S$. We assume that this mixed strategy equilibrium is isolated and that all
strategies not in the support are strictly worse. Consider the branch of the
QRE graph ending in $p^{\ast}$. For $\beta$ large enough, the QRE equilibria
on that branch put negligible weight outside $S$, and strictly positive weight
at least equal to $\underline{p}>0$ on strategies in $S$.

We show that for $\beta$ large enough, the QRE equilibria on that branch are
locally unstable. As a corollary, we obtain that generically, if a game has no
pure strategy equilibria, limit precision is necessarily bounded. It also
implies that, generically, a limit distribution cannot approach a mixed
strategy equilibrium.

Let $V=(v_{kh})_{k,h\in\overline{S}}$ be the payoff matrix. Let $V_{k}%
=(v_{kh})_{h\in S}$ the raw vectors for $k\in\overline{S}$. Let $M_{k}\equiv
V_{k}-V_{K}$, where $K\in S$. Our mixed strategy equilibrium $p^{\ast}$ solves
$M_{k}.p^{\ast}=0$ for all $k\in S$, $\sum_{k\in S}p_{k}=1$ (and
$M_{k}.p^{\ast}<0$ for all $k\notin S$).

Next define $X=\{x\in\mathcal{R}^{S},\sum_{k\in S}x_{k}=0$ and $\sum_{k\in
S}|x_{k}|=1\}$ and $\varpi\equiv\min_{x\in X}\max_{k\in S\backslash K}%
|M_{k}.x|$. If $\varpi=0$, there exists $x\in X$ s.t. $M_{k}.x=0$ for all
$k\in S$. This implies that any $p^{\ast}+\delta x$ is an equilibrium for any
$\delta$ sufficiently small, contradicting the assumption that $p^{\ast}$ is
isolated. So $\varpi>0$.

Consider now $\beta_{0}$ large and the QRE branch starting at $p^{0}\in
\Sigma_{\beta_{0}}$ and ending at $p^{\ast}$. We examine the logit response
$p^{\prime}=\phi_{\beta_{0}}(p)$ for $p=p^{0}+\delta x$ where $x\in X$ and
$\delta$ small. We choose $\delta_{0}$ small enough, and $\beta_{0}$ large
enough so that $p^{\prime}$ puts negligible weight on strategies outside $S$
for all $\delta\leq\delta_{0}$ and $x\in X$, and weight at least equal to
$\underline{p}>0$ on any strategy in $S$ along the branch.

Let $x^{\prime}=p^{\prime}-p^{0}$. For $\delta$ very small, $p^{\prime}$ is
close to $p^{0}$, so we have, using the definition of $\phi_{\beta_{0}}$,
\[
\frac{x_{k}^{\prime}}{p_{k}^{0}}-\frac{x_{K}^{\prime}}{p_{K}^{0}}\simeq
\ln\frac{p_{k}^{\prime}}{p_{k}^{0}}-\ln\frac{p_{K}^{\prime}}{p_{K}^{0}}%
=\ln\frac{p_{k}^{\prime}}{p_{K}^{\prime}}-\ln\frac{p_{k}^{0}}{p_{K}^{0}}%
=\beta_{0}M_{k}.(p^{\prime}-p^{0})=\beta_{0}\delta M_{k}.x
\]
implying $\beta_{0}\delta\varpi\leq(|x_{k}^{\prime}|+|x_{K}^{\prime
}|)/\underline{p}$, hence $|x_{k}^{\prime}|\geq\beta\delta\varpi
\underline{p}/2$ for some $k\in S$. Choosing $\beta_{0}>2/(\varpi
\underline{p})$, the logit-response dynamics cannot converge as for any $p$ at
$L1$-distance $\delta\leq\delta_{0}$ from $p^{0}$, $\phi_{\beta_{0}}(p)$ is at
a larger $L1-$distance from $p^{0}$.

\end{document}